\documentclass[12pt,a4paper]{article}
\usepackage{geometry}
\usepackage[round]{natbib}
\usepackage{mathtools}
\allowdisplaybreaks
\usepackage{mathrsfs}
\usepackage{amsfonts}
\usepackage{amssymb}
\usepackage{graphicx}
\usepackage{placeins}
\usepackage{stmaryrd}
\usepackage{lipsum}  
\usepackage[labelfont=bf]{caption}
\usepackage[skip=-0.25\textwidth]{subcaption}
\usepackage{enumitem}
\usepackage{titlesec}
\titleformat{\section}
{\normalfont\fontsize{12}{15}\bfseries}{\thesection .}{0.3em}{}
\titleformat{\subsection}
{\normalfont\fontsize{12}{15}\itshape\bfseries\centering}{\upshape\thesubsection .}{0.3em}{}
\titleformat{\subsubsection}
{\normalfont\fontsize{12}{15}\itshape\bfseries}{\upshape\thesubsubsection .}{0.3em}{}

\usepackage{color}
\usepackage[titletoc,title]{appendix}
\usepackage{xpatch}
\usepackage{placeins}
\renewenvironment{abstract}  %To change abstract width
{\normalfont
	\begin{center}
		\bfseries \abstractname\vspace{-.5em}\vspace{0pt}
	\end{center}
	\list{}{%
		\setlength{\leftmargin}{0mm}% <---------- CHANGE HERE
		\setlength{\rightmargin}{\leftmargin}%
	}%
	\item\relax}
{\endlist}  %To change abstract width
\def\equationautorefname~#1\null{%
	(#1)\null
}   % do \autoref{eq:label_name} to get no equation~ before number
\xpretocmd{\eqref}{equation\,}{}{}

\usepackage{authblk}

\title{\textbf{Electrorheology  of a  dilute emulsion of  surfactant-covered drops}}
\author[1]{Antarip Poddar}
\author[1,2]{Shubhadeep Mandal}
\author[1]{Aditya Bandopadhyay\thanks{Email: aditya@mech.iitkgp.ernet.in}}
\author[1]{Suman Chakraborty\thanks{Email: suman@mech.iitkgp.ernet.in}}
\affil[1]{Department of Mechanical Engineering, Indian Institute of Technology Kharagpur, Kharagpur, West Bengal - 721302, India}
\affil[2]{Max Planck Institute for Dynamics and Self-Organization, Am Fassberg 17, D-37077 G\"{o}ttingen, Germany}
\date{}         
\begin{document}
			\maketitle
	\begin{abstract}
\noindent
The effects of surfactant coating on a deformable viscous drop under the combined action of a shear flow and a uniform electric field, are investigated by solving the coupled equations of electrostatics, fluid flow and surfactant transport. Employing a comprehensive three-dimensional solution technique, the non-Newtonian shearing response of the bulk emulsion is analyzed in the dilute suspension regime. The present results reveal that the surfactant non-uniformity creates significant alterations in the flow disturbance around the drop, thereby influencing the  viscous dissipation from the flowing emulsion. This, in effect, triggers changes in the bulk shear viscosity. It is striking to observe that the balance between  electrical and hydrodynamic stresses is affected in such a way that surface tension gradient on the drop surface vanishes for some specific shear rates and the corresponding effective change in the bulk viscosity becomes negligible too. This critical condition hugely depends on the electrical permittivity and conductivity ratios of the two fluids and orientation of the applied electric field. Also the physical mechanisms of charge convection of surface deformation play their role in determining this critical shear rate. The charge convection instigated shear thinning or shear thickening behavior of the emulsion gets reversed due to a coupled interaction of the charge convection and Marangoni stress. In addition, the electrically created anisotropic normal stresses in the bulk rheology,  get reduced due to the presence of surfactants, especially when the drop viscosity is much lesser than the continuous fluid. A thorough description of the drop-level flow physics and its connection to the bulk rheology of a dilute emulsion, may provide a fundamental understanding of a more complex emulsion system. 
	\end{abstract}
\par\noindent\quad\rule{0.95\textwidth}{0.4pt}

\section{Introduction}

An immiscible dispersion of liquid drops in another liquid medium is ubiquitous in diverse domains of practical interest, ranging from food and material processing,  pharmaceutical industries to petroleum refinery \citep{Eow2002,Barnes1994}.  These dispersions, commonly known as emulsions, can also be generated in the controlled environments of the state-of-the art microfluidic techniques \citep{Anna2003,Haliburton2017}.   
Variations in the microstructure and morphology of the individual drops in the dispersed phase may lead to drastic alterations in the bulk physicochemical properties of the emulsion \citep{Tucker2002,Puyvelde2001,Lequeux1998}.
In addition, the  application of external stresses has proved to be immensely useful in modifying the distribution, deformation and motion of the individual constituents of the emulsion. 

In suspension mechanics, the bulk scale description of the dispersion is often provided by considering a continuous effective fluid. Now the disturbance in flow due to the stresses exerted on  the drops, affects the dissipation of energy by the continuous fluid. Consequently the stress-response of the bulk mixture deviates from that of a Newtonian fluid \citep{Guazzelli2011}.   \citet{Taylor1932} was the first to theoretically predict the contribution of drop-phase in increasing  the macroscale viscosity of a dilute emulsion under an imposed linear flow. Later, it was found that the  deformable nature of the suspended drops can not only impart an elastic nature to the emulsion rheology \citep{Schowalter1968} but also makes it exhibit a shear-thinning property \citep{Barthes-Biesel1973}. The nature of drop break-up and deformation under different linear flow fields was also characterized  \citep{Rallison1984,Bentley1986}.
It was observed that under a two dimensional linear shear flow, the drop makes an angle of $ \pi/4 $ with the imposed flow direction and deforms into an ellipsoidal shape when the  of surface tension is high compared to the flow strength. Further, the rotational component of the shear flow tries to reorient the drop.

One of the most promising  techniques regarding the  manipulation of rheological  properties of an emulsion is the application of electric field \citep{Pan1997,Ha2000,Na2009}. Electrohydrodynamic effects can deform a drop in an oblate or prolate spheroidal shape, depending on the combination of the electrical properties of the dispersed and  continuous phase \citep{Taylor1966,Xu2006,Esmaeeli2011,Thaokar2012,Lanauze2015}. In the case of leaky dielectric fluids, the electric field also induces a tangential flow around the drop \citep{Saville1997}. The electrohydrodynamic flow and drop deformation get further affected by the presence of finite charge convection at the interface \citep{Feng1999,Xu2006a,Yariv2016,Das2016}.  The significance of a tilted electric field configuration in breaking the fore-aft symmetry of a  sedimenting drop was  experimentally shown by \citet{Bandopadhyay2016}. Later, they also observed that a similar phenomenon influences  the migration characteristics of a drop in Poiseuille flow \citep{Mandal2016}.  The coupling of an external flow field and electrohydrodynamics can exhibit various non-intuitive flow physics in and around an individual drop \citep{Allan1962,Maehlmann2009} and thus has the ability to drastically alter the bulk rheology of emulsions \citep{Ha2000,Fernandez2008}. 
\citet{Vlahovska2011} established a theoretical understanding of the impact of drop-level flow dynamics and deformation on the electrorheology of a dilute emulsion. They considered a uniform electric field in the direction of velocity gradient and analyzed the combined interplay among the flow and electric field in the limits of weak flow and highly viscous drops. The charge convection was neglected in the weak flow analysis. The orientation of the deformed drop along the flow direction was found responsible for the shear thinning effect due to reduction in the flow resistance. Similarly the emulsion shows shear thickening property when the drop elongates perpendicular to the direction as mentioned above. Apart from the shear rate dependent rheology, the electric field induced compression or elongation of drops causing an extra tension along the streamlines. As a result, the suspension possess normal stress differences, similar to polymeric liquids \citep{Bird1987}. Later \citet{Mandal2017b} brought out the role of charge convection using an asymptotic analysis. 
Subsequently they showed \citep{Mandal2017a}  that application of an electric field in the shearing plane and along a tilted direction to that of the flow, can give rise to many non-intuitive bulk rheological properties of the emulsion, especially the existence of an electrical component of effective shear viscosity in the absence of drop deformation, which makes it deviate significantly from the result of \citep{Taylor1932}. 

In many industrial applications the emulsion stabilizers act as a surface-active agents and reduce the interfacial tension after being adsorbed at the interface \citep{Fischer2007}. Under dynamic conditions, these surface active agents or surfactants bring Marangoni stress in the picture and complicates the interfacial stress balance. Experiments in the literature  have revealed the significant role of surfactants on the deformation of drops and the bulk rheology of an emulsion  \citep{Velankar2001,Velankar2004,Jeon2003,Hu2003}. To analyze the specific physical phenomena responsible for such behavior, both analytical and numerical investigations have been performed \citep{Stone1990a,Milliken1993,Li1997,Vlahovska2005,Vlahovska2009}. These analyses show that the surfactant effects are mostly governed by the  competing processes of surface convection,  dilution and tip streaming. Depending on these mechanisms, the local surface tension on the drop varies up to a great extent, resulting in either suppression or enhancement of drop deformation. Several factors, such as the viscosity contrast of the drop-matrix fluid pair, local gradient in surface tension, initial surface tension and the flow strength, control the dominance among the said mechanisms. 

The effect of surfactants on the electrically actuated drops, is a less studied domain in the literature \citep{Ha1995,Teigen2010,Nganguia2013,Poddar2018,Ervik2018}. The electrorheology of concentrated emulsions  in the presence of non-ionic surfactants has only been experimentally investigated by \citet{Ha1999}. However, a theoretical understanding of the physical processes associated with drop scale flow physics and its connection to the bulk rheology of the effective fluid mixture, are missing in the literature.  Thus in the present work we attempt to unveil the complicated interconnection between the surfactant covering on the drop surface, the electric field and the imposed linear shear flow, towards influencing the effective rheology. Considering non-ionic, bulk insoluble surfactant coating on the surface of a deformable  viscous drop, dispersed in a electrically actuated emulsion, we adopt a  comprehensive three-dimensional framework and analyze the problem through an asymptotic approach.  Our results indicate that under the combined action of a tilted electric field and linear shear, the surfactant-induced Marangoni stresses modify the flow field around the drop in a diverse fashion, which cannot be predicted by a simple linear superposition of the governing processes. Consequently the bulk emulsion viscosity can either increase or decrease, depending on the choice of the ratio of dielectric permittivity and electrical conductivity combination of the two mediums. We also observed the unique existence of a critical shear rate relative to the electrical stresses, where the Marangoni effects vanish. Moreover, depending on the interplay between the charge convection and Marangoni effects,  the emulsion rheology may become either shear thinning or shear thickening. Both the mechanisms of charge convection and small shape deformation of drops, have the ability to shift the critical shear  rate observed in the leading order of case. Finally, depending on the tilt angle of the applied electric field, the elastic behavior of the emulsion, interpreted through normal stress differences, gets significantly altered due to the presence of interfacial tension gradients brought in by the surface surfactant distribution.

\section{Physical description}
\label{sec:physical}
\begin{figure}[!htb]	
	\centering
	\hspace{-5ex}
	\includegraphics[width=0.80\textwidth]{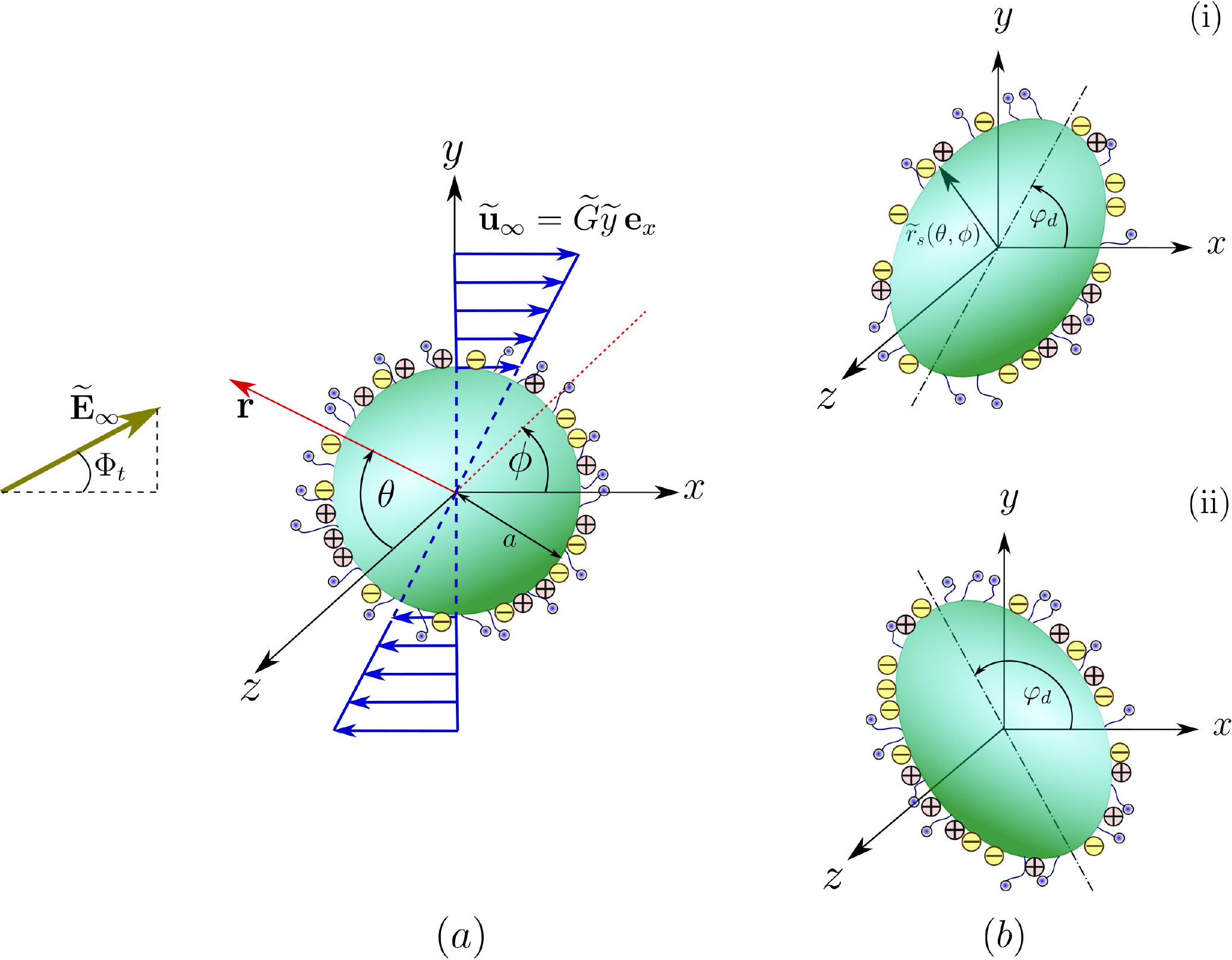}
	\vspace{-0pt} 
	\caption{Schematic description of a deformable viscous drop being acted upon by a tilted electric field and background linear shear flow. The drop is covered with non-ionic surfactant molecules and interfacial charges accumulate on its surface. In sub-figure (a) the initial undeformed condition of the drop is shown. Under the action of the external fields the drop may deform with an inclination angle $ \varphi_d $ as shown in sub-figure (b). Two of the many possible configurations with $ \varphi_d < \pi/2 $ and $ \varphi_d>\pi/2 $ are shown as configuration (i) and (ii), respectively. } 
	\label{fig:schematic}
\end{figure}

In the present work, we consider a model emulsion where a drop of radius `$ a $' is dispersed in a background linear shear flow of the continuous phase along the $ x $ direction given by $ \widetilde{u}_\infty=\widetilde{G}\widetilde{y} $. Here $ \widetilde{G} $ is the uniform shear rate on the external fluid flow in the far stream.  The drop is being acted upon by a uniform magnitude DC electric field $ \mathbf{\widetilde{E}}_\infty $ with a tilt angle $\Phi_t$ in the plane of shear. In addition, the drop surface is covered with non-ionic surfactant molecules.  As a first step towards understanding the drop scale flow dynamics, we neglect the interaction between the drops and assume that the drop phase volume fraction $ \nu $ to be in the dilute limit $ (\nu \ll 1$, where $ \nu $ is the drop phase concentration in the emulsion) and analyze the problem in a steady state condition. The scenario is shown schematically in figure~\ref{fig:schematic}(a). When the drop surface is uncontaminated the drop-matrix interface has a uniform surface tension  $ \widetilde{\gamma}_{cl}$. The presence of surface active agents on this interface lowers this surface tension and in the quiescent condition the surfactants has a uniform surface concentration of $ \widetilde{\Gamma}_{eq}$. Under the simultaneous action of the background flow and the electric field the equilibrium distribution of the surfactants gets disrupted and the concentration distribution in the dynamic condition is given as $  \widetilde{\Gamma}(\theta,\phi)$. Depending on the flow behavior and its surface properties, the drop surface may get deform to an ellipsoidal shape with its major axis being aligned with the $ x $ direction at an angle $\varphi_d$. Two of the many possible configurations of the deformed drop in the $ x-y $ plane are shown in figure \ref{fig:schematic}(b). A body fitted coordinate system fixed at the drop centroid is also shown in the figure. The drop phase physical properties are denoted as: viscosity ($ \mu_d$), electrical conductivity ($ \sigma_d$) and dielectric permittivity ($ \epsilon_d $) while the corresponding quantities for the outer continuous phase are denoted with the subscript $ `d $'  replaced with `$c$'. We assume that the drop is neutrally buoyant in the continuous fluid i.e. the densities of the two mediums are nearly equal ($ \rho_d\approx \rho_c $).

\section{Mathematical description }
\label{sec:Mathematical}
In order to derive the dimensionless governing differential equations depicting the above physical problem we adopt the following characteristic values as reference scales for the different variables:
(i) length $ \sim a$, (ii) velocity $ \sim \widetilde{G}a$, (iii) electric field $ \sim \widetilde{E}_\infty $ and (iv) surfactant concentration   $ \sim \widetilde{\Gamma}_{eq} $. The relative importance of the characteristic electric stress $( \sim \epsilon_c \widetilde{E}_\infty^2) $ and hydrodynamic stress $(\sim \mu_c \widetilde{G})$ is embodied in the dimensionless quantity defined as the Mason number, $ M=\dfrac{\epsilon_c \widetilde{E}_\infty^2}{\mu_c \widetilde{G}}$. Due to this non-dimensionlization scheme we obtain the following important property ratios appearing in the problem formulation: electrical conductivity ratio, $R=\sigma_d/\sigma_c $, dielectric permittivity ratio, $ S=\epsilon_d/\epsilon_c$ and viscosity ratio, $ \lambda=\mu_d/\mu_c$. 
\subsection{Governing equations and boundary conditions}
\label{ssec:math_gov_cq}
We assume that the concentration of the adsorbed surfactants on the drop interface is small so that  it does not deviate significantly from the equilibrium concentration. Thus the surface tension, $ \widetilde{\gamma} $ can be related to the local surfactant concentration by the relation  \citep{Leal2007,Stone1990}
\begin{equation}
\label{eq:linear}
\widetilde{\gamma}(\widetilde{\Gamma})=\widetilde{\gamma}_{cl}-R_gT\,\widetilde{\Gamma},
\end{equation}
where $ \widetilde{\gamma}_{cl}$ is the surface tension of a clean (surfactant-free) drop, $ T $ stands for the absolute temperature and $ R_g $ denotes the ideal gas constant. Further,  $ \widetilde{\gamma}_{cl}$ is related to the equilibrium interfacial tension as  
 $\widetilde{\gamma}_{eq}=\widetilde{\gamma}_{cl}-R_gT\,\widetilde{\Gamma}_{eq}$. 
Hence \eqref{eq:linear} takes the alternative form 
\begin{equation}
\label{eq:mod_linear}
\widetilde{\gamma}(\widetilde{\Gamma})=\widetilde{\gamma}_{eq}+R_gT(\widetilde{\Gamma}_{eq}-\widetilde{\Gamma}).
\end{equation}  
Here onwards we denote the variables in their dimensionless forms and drop the `\,$ \,{\widetilde{}}\, $\,' symbol from the notations.
Now based on the present considerations of a steady state process and bulk insoluble surfactant molecules, the surfactant transport is governed by the following convection-diffusion equation \citep{Leal2007,Mandal2016a}
\begin{equation}
Pe_{\!_S}\nabla_S\cdot({\mathbf{u}_{\!_{\,S}}}\Gamma)=\nabla^2_{\!_S}\Gamma,
\label{eq:surf_transport}
\end{equation}
where $ Pe_{\!_S}=a^2\,\widetilde{G}/D_{\!_S} $ is the surface P\'eclet number which represents the relative importance of the convection and diffusion at the drop interface.
In order to model the electrical effects  we consider the situation when  both the drop and continuous phases are leaky dielectric in nature and adopt the Taylor-Melcher leaky dielectric model  \citep{Melcher1969,Saville1997,Vlahovska2019}. 
According to this model   free charge accumulation at the interface takes place due to finite Ohmic conductivities of the constituent fluidic phases. Now the time scale   associated with the charge relaxation on account of electrical conduction from the bulk fluid to the drop-carrier fluid interface  $(\tau_{CR}={\epsilon}/{\sigma})$ turns out to be significantly shorter than the time scale for convective process. As a consequence the volumetric charges may be neglected, allowing the momentum equation to be decoupled from electrostatics. However the finite electrical conductivities of  fluid phases  allow the free charge accumulation at the interface. This effectively reduces the volumetric free charge density to a boundary condition on the drop surface. Similarly, the ionic diffusion time scale turns out to be considerably higher than $ \tau_{CR}$ leading to the assumption of zero bulk charge density.
%In order to obtain the electrostatic potential distribution we adopt the Taylor-Melcher leaky dielectric model 
%\citep{Saville1997}. Accordingly we consider no volumetric charge density and accumulation of charges at the interface is modeled through  the electrical Maxwell stress. 
Thus the electric potential in the drop and continuous phase $ (\psi_d, \psi_c) $ satisfy   
\begin{equation}\label{eq:laplace}
\nabla^2\psi_{d,c}=0 
\end{equation}
Apart from the usual boundedness of potential in and around the drop, the matching condition of at the drop interface and the condition of  uniformly imposed  electric field at the far stream   $(\text{i.e.}\,r\to\infty,$\,\,$\nabla\psi_c=-\mathbf{E}_\infty )$, the interfacial charge balance turns out to an important condition for the present problem. In the quasi-steady state this reads \citep{Xu2006a}
\begin{equation}
\textrm{\\at} \,\, r=r_{\!_S}(\theta, \phi),\quad \mathbf{n}\cdot (R\nabla \psi_d-\nabla\psi_c)=-Re_E\nabla_{\!_S}\cdot(q\,_{\!_S}\mathbf{u}_{{\,\!_S}}),
\label{eq:cc}
\end{equation}
where $ \mathbf{n} $ is the unit vector in the normal direction at the drop surface and $\nabla_{\!_S}$ is the surface gradient operator $ \left(\nabla_{\!_S}=(\mathbf{I}-\mathbf{n}\mathbf{n})\cdot\nabla\right)$.
Here $ Re_E={\epsilon_c \widetilde{G}}/{\sigma_c} $ is the electrical Reynolds number quantifying the relative importance of the charge relaxation time scale and charge convection time scale. Also  the charge density is calculated as  
\begin{equation}
\label{eq:charge-formula}
q_{\!_{\,S}}(\theta,\phi)=\mathbf{n}\cdot (S\nabla \psi_d-\nabla\psi_c)\big|_{r=r_{\!_S}}^{}.
\end{equation}
The charge convection  \eqref{eq:cc} suggests that the charge distribution and in effect the electric potential are coupled with the velocity field. Now since the velocity field is affected by the surfactant non-uniformity, due to the coupling as suggested by the \eqref{eq:cc}, the electrical potential is also influenced by the surfactant. 

Neglecting the inertial effect in the creeping flow limit $ (Re \ll 1) $  the hydrodynamics can be described by the following equations \citep{Happel1981}
\begin{equation}
\begin{aligned}
-\nabla{p}_d&+\lambda\nabla^2\mathbf{u}_d=0, &&\nabla\cdot{\mathbf{u}_d}=0\\
-\nabla{p}_c&+\,\,\,\,\nabla^2\mathbf{u}_c=0,        &&\nabla\cdot{\mathbf{u}_c}=0
\end{aligned}
\label{eq:momentum}
\end{equation}
Similar to the electric potential the pressure and velocity both in the drop as well as in the outer region must be bounded and they are continuous at the drop interface. Also the fluid velocity at the far stream matches with the applied background flow velocity i.e.  as
$r\to\infty,$\quad$\mathbf{u}_c=\mathbf{V}_\infty$. Again in the quasi-steady state the normal velocity at the interface becomes zero i.e. 
 $\textrm{\\at} \,\, r=r_{\!_S}(\theta, \phi)\quad \mathbf{u}_d\cdot{\mathbf{n}}=\mathbf{u}_c\cdot{\mathbf{n}}=0$. 
Finally the interface stress balance can be expresses as:
	\begin{equation}
\\\textrm{at } \,r=r_{\!_S}(\theta,\phi),\quad \llbracket\boldsymbol{T}^H\rrbracket+M\:\llbracket\boldsymbol{T}^E\rrbracket=\frac{1}{Ca}(\nabla\cdot{\mathbf{n}})\mathbf{n}+\underbrace{Ma\,[(1-\Gamma)(\nabla\cdot{\mathbf{n}})\mathbf{n}+\nabla_{\!_S}\Gamma]}_\text{Marangoni stress},
\label{eq:stress_bal}
\end{equation}
where $\llbracket\,\boldsymbol{\chi}\,\rrbracket$ denotes the jump of a variable $ \boldsymbol{\chi} $ at the interface (e.g.\ $\llbracket\boldsymbol{\chi}\rrbracket=\boldsymbol{\chi}_c-\boldsymbol{\chi}_d$); $\boldsymbol{T}^{H,E}=
{\mathbf{n}}\cdot\boldsymbol{\tau}^{H,E}$ are the viscous (superscript-$ H $) and electrical (superscript-$ E $) traction vectors at the interface. The Marangoni number  $ (Ma) $ defined as  $ Ma=R_gT\,\widetilde{\Gamma}_{eq}/\mu_c \, a \, \widetilde{G} $, as appeared in \eqref{eq:stress_bal}, represents the relative importance of the characteristic Marangoni stress due to surfactants and the characteristic viscous stress. Also $ Ca $ is the Capillary number  denoting the ratio of the relative magnitudes  of viscous and capillary stresses at the interface. It is a non-dimensional representation of the deformability of the drop surface.   $ Ma $
 can be related to $ Ca $ by the relation $Ma=\dfrac{\beta}{Ca(1-\beta)}$ where elasticity number, $ \beta $ is defined in the form
 $\beta=
 -\dfrac{d(\widetilde{\gamma}/\widetilde\gamma_c)}
 {d(\widetilde\Gamma / \widetilde\Gamma_{eq})}
 =\dfrac{R_gT\,\widetilde\Gamma_{eq}}{\widetilde\gamma_c}$. The parameter $\beta$ signifies the sensitivity surface tension on the concentration of surfactant molecules on the drop interface. For most of the practical purposes $ \beta $ varies between 0 and 0.8 \citep{Stone1990,Li1997}. The effects of Marangoni stress on tangential and normal stress balance at the drop interface can be visualized when the two components of the stress balance are considered separately as:

\begin{equation}
\textrm{Tangential stress balance: at } \,r=r_{\!_S}(\theta,\phi),\quad 
\llbracket{T}^H_t\rrbracket+M\:\llbracket{T}^E_t\rrbracket=\underbrace{\frac{\beta}{Ca(1-\beta)}\,(\nabla_{\!_S}\Gamma)\cdot \mathbf{t}}_{\textrm{ Marangoni effect}}
\label{eq:t_stress_bal}\,,
\end{equation}
and
\begin{equation}
\textrm{Normal stress balance: at } \,r=r_{\!_S}(\theta,\phi),\quad \llbracket{T}^H_n\rrbracket+M\:\llbracket{T}^E_n\rrbracket=\frac{1}{Ca}(\nabla\cdot{\mathbf{n}})+\underbrace{\frac{\beta}{Ca(1-\beta)}\,[(1-\Gamma)(\nabla\cdot{\mathbf{n}})]}_\text{Marangoni effect}\,.
\label{eq:n_stress_bal} 
\end{equation}
Here $ \mathbf{t} $ is the unit vector in the tangential direction at the drop surface. 
\subsection{Asymptotic solution}
\label{ssec:solution}
The governing equations depicting the present physical problem are coupled in such a manner that an exact solution for arbitrary values of the dimensionless parameters is not possible. To circumvent this problem we consider a physically realistic condition for which the charge convection effect is weak in comparison to the Ohmic conduction (i.e. $ Re_E \ll 1 $) and the drop deviates only slightly from the initial spherical shape (i.e. $ Ca \ll 1 $) \citep{Xu2006a,Mandal2016}. In addition we consider that the surface P\'eclet number $ (Pe_{\!_S}) $ is low (i.e. $ Pe_{\!_S} \ll 1 $) \citep{Pak2014,Mandal2017c,Ha1995}, depicting the condition when the interfacial transport is mainly governed by the diffusion mechanism rather  than convection. Thereafter  a dimensionless physicochemical constant, $ k $ can be defined as \citep{Stone1990,Mandal2016a}
 \begin{equation}\label{eq:k-def}
k=\frac{Pe_{\!_S}}{Ca}=\frac{\widetilde\gamma_{eq}\,a}{\mu_c\,D_{\!_S}}.
\end{equation}
In this context, the physical condition of $ k=0 $ stands for a uniformly contaminated drop \citep{Stone1990}.
 Now for the assumed situation of $ Ca \ll 1, Pe_{\!_S} \ll 1  $, the parameter $ k $ turns out to be $ \sim O(1). $ These considerations let us adopt a regular perturbation scheme which for any generic field variable  $ \chi $ reads 
\begin{equation}
\chi=\chi^{(0)}+Ca\,\chi^{(Ca)}+Re_E\,\chi^{(Re_E)}+O(Ca^2,CaRe_E,Re_E^2).
\label{eq:general_perturb}
\end{equation}
Since the pressure and stress inside the drop must balance the capillary pressure in the static condition, they are to be expanded as 
 \citep{Chan1979,Bandopadhyay2016}:
\begin{equation}
\label{eq:cap-press}
\left.
\begin{split}
p_d &=\frac{1}{Ca}p^{(1/Ca)}_d+p^{(0)}_d+Ca\,p^{(Ca)}_d+Re_E\,p^{(Re_E)}_d+O(Ca^2,CaRe_E,Re_E^2)\\
\boldsymbol{\tau_d} &=\frac{1}{Ca}\boldsymbol{\tau_d}^{(1/Ca)}+\boldsymbol{\tau_d}^{(0)}+Ca\,\boldsymbol{\tau_d}^{(Ca)}+Re_E\,\boldsymbol{\tau_d}^{(Re_E)}+O(Ca^2,CaRe_E,Re_E^2)
\end{split}
\right \}
\end{equation} 
Solving the normal stress  equation \eqref{eq:n_stress_bal}  in different order of perturbation the drop shape is obtained in the  form
\begin{equation}
r_{\!_S}=1+Ca\,f^{(Ca)}+CaRe_E\,f^{(CaRe_E)}+Ca^2\,f^{(Ca^2)}+\ldots
\label{eq:shape_perturb}
\end{equation} 
In a similar fashion the  interfacial  surfactant concentration $ \Gamma $ can be expressed as \citep{Stone1990,Ha1995}
\begin{equation}
\Gamma=\Gamma^{(0)}+Ca\,\Gamma^{(Ca)}+CaRe_E\,\Gamma^{(CaRe_E)}+Ca^2\Gamma^{(Ca^2)}+\ldots.
\label{eq:surf_perturb}
\end{equation}
The local surfactant concentration also has to satisfy conservation of  mass for the surfactant molecules which is mathematically denoted as
\begin{equation}
\label{eq:mass-constraint}
\int_{\phi=0}^{2\pi}\int_{\theta=0}^{\pi} \Gamma(\theta, \phi)\, r^2_{\!_S}(\theta,\phi)\sin{\theta} \,d\theta\,d\phi=4 \pi.
\end{equation}

Based on the above perturbation scheme the linearized hydrodynamic equations are solved using the generalized Lamb solution technique \citep{Happel1981,Bandopadhyay2016}. Accordingly the velocity and pressure field inside $ (\mathbf{u}_d, p_d) $ and outside $ (\mathbf{u}_c, p_c) $ the drop can be expressed a series of solid spherical harmonics given as \citep{Haber1971,Haber1972}
\begin{equation}
\mathbf{u}_d=\sum_{n=1}^{\infty} \left [\nabla\times (\mathbf{r}\chi_n)+\nabla \Phi_n+\frac{n+3}{2(n+1)(2n+3)\lambda}r^2\nabla p_n-\frac{n}{(n+1)(2n+3)\lambda}\mathbf{r}p_n \right ]
\label{eq:in-vel-lamb}
\end{equation}
\begin{equation}
\text{and }\quad p_d=\sum_{n=1}^{\infty}p_n
\label{eq:in-press-lamb}
\end{equation}
 
\begin{equation}
\mathbf{u}_c=\mathbf{V}_\infty+\sum_{n=1}^{\infty} \left [\nabla\times (\mathbf{r}\chi_{-n-1})+\nabla \Phi_{-n-1}-\frac{n-2}{2n(2n-1)}r^2\nabla p_{-n-1}+\frac{n+1}{n(2n-1)}\mathbf{r}p_{-n-1}  \right ]
\label{eq:out-vel-lamb}
\end{equation}
\begin{equation}
\text{and }\quad p_c=\sum_{n=1}^{\infty}p_{-n-1}
\label{eq:out-press-lamb}
\end{equation}
The different growing and decaying harmonics appearing in the above equations can be expressed as 
\begin{equation}
\left.\begin{split}
p_n=\lambda r^n\sum_{m=0}^{n}\left [ A_{n,m}\cos(m \phi)+\hat{A}_{n,m}\sin(m \phi) \right ]P_{n,m}(\eta)
\\ \Phi_n=r^n\sum_{m=0}^{n}\left [ B_{n,m}\cos(m \phi)+\hat{B}_{n,m}\sin(m \phi) \right ]P_{n,m}(\eta)
\\ \chi_n=r^n\sum_{m=0}^{n}\left [ C_{n,m}\cos(m \phi)+\hat{C}_{n,m}\sin(m \phi) \right ]P_{n,m}(\eta)
\end{split}\right\}
\label{eq:growing-Hs}
\end{equation}
\\	
\begin{equation}
\left.\begin{split}
p_{-n-1}= r^{-n-1}\sum_{m=0}^{n}\left [ A_{-n-1,m}\cos(m \phi)+\hat{A}_{-n-1,m}\sin(m \phi) \right ]P_{n,m}(\eta)
\\ 
\Phi_{-n-1}= r^{-n-1}\sum_{m=0}^{n}\left [ B_{-n-1,m}\cos(m \phi)+\hat{B}_{-n-1,m}\sin(m \phi) \right ]P_{n,m}(\eta)
\\ 
\chi_{-n-1}= r^{-n-1}\sum_{m=0}^{n}\left [ C_{-n-1,m}\cos(m \phi)+\hat{C}_{-n-1,m}\sin(m \phi) \right ]P_{n,m}(\eta)
\end{split}\right\},
\label{eq:decaying-Hs}
\end{equation}
where  $ P_{n,m}(\eta) $ stands for the associated Legendre polynomial of degree $ n $ and order $ m $  with an argument $ \eta=\cos(\theta) $. The various arbitrary constants $ (A_n, B_n, C_n, A_{-n-1}, B_{-n-1},  C_{-n-1}$, $\hat{A}_n, \hat{B}_n, \hat{C}_n, \hat{A}_{-n-1}, \hat{B}_{-n-1}$ $ \mathrm{\;and\;} \hat{C}_{-n-1})$ are to be determined by simultaneously solving the surfactant transport equation and the appropriate form of the hydrodynamic boundary conditions. The detailed procedure regarding the treatment of these boundary conditions is tedious and can be found in earlier studies \citep{Haber1972,Bandopadhyay2016}.

The governing equations electrical potential (\eqref{eq:laplace}) in both the drop and continuous  phase are solved by expanding the potential in the form
\begin{equation}
\label{eq:potential-general}
\left.
\begin{split}
\psi_d=\sum_{n=0}^{\infty}r^n\sum_{m=0}^{n}\left [ a_{n,m}\cos(m \phi)+\hat{a}_{n,m}\sin(m \phi) \right ]P_{n,m}(\cos(\theta))\qquad\qquad\\
\psi_c=\psi_\infty+\sum_{n=0}^{\infty}r^{-n-1}\sum_{m=0}^{n}\left [ b_{-n-1,m}\cos(m \phi)+\hat{b}_{-n-1,m}\sin(m \phi) \right ]P_{n,m}(\cos(\theta))
\end{split}
\right \},
\end{equation}
%Here $ P_{n,m}(\eta) $ denotes the associated Legendre polynomial of degree $ n $ and order $ m $ with an argument $ \eta=\cos(\theta) $ and $ \varphi_\infty $ is the unperturbed electric potential in the far-stream representing the externally applied electric potential by means of a tilted electric field $ \mathbf{E}_\infty $. 
where $ \psi_\infty $ takes care of the far field  condition of a uniform tilted electric field.

We seek solutions for the shape function $ f(\theta,\phi) $  and the surfactant concentration, $ \Gamma(\theta,\phi) $ by expanding them in terms of surface harmonics as:
\begin{equation}
f=\sum_{n=0}^{\infty}\sum_{m=0}^{n}\left [ L_{n,m}\cos(m \phi)+\hat{L}_{n,m}\sin(m \phi) \right ]P_{n,m}(\eta)
\end{equation} and 
\begin{equation}
\Gamma=\sum_{n=0}^{\infty}\sum_{m=0}^{n}\left [ \Gamma_{n,m}\cos(m \phi)+\hat{\Gamma}_{n,m}\sin(m \phi) \right ]P_{n,m}(\eta).
\end{equation}

\subsection{Emulsion rheology}
\label{ssec:theo-suspension-rheo}
In the present physical situation, the simultaneous action of the imposed shear flow and a uniform electric field significantly alters the flow field around the drop and also modifies the  drop shape. The presence of surfactant induced Marangoni stress further complicates the scenario. Thus the effective rheology of the emulsion containing the drops is expected to deviate from the condition when the surrounding fluid is considered alone. The contribution of micro level flow physics  in the dispersed phase to the effective macro-scale rheology of a dilute emulsion can be calculated by determining a volume averaged  stress of the emulsion \citep{Schowalter1968,Batchelor1970,Guazzelli2011}: . 
  %\cite{Batchelor1970,Barthes-Biesel2012,Kim1991}
\begin{equation}\label{eq:eff-stress}
\left \langle \boldsymbol{\tau} \right \rangle =-\left \langle p \right \rangle\mathbf{I}+2\mathbf{D}_\infty+\frac{\nu}{V_d}\mathbf{S}.
\end{equation}
Here $ \mathbf{D}_\infty=\left({\nabla \mathbf{V}_\infty+\left(\nabla \mathbf{V}_\infty\right)^T}\right)/{2} $
 is the straining component of the externally imposed linear shear flow and $ 
\mathbf{S} $ is the stresslet defined as 
\begin{equation}\label{eq:stresslet}
\mathbf{S}=\int_{\phi=0}^{2\pi} \int_{\theta=0}^{\pi} \left[ \frac{1}{2} \left( (\boldsymbol{\tau}^H_c \cdot \mathbf{n}) \mathbf{r}+\left((\boldsymbol{\tau}^H_c \cdot \mathbf{n}) \mathbf{r}\right)^T-\frac{2}{3}\mathbf{I}(\boldsymbol{\tau}^H_c \cdot \mathbf{n}) \mathbf{r} \right)- \left( \mathbf{u_c n} +(\mathbf{u_c n})^T \right)  \right] r_S^2 \sin(\theta) d\theta d\phi.
\end{equation}
The stresslet represents the change in the total stress brought in by the velocity and stress modifications in the continuous phase by the  presence of the drop phase. Following the presently adopted Lamb solution technique mentioned in section \ref{ssec:solution}, the stresslet  can be alternatively  expressed in a more compact form as a function of the  decaying solid spherical harmonic $p_3$ as:
\citep{Happel1981,Kim1991}
\begin{equation}\label{eq:stresslet-expression}
\mathbf{S}=-\frac{ \nu}{2}\nabla \nabla \left( r^5 p_{-3}\right).
\end{equation}
The heterogeneity in the emulsion is quantified by defining an effective shear viscosity of emulsion, the dimensionless form of which is given as 
%\begin{equation}\label{eq:eff_vis}
%\mu_\text{eff}=\frac{\widetilde{\left \langle \mathbf{ \tau  }\right \rangle}_{xy}}{G}
%\end{equation}
\citep{Schowalter1968,Bird1987}
\begin{equation}\label{eq:eff_vis}
\eta_\text{eff}=\frac{\mu_\text{eff}  }{\mu_c}=\left \langle \boldsymbol{ \tau  }\right \rangle_{xy}.
\end{equation}

The isotropy of normal stress is disturbed in  a sheared emulsion \citep{Guazzelli2011}. This anisotropy  is quantified by the first and second normal stress differences which are calculated  as:
\begin{equation}\label{eq:first-normal_stress_diff}
\text{first normal stress difference:}\quad N_1=\frac{{\left \langle \widetilde{\boldsymbol{ \tau  }}\right \rangle}_{xx}-{\left \langle \widetilde{\boldsymbol{ \tau  }}\right \rangle}_{yy}}{\mu_c\, \widetilde{G}}=\left \langle {\boldsymbol{ \tau  }}\right \rangle_{xx}-\left \langle {\boldsymbol{ \tau  }}\right \rangle_{yy}
\end{equation}
and 
\begin{equation}\label{eq:second-normal_stress_diff}
\text{second normal stress difference:}\quad N_2=\frac{{\left \langle \widetilde{\boldsymbol{ \tau  }}\right \rangle}_{yy}-{\left \langle \widetilde{\boldsymbol{ \tau  }}\right \rangle}_{zz}}{\mu_c \, \widetilde{G}}=\left \langle {\boldsymbol{ \tau  }}\right \rangle_{yy}-\left \langle {\boldsymbol{ \tau  }}\right \rangle_{zz}.
\end{equation}

We proceed to  calculate the electric potential, flow field and the surfactant distribution in the different orders of perturbation, namely the leading order $ O(1) $, the $ O(Re_E) $ and the $ O(Ca) $.  Next the effective emulsion stress and the normal stress difference expressions  of orders $ O(\nu),O(\nu Re_E) $ and $ O(\nu Ca) $ are determined by considering a dilute emulsion $ (\nu \ll 1) $ where drop-drop interactions are not significant. The detailed solution strategy at different perturbation orders are given in earlier studies \citep{Bandopadhyay2016,Mandal2016} and hence in the following sections we only provide the expressions of different  physical variables which are important for the present study.

\subsection{Description at leading order}
\label{ssec:detailed-solution-leading}
The leading order description of different variables are obtained by considering an undeformed drop for which the charge convection at the surface is too small to be taken into account. Thus in this order of perturbation the electrical potential distribution becomes independent of any flow induced effect and can be solely expressed in terms of electrical property ratios as given below:
\begin{subequations}
	\label{eq:leading-pot}
	\begin{equation}
	\psi_d^{(0)}=-\frac{3\,r}{R+2}\cos(\Phi_t-\phi)P_{1,1}(\eta),
	\end{equation}
	\begin{equation}
	\psi_c^{(0)}=\left(\frac{1}{r^2}\frac{R-1}{R+2}-r\right)\cos(\Phi_t-\phi)P_{1,1}(\eta).
	\end{equation}
\end{subequations}
Similarly the leading order interfacial charge density becomes
\begin{equation}
\label{eq:leading-charge}
q_S^{(0)}=\frac{3(R-S)}{R+2}\cos(\Phi_t-\phi)P_{1,1}(\eta).
\end{equation}

%section~\ref{ssec:solution}
The  surfactant distribution up to $ O(Ca) $ can be expressed as 
\begin{equation}
\label{eq:leading-surf-distr}
\Gamma=\Gamma^{(0)}+\Gamma^{(Ca)}= \Gamma^{(0)}+ \Gamma_{2,0}^{(Ca)} P_{2,0}(\eta)+\Gamma_{2,2}^{(Ca)}\cos(2 \phi)P_{2,2}(\eta)+\hat{\Gamma}_{2,2}^{(Ca)}\sin(2 \phi)P_{2,2}(\eta),
\end{equation}
where different harmonics of surfactant concentration are obtained as
\begin{equation}
\label{eq:leading-surf-harmonics}
\begin{split}
&\Gamma^{(0)}=1; \\
& \Gamma^{(Ca)}_{2,0}=-\frac{3}{2}{\frac {M   ( 1-\beta) (R-S) {\it k}}{ \left( R+2 \right) ^{2} \left(  \left( {\it k}
		-5\lambda-5 \right) \beta+5\lambda+5 \right) }};\\
& \Gamma^{(Ca)}_{2,2}=\frac{3}{4}{\frac {M \cos(2\Phi_t) ( 1-\beta) (R-S) {\it k}}{ \left( R+2 \right) ^{2} \left(  \left( {\it k}
		-5\lambda-5 \right) \beta+5\lambda+5 \right) }};\\
& \hat{\Gamma}^{(Ca)}_{2,2}=\frac{5}{12}{\frac { \left(1-\beta \right) {\it k}}{ \left( {\it k}-5
		\lambda-5 \right) \beta+5\lambda+5}}+\frac{3}{4}{\frac {M\,\sin(2\Phi_t) \left(1- \beta \right)  \left( R-S
		\right) {\it k}}{ \left( R+2 \right) ^{2} \left(  \left( {\it k}
		-5\lambda-5 \right) \beta+5\lambda+5 \right) }}.
\end{split}
\end{equation}

The leading order velocity and pressure fields, both inside the drop $(\mathbf{u}_d^{(0)},p_d^{(0)}) $ and in the matrix phase  $(\mathbf{u}_c^{(0)},p_c^{(0)}) $, can be fully expressed in terms of the non-zero solid harmonics up to $ n=2$ and all the other term vanish. Of particular interest is the leading order interfacial fluid velocity which can be expressed in the spherical coordinate system as $ \mathbf{u}_S^{(0)}= u^{(0)}_{S,\theta}\, \mathbf{i}_\theta+u^{(0)}_{S,\phi}\, \mathbf{i}_\phi$, where the scalar components are given in the equation~\ref{eq:leading_us}.

The leading order effective viscosity  becomes a function of the arbitrary constant  $ \hat{A}_{-3,2}^{(0)} $ which is present in the decaying harmonics in \eqref{eq:decaying-Hs} and is given by \citep{Mandal2017a}
\begin{equation}\label{eq:eff_vis_A32_leading}
\eta_\text{eff}^{(0)}=1-3 \nu \hat{A}_{-3,2}^{(0)}.
\end{equation}
Finally $ \eta_\text{eff}^{(0)} $ can be written in the form
	\begin{equation}
\label{eq:leading--eta}
\begin{split}
 \eta_\mathrm{eff}^{(0)}=& \underbrace{1+\nu{\frac {\left( 5\,\lambda+2 \right) }{\left(2\,\lambda+2 \right)}
}}_{\eta^{(0)}_\text{h,clean}}+ \underbrace{\frac{3}{2}{\frac {\nu k\,\beta}{   \left( 
		\left( k-5\,\lambda-5 \right) \beta+5\,\lambda+5 \right)\left( 1+\lambda \right) }}}_{\text{surfactant contribution}\, \left( \eta^{(0)}_\text{h,surf}\right)}\\
& 
\underbrace{-{\frac {27 M \nu \sin(2\Phi_t) \left( R-S \right) }{ 10\left( \lambda+1 \right)  \left( R+2 \right) ^{2}}}}_{\eta^{(0)}_\text{e,clean}}\times \underbrace{{\frac { \left(1-\beta \right)  \left( 1+\lambda \right) }{
			\left( {\it k}-5\lambda-5 \right) \beta+5\lambda+5}}.}_{\text{surfactant contribution}\, \left( \eta^{(0)}_\text{e,surf}\right)}
\end{split}
\end{equation}
It is noteworthy to  observe that surfactant induced Marangoni stress  affects both the hydrodynamic $ (\eta^{(0)}_\mathrm{h,surf}) $ and electrical $ (\eta^{(0)}_\mathrm{e,surf}) $ contributions to the leading order effective viscosity.
Moreover, the effect on the electrical component is of a  complex functional form which is in contrast to the  linear superposition similar to the hydrodynamic contribution as shown above.

In a similar manner the first and second normal stress differences are obtained as 
\begin{subequations} 
	\begin{equation}
	\label{eq:leading-N1}
	N_{1}^{(0)}=-6 \, \nu A_{-3,2}^{(0)}=-{\frac {27\nu M \cos(2\Phi_t) 
			\left( R-S \right) }{ \left(\lambda+1 \right)  \left( R+2
			\right) ^{2}}}\times \underbrace{{\frac { (1-\beta)  \left( 1+\lambda \right) }{
				\left( {\it k}-5 \lambda-5 \right) \beta+ 5(\lambda+1)}}
	}_{\mathcal{N}^{(0)}_{1,surf}=\text{ surfactant contribution}}
	\end{equation}
	and
\begin{equation}
\label{eq:leading-N2}
N_{2}^{(0)}=\nu \left(3 \, A_{-3,2}^{(0)}+ \frac{3}{2} A_{-3,0}^{(0)}  \right )=-{\frac {27 \nu M \sin^2(\Phi_t)
		\left( R-S \right) }{ \left( \lambda+1 \right)  \left( R+2
		\right)^{2}}}\times \underbrace{{\frac { (1-\beta)  \left( 1+\lambda \right) }{
			\left( {\it k}-5\lambda-5 \right) \beta+5(\lambda+1)}}
}_{\mathcal{N}^{(0)}_{2,surf}=\text{ surfactant contribution}},
\end{equation}
	\label{eq:leading_N1N2}
\end{subequations}
where both the expressions contain a common surfactant contribution term, 

\begin{equation}\label{eq:corr_N1_N2}
\mathcal{N}^{(0)}_{1,surf}=\mathcal{N}^{(0)}_{2,surf}=\mathcal{N}^{(0)}_{surf}=\dfrac { (1-\beta)  \left( 1+\lambda \right) }{
	\left( {\it k}-5\lambda-5 \right) \beta+5(\lambda+1)}.
\end{equation}

\subsection{Order $\boldsymbol{Re_E}$ description}
\label{ssec:sol_Re_E}
In this case we consider an undeformed drop but take into account of the charge convection at the drop surface.
In contrast to the leading order, the interfacial charge distribution of $ O(Re_E) $ gets affected by the flow field and the surfactant effects and subsequently  we get $ q_S^{Re_E}= q_S^{Re_E}\left(\theta,\phi \,; R,S,\lambda,M,\Phi_t,\beta,k\right)$ (see the supplementary material for the full expression of $ q_S^{(Re_E)} $). This behavior can be understood by looking into the charge convection equation applicable for  $ O(Re_E) $ given as  
\begin{equation}
\textrm{\\at} \,\, r=1\quad R\frac{\partial\psi_d}{\partial r}-\frac{\partial\psi_c}{\partial r}=-\nabla_{\!_S}\cdot(q\,_{\!_S}^{(0)}\mathbf{u}_{{\,\!_S}}^{(0)}).
\label{eq:cc-ReE}
\end{equation}

The surfactant transport equation in this order of perturbation takes the form 
\begin{equation}\label{eq:ReE_surf}
\textrm{\qquad at} \,\,r=1,\quad 
k\nabla_{\!_S}\cdot\left({\mathbf{u}^{(Re_{\!_E})}_{\!_S}}\right)=\nabla^2_{\!_S}\Gamma^{(CaRe_{\!_E})},
\end{equation}
where the surfactant concentration 
distribution is obtained as 
\begin{equation}
\label{eq:O(ReE)-surf-distr}
\Gamma^{(CaRe_E)}=\sum_{n=2}^{4}\sum_{m=0}^{n} \left(\Gamma_{n,m}^{(CaRe_E)}\cos(m \phi)+\hat{\Gamma}_{n,m}^{(CaRe_E)}\sin(m \phi)\right)P_{n,m}(\eta).
\end{equation}
Again the harmonic constants $ \left(\Gamma_{n,m}^{(CaRe_E)},\hat{\Gamma}_{n,m}^{(CaRe_E)}\right) $ can be found in the Supplementary Material. 
Here the non-zero solid harmonics upto $ n=4 $ are required to fully describe the pressure and velocity fields $ \left( \mathbf{u}_d^{(Re_E)},p_d^{(Re_E)} , \mathbf{u}_c^{(Re_E)},p_c^{(Re_E)}\right) $. The full expressions are not mentioned due to their excessive lengths but can be found in the supplementary material

The effective viscosity is obtained as a correction to the leading order term, i.e. 
$ \eta_\text{eff}=\eta_\text{eff}^{(0)}+Re_E \, \eta_\text{eff}^{(Re_E)}.$
In terms of the arbitrary constants present in the decaying velocity field the  $ \eta_\text{eff}^{(Re_E)} $ can be written as \citep{Mandal2017a}
\begin{equation}\label{eq:eff_vis_A32_ReE}
\eta_\text{eff}^{(Re_E)}=-3 \nu \hat{A}_{-3,2}^{(Re_E)}.
\end{equation}

%\subsection{Order $\boldsymbol{Re_E}$  normal stress difference}
%\label{ssec:N1_N2_Re_E}

Unlike the leading order, the surfactant contribution parameters of order $Re_E$ are not equal, i.e. $(\mathcal{N}^{(Re_E)}_1 \ne \mathcal{N}^{(Re_E)}_2)$. In addition to being functions of the surfactant parameters $\beta$ and $k$, they are also strongly dependent on electrical parameters, namely the   tilt angle ($\Phi_t$), the Mason number ($M$) and the electrical property ratios ($R,S$).  This again suggests a coupled nature of the Marangoni effects on the electrical Maxwell stress.

\subsection{Order $\boldsymbol{Ca}$ description}
\label{ssec:sol_Ca}
In this section we consider the deformable nature of the drop and disregard the charge convection effect. This is mathematically represented with the help of deformed drop radius, $ r_S(\theta,\phi)=1+Ca f^{(Ca)}$.  
Now the value of any variable on the deformed surface can be expressed as a correction to the leading order solution, calculated using the Taylor series expansion as  
\begin{equation}\label{eq:taylor-expansion}
\left[\chi\big|_{ \\r=r_{\!_S} }^{}\right]^{(Ca)}={\chi}^{(Ca)}\big|_{ \\r=1 }^{}+f^{(Ca)}\frac{\partial\chi^{(0)}}{\partial r}\bigg|_{r=1},
\end{equation} 
where the shape function $ f^{(Ca)} $ is obtained by solving the  normal stress balance  at the leading order:
\begin{equation}
\textrm{\\at} \,r=1,\quad \llbracket{T}^{H(0)}_{r}\rrbracket+M\:\llbracket{T}^{E(0)}_{r}\rrbracket=-(2f^{(Ca)}+\nabla^2f^{(Ca)})-\frac{2\beta}{1-\beta}\Gamma^{(Ca)}
\label{eq:leading-order-normal}.
\end{equation}
 The shape function has a form 
\begin{equation}\label{eq:leading-shape}
f^{(Ca)}=\sum_{m=0}^{2} \left(L^{(Ca)}_{2,m}\cos(m \phi)+\hat{L}^{(Ca)}_{2,m}\sin(m \phi)\right)P_{2,m}(\eta),
\end{equation}
where the different non-zero deformation modes are obtained as
	\begin{equation}
\label{eq:sol-leading-shape}
\begin{split}
& L^{(Ca)}_{2,0}=L^{(Ca)}_{2,0}\big{|}_\text{Clean}-\frac{3}{40}{\frac {\,M k\,\beta\,   \left( \lambda+4 \right)  \left( R-S \right) }{ \left( \, 
		\left( -5\,\beta+5 \right) \lambda+5+\, \left( k-5
		\right) \beta \right)  \left( R+2 \right) ^{2} \left( \lambda+1
		\right) }};\\
& L^{(Ca)}_{2,2}=L^{(Ca)}_{2,2}\big{|}_\text{Clean}+\frac{3}{80}{\frac {M k\,\beta\, \left( \lambda+4 \right)  \left( R-S
		\right) \cos \left( 2\,\Phi_{{t}} \right) }{ \left( \lambda+1
		\right)  \left(  \left( -5\,\beta+5 \right) \lambda+5+ \left( k-5 \right) \beta \right)  \left( R+2 \right) ^{2}}};\\
& \text{and} \quad \hat{L}^{(Ca)}_{2,2}=\hat{L}^{(Ca)}_{2,2}\big{|}_\text{Clean}
+\frac{1}{48}{\frac {k\,\beta\, \left( \lambda+4 \right) }{ \left( \lambda+
		1 \right)  \left(  \left( -5\,\beta+5 \right) \lambda+5+ \left( k-5 \right) \beta \right) }} \\
&\qquad \qquad \qquad \qquad \qquad \; \; +\frac{3}{80}{\frac {k\,\beta\, \left( \lambda+4 \right) M \sin(2\Phi_t)
		\left( R-S \right) }{ \left( \lambda+1 \right)  \left(  \left( -5\,
		\beta+5 \right) \lambda+5+ \left( k-5 \right) \beta \right) 
		\left( R+2 \right) ^{2}}}.
\end{split}
\end{equation}
In each of the above expressions the second term corresponds to the modification in the drop shape  brought in by the non-uniform surfactant distribution. Interestingly, for the deformation modes $ {L}^{(Ca)}_{2,0},{L}^{(Ca)}_{2,2} $ the surfactant only has its role only if there is an applied electric field. On the other hand the form of $ \hat{L}^{(Ca)}_{2,2}$ suggests that the deformation  arising from the hydrodynamic origin is itself influenced by the Marangoni effect.

The surfactant transport equation takes a slightly different form than the previously discussed perturbation orders and is given by
\begin{equation}\label{eq:ca-st}
\textrm{at} \,\,r=1+Ca f^{(Ca)},\quad 
k\nabla_{\!_S}\cdot({\mathbf{u}^{(0)}_{\!_S}}\Gamma^{(Ca)}+{\mathbf{u}^{(Ca)}_{\!_S}})=\nabla^2_{\!_S}\Gamma^{(Ca^2)}.
\end{equation}
Determination of the complete  $O(Ca)$ flow field and surfactant distribution becomes too cumbersome to perform. However the main objective of the present work being the determination of the bulk rheology characteristics of the emulsion, it is required to calculate only the arbitrary constants $  \hat{A}_{-3,2}^{(Ca)}, A_{-3,2}^{(Ca)} \text{\;and\;} A_{-3,0}^{(Ca)} $ which are present in the  $ O(Ca) $ flow field. Using these constants we can express the  $ O(Ca) $ effective viscosity and normal stress differences as \citep{Mandal2017a}:

\begin{equation}
\label{eq:eff_vis_A32_Ca}
\eta_\text{eff}^{(Ca)}=-3 \nu \hat{A}_{-3,2}^{(Ca)}.
\end{equation}

\begin{equation}\label{eq:Ca-N1}
N_{1}^{(Ca)}=-6 \, \nu A_{-3,2}^{(Ca)}
\end{equation}
and
\begin{equation}\label{eq:Ca-N2}
N_{2}^{(Ca)}=\nu \left(3 \, A_{-3,2}^{(Ca)}+ \frac{3}{2} A_{-3,0}^{(Ca)}  \right ).
\end{equation}
The final forms of the $O(Ca)$ bulk rheological parameters (please refer to the Supplementary Material for full expressions) again suggest a strong coupling between the Marangoni stress, hydrodynamic and  electrical Maxwell  stresses under the condition of a deformed drop surface.

\section{Results and Discussions}
In the above  formulation of the coupled governing equations and boundary conditions,  we have finally represented various physical variables in terms of a set of dimensionless parameters, given as $ Ca, M, Re_E, R, S, \lambda, \Phi_t , \beta$ and $k$. To ensure the correctness of the present mathematical analysis, we consider different limiting conditions of the present study and compare the key results with those of the existing works in the literature (see section~\ref{sec:validation} for details).  In the following discussions we will first analyze the effects of various electrohydrodynamic  parameters  on the interfacial distribution of the surfactant molecules. Next we discuss the impact of the coupled interplay of the surfactant parameters $ \beta,k $ and electrical effects (characterized by the parameters $ R, S, \Phi_t$) on the bulk emulsion rheology when the drop is undeformed and not affected by the charge convection. In the consequent subsections we delve deeper to investigate the individual effects of charge convection and small shape deformation on the emulsion rheology.  The theoretical model presented above considers a parametric range of as $ Ca\ll1,Re_E\ll1,M\sim O(1)$. In addition we assume that the surface diffusion is the dominant  transport mechanism for the surface active agents, $ Pe_{\!_S} \ll1 $ (or $ k\sim O(1)$). 

Here we demonstrate the results by considering  that both the dispersed and continuous  phases are leaky dielectric in nature.  To find the practical relevance of these assumptions
we consider a model emulsion where a silicone oil drops $(\mu_d=1.29 \;\mathrm{Pa\,s},\;\sigma_d=5.8\times 10^{-12}\;\mathrm{S/m}, \;\epsilon_d=2.7\epsilon_0) $ is dispersed in the continuous phase of castor oil $(\mu_c=1.29 \;\mathrm{Pa\,s},\;\sigma_c=5.8\times 10^{-10}\;\mathrm{S/m},\;\epsilon_c=3.8\epsilon_0) $ \citep{Ha1999}. Also the drop surface is covered with nonionic surfactants such as PL64 or Tween 60.  The interfacial tension of uncontaminated drop is taken as $ \widetilde{\gamma}_{cl}=4.6\;\mathrm{mN/m}$ \citep{Ha2000}. Considering $ \beta=0.8$ and using the relation $ \widetilde{\gamma}_{eq}=(1-\beta)\widetilde{\gamma}_{cl}$ we get the equilibrium surface tension for a surfactant coated drop in this case as $ \widetilde{\gamma}_{eq}=0.92\;\mathrm{mN/m}$. Now if we choose a typical shear rate $ \widetilde{G}=10\;\mathrm{s^{-1}} $, drop diameter of $ 30 \, \mathrm{\mu m}$ (e.g.  in the experiments of \citet{Ha1999}), characteristic electric field strength of $ \widetilde{E}_\infty=0.8\;\mathrm{kV/mm}$ and a  surface diffusivity value of $ D_S=10^{-8}\; \mathrm{\,m^2/s}$ \citep{Valkovska2000}  then the various dimensionless parameters relevant to the present problem can be calculated as $M=1.35,Ca=0.26,Re_E=0.29 $ and $ Pe_{\!_S}=0.23$. Hence the perturbation scheme presented in section \ref{ssec:solution} can predict a practical experimental condition.
 The set of electrohydrodynamic property ratios $ (R=0.005,S=0.71,\lambda=0.81 )$, discussed so far, is named as \textit{system-A} in the demonstrative examples to follow. Along with this, to bring out the contrast in results which the different combination of drop-matrix properties can exhibit, we have chosen two other combinations of the electrohydrodynamic property ratios given as  \textit{system-B}: $R=2.5, S=0.1$, $\lambda= 0.1$ \citep{Mandal2017}; and  \textit{system-C}: $R=0.01, S=1.7$, $\lambda= 0.03$ \citep{Xu2006a}.

\subsection{Nonuniform surfactant distribution}
\label{ssec:surf_dis}
\begin{figure}[!htb]	
	\centering
	\includegraphics[width=0.75\textwidth]{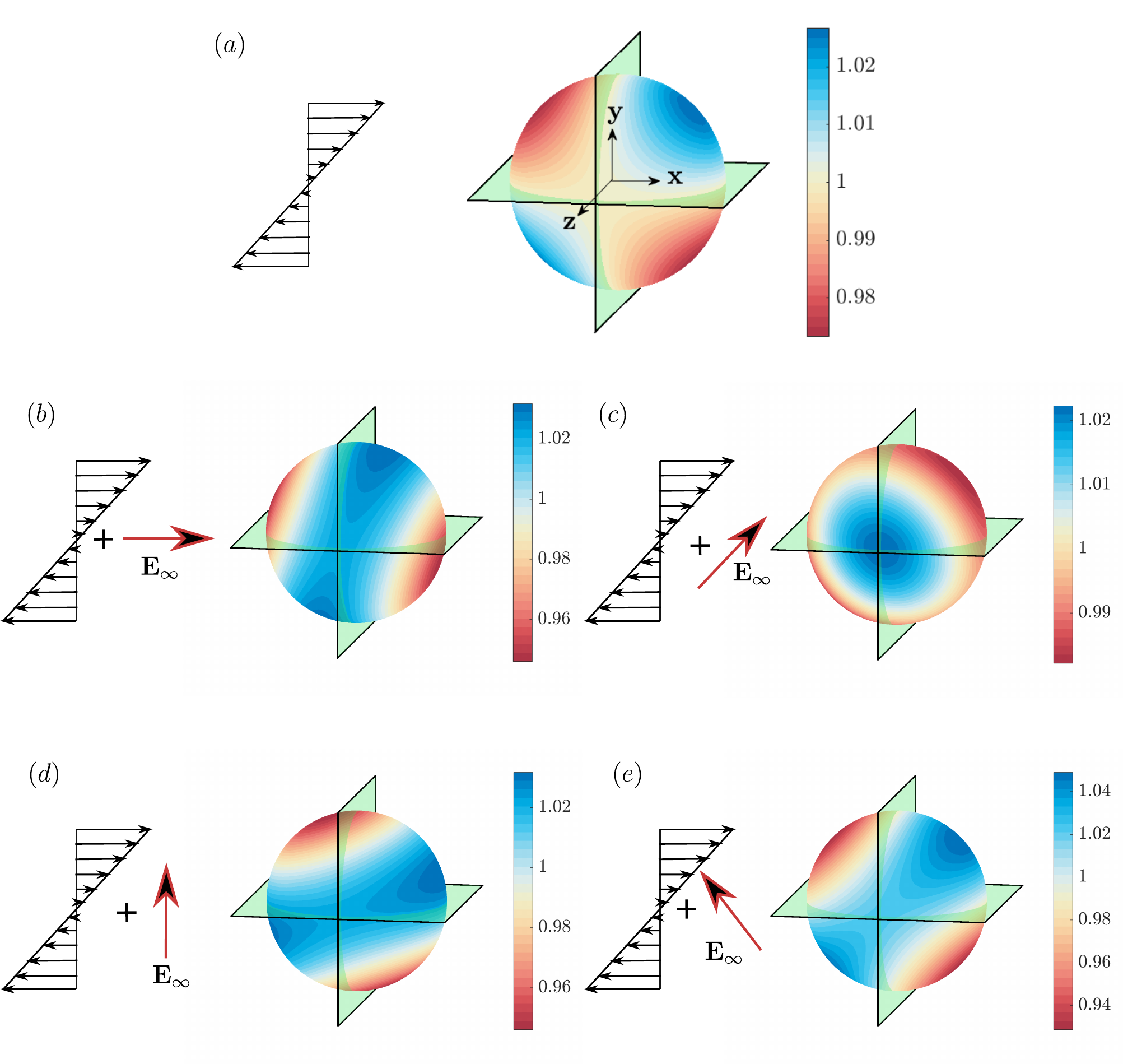}
	\vspace{-10pt} 
	\caption{Effect of electric field on the surfactant concentration distribution on the drop surface $\Gamma=\Gamma^{(0)}+Ca \, \Gamma^{(Ca)}$. In subplot (a) only the effect of background shear flow is considered. In the subplots (b)-(e) the electrical effects are shown by varying the tilt angle  of the applied electric field as $ \Phi_t=0,\pi/4,\pi/2\,\text{and}\,3\pi/4$, respectively. Here the other electrohydrodynamic parameters are chosen as per system-A. The other parameters are: Capillary number, $ Ca=0.2 $; Mason number, $ M=5$; the elasticity parameter, $\beta=0.8$ and physicochemical constant, $k=1$.} 
\label{fig:surf_M_tilt}
\end{figure}
In the absence of any electric field, when only a linear shear flow is imposed, the surfactant molecules obtains a non-uniform but anti-symmetric distribution $ (\Gamma(\theta,\phi)) $ as shown in figure~\ref{fig:surf_M_tilt}(a). The $ \Gamma(\theta,\phi) $  follows the straining axis of the flow which is inclined at an angle $ \phi=\pi/4$ with the background flow  direction \citep{Bentley1986}.  In that case, the surfactant molecules are observed to accumulate more towards the polar regions of this axis and a dominant depletion takes place orthogonal to these locations. But this anti-symmetric pattern changes when an electric field is applied (shown in  figures~\ref{fig:surf_M_tilt} (b)-(e)). With the application of an external field the  surfactant distribution does not follow the  axisymmetric deposition along $\phi=\pi/4$. While for the tilt angles $ \Phi_t=0,\pi/2 $ and $ 3\pi/4$ this common trend is followed, a completely different characteristic of surfactant distribution emerges for $ \Phi_t=\pi/4 $. In this case the maximum surfactant concentration points are no longer in the plane of shear ($ x-y  $ plane in this case) and the maxima points shift to $ y-z $ plane which is orthogonal to the plane of shear. Such undulations in the local surfactant concentration trigger corresponding changes in the interfacial tension $ \gamma (\theta,\phi) $ and gradients in surface tension $ (|\gamma_\text{max}-\gamma_\text{min}|) $ along the drop surface. 
Along with the tilt angle of the applied electric field, the Mason number ($ M $), indicating the relative importance of the electrical stress in comparison to the hydrodynamic stress, also deeply influences the surface tension distribution along the interface. This influence is depicted in figure~\ref{fig:gamma_vs_Phi_vary_G} for the configuration of vertically applied electric field. It is observed that the symmetric deviation of the maximum and minimum surface tension about the equilibrium condition in the case of simple shear flow, is not maintained when electric field is applied. Also the gradients in surface tension $ (|\gamma_\text{max}-\gamma_\text{min}|) $ become progressively higher as the influence of electrical stresses increase relative to their hydrodynamic counterparts.

The above characteristics of $\Gamma (\theta,\phi)$ under different conditions become clear by looking  into the  \eqref{eq:leading-surf-harmonics}. It shows that different modes in $\Gamma^{(Ca)}$ contain the information of various aspects governing the flow physics. While $  \Gamma^{(Ca)}_{2,0} $ and $  \Gamma^{(Ca)}_{2,2} $ are only present when an external electric field is applied, the harmonic $  \hat{\Gamma}^{(Ca)}_{2,2} $ has a non-zero component arising solely from hydrodynamic origin. Again, the surfactant harmonics are strongly dependent on the tilt angle of the applied electric field. $  \Gamma^{(Ca)}_{2,0} $ appears for any configuration of the electric field, but $  \Gamma^{(Ca)}_{2,2} $ vanishes completely when the tilt angle becomes $ \Phi_t=\pi/4$ or $ 3\pi/4 $. Similarly the electric component of $\hat{\Gamma}^{(Ca)}_{2,2}$ becomes zero when the electric field is applied in either horizontal $ (\Phi_t=0) $ or vertical $(\Phi_t=\pi/2)$ configuration.
In order to investigate the physics behind the coupled behavior  of imposed linear shear and electric field we appeal to the flow characteristics around the drop originating from the individual effects \citep{Mandal2017a,Vlahovska2011}. The linear shear flow can be decomposed as a combination of a pure straining and pure rotational flow. On the other hand, the externally applied tilted electric field can give rise to either a uniaxial (for $ R>S $) or biaxial straining (for $ R<S $) flow around the drop. Hence, under the simultaneous action of shear flow and electrically induced flow, the surfactant molecules assume diverse redistribution patterns, the final form of which is determined by a combined interfacial convection and diffusive transport.
\subsection{Leading order bulk rheology}
\label{ssec:bulk-sph}
\begin{figure}[!htb]
	\centering
\begin{subfigure}[!htb]{0.3\textwidth}
		\centering
		\includegraphics[width=1.03\textwidth]{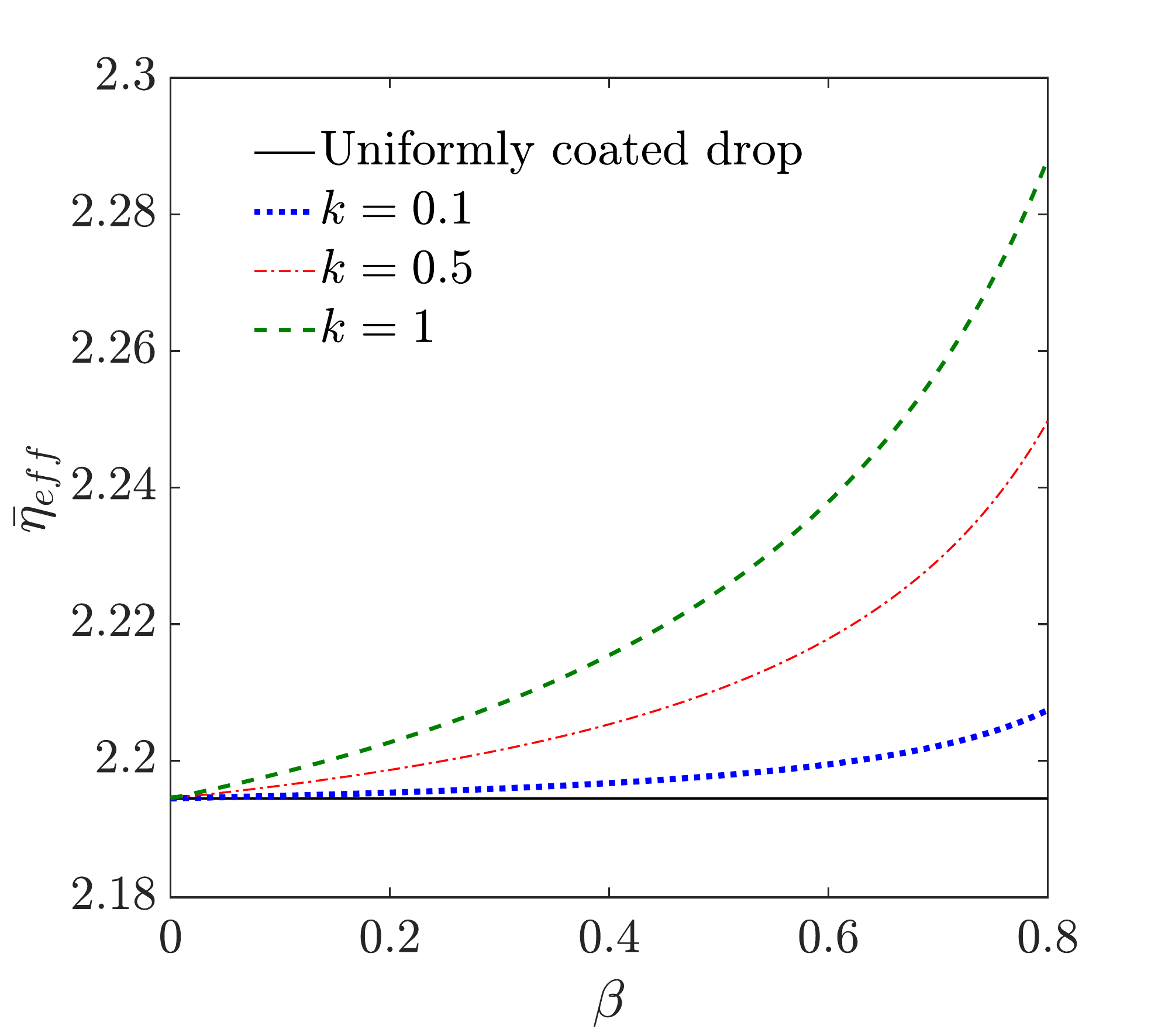}	
		\vspace{3ex}
		\caption{System-A}
		\label{fig:eff_vs_beta_vary_k_HY_A}
\end{subfigure}
\begin{subfigure}[!htb]{0.3\textwidth}
		\centering
		\hspace{-12pt}
		\includegraphics[width=1.115\textwidth]{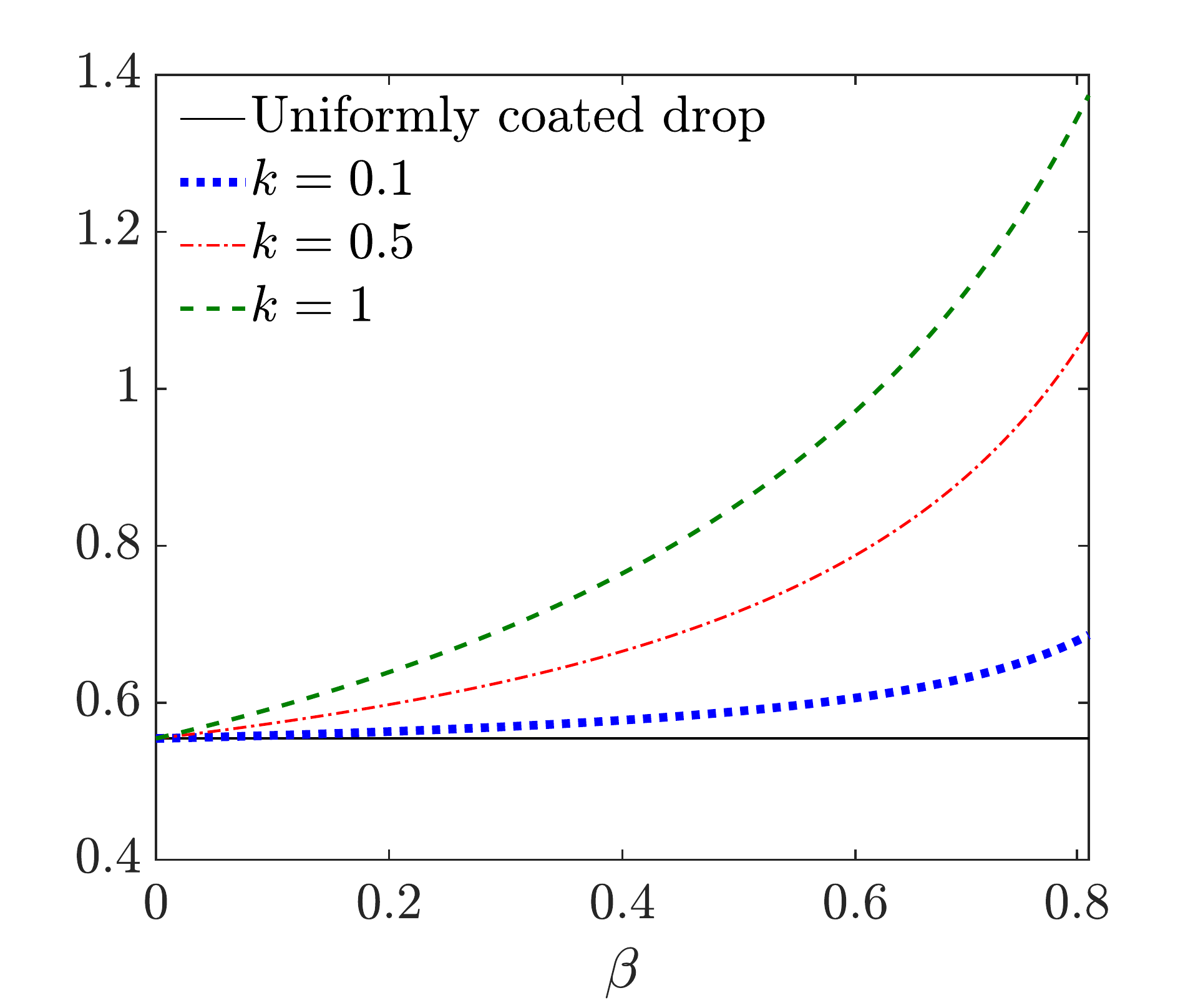}
		\vspace{3ex}
		\caption{System-B}
		\label{fig:eff_vs_beta_vary_k_B}
\end{subfigure}
\begin{subfigure}[!htb]{0.3\textwidth}
	\centering
	\includegraphics[width=1.077\textwidth]{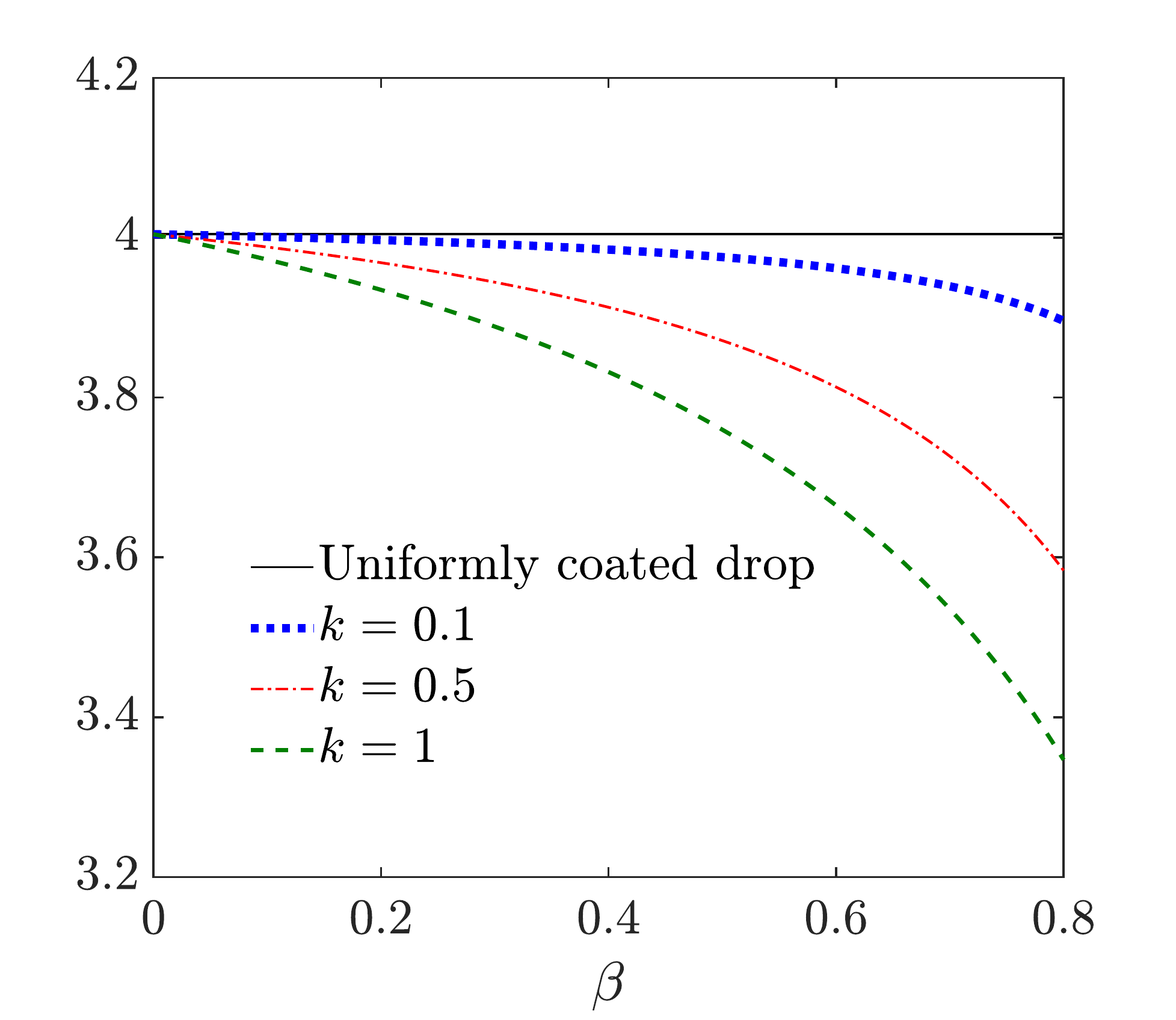}
	\vspace{3ex}
	\caption{System-C}
	\label{fig:eff_vs_beta_vary_k_C}
\end{subfigure}
\vspace{-1.5ex}
	\caption{Leading order normalized effective shear viscosity, $ \bar{\eta}_\text{eff}= (\eta^{(0)}_\text{eff}-1)/\nu $ vs. elasticity parameter for different physicochemical constant, $ k $. The electrohydrodynamic parameters in subplots (a),(b) and (c) correspond to system-A, B and C, respectively.  In both the figures a Mason number of $ M=2$ is chosen.} 
	\label{fig:eff_vs_beta_vary_k}
\end{figure}
In figures~\ref{fig:eff_vs_beta_vary_k} (a) to (c) the contribution of non-uniform surfactant distribution on the drop surface, in modulating the leading order normalized effective shear viscosity $ (\bar{\eta}_\text{eff}^{(0)}= (\eta^{(0)}_\text{eff}-1)/\nu) $, is described. It is observed from figures~\ref{fig:eff_vs_beta_vary_k} (a) and (b)  that  increase in either of the parameters $ \beta $  and $ k $ causes an enhancement in the effective viscosity.  In contrast,  figure~\ref{fig:eff_vs_beta_vary_k} (c) depicts that the surfactant characterization parameters have a completely reverse effect on $ \bar{\eta}_\text{eff} $.   To understand the effects of the surfactant characterization parameters $ \beta $ and $ k $, we appeal to figure~\ref{fig:SC_ST_var_beta_k} where the variations in surfactant concentration and surface tension are shown. The surface elasticity number $ \beta $ signifies the sensitivity of the surface tension on the local surfactant concentration. From figures~\ref{fig:SC_ST_var_beta_k}(a) and (b) we observe that an increase in $ \beta $ results in an enhanced gradient in surface tension $ (|\gamma_\text{max}-\gamma_\text{min}|)$, 
but 
the deviations in the surfactant concentration from the equilibrium condition $ (|\Gamma_\text{max}-\Gamma_\text{min}|)$ get suppressed with increasing $\beta$. This behavior is consistent with the work of \citet{Li1997} where the surfactant effects had been analyzed for the case of a drop in simple shear flow. Greater amount of surface tension gradient leads to an enhancement in Marangoni convection which in turn opposes  the convective transport of surfactant. As a result the distribution of $ \Gamma $ reaches towards equilibrium. An increase  in $k$, on the other hand, signifies a corresponding increase in the convective transport mechanism of the surfactants (please refer to \eqref{eq:k-def}).  Thus for a fixed $ \beta $, when $ k $ is increased, the  convective transport of surfactants  dominates over the effect due to increased surface tension gradient, resulting in greater undulations in both $ \Gamma $ and $ \gamma $ from the equilibrium condition as portrayed in figures~\ref{fig:SC_ST_var_beta_k}(c) and (d), respectively. Now the Marangoni effect due to non-uniformity in surfactant distribution  modifies the balance between the tangential component of the hydrodynamic and electrical Maxwell stresses at the interface (please refer to \eqref{eq:t_stress_bal}). 

\begin{figure}[!htb]
	\centering
		\begin{subfigure}{0.45\textwidth}
		\centering
		\includegraphics[width=01.2\textwidth]{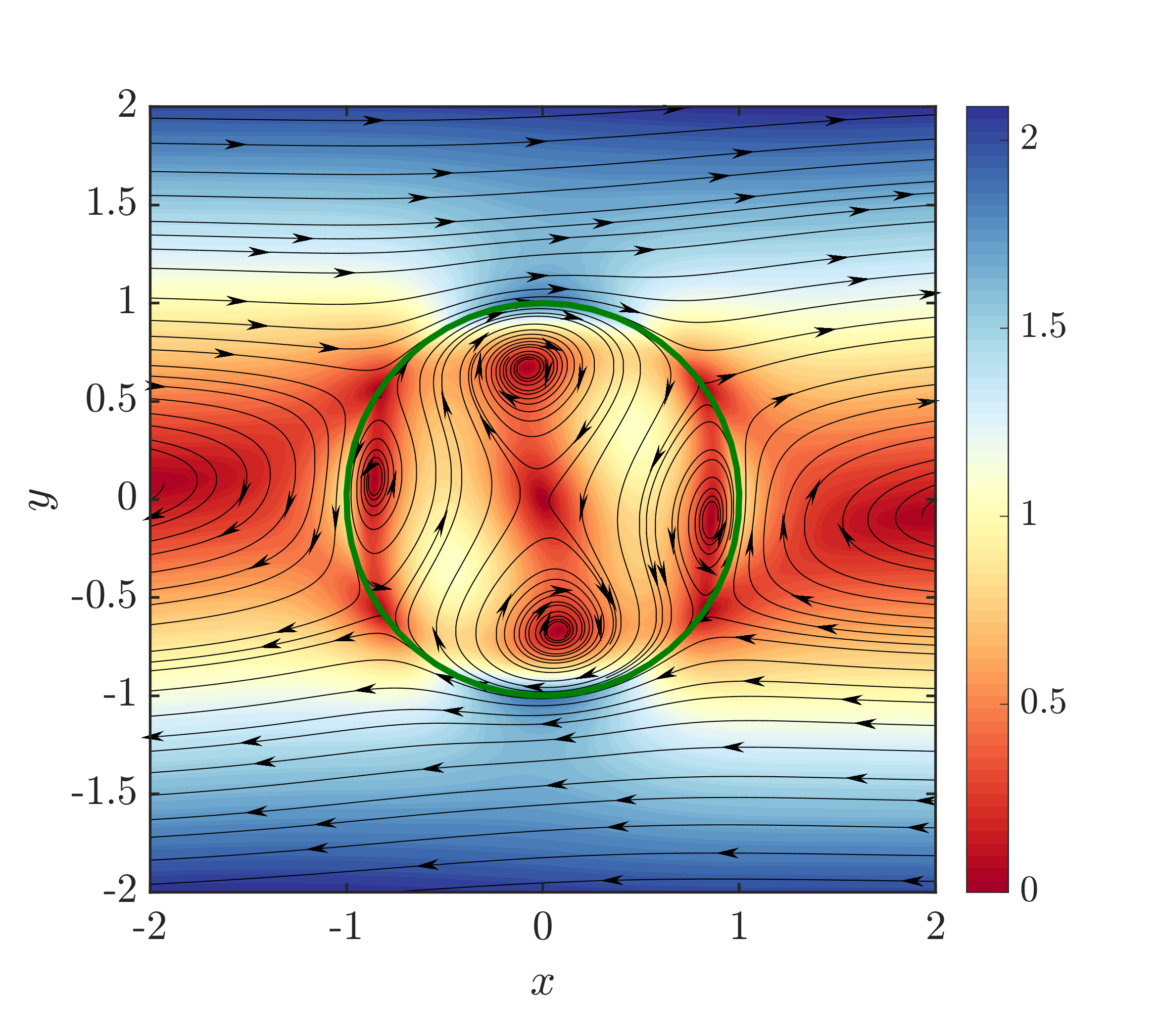}
		\vspace{5.5ex}
		\caption{Clean, system-B}
		\label{fig:stream_B_M_5_tilt_piby4_clean}
	\end{subfigure}	
	\qquad
	\begin{subfigure}{0.45\textwidth}
		\centering
		\includegraphics[width=1.2\textwidth]{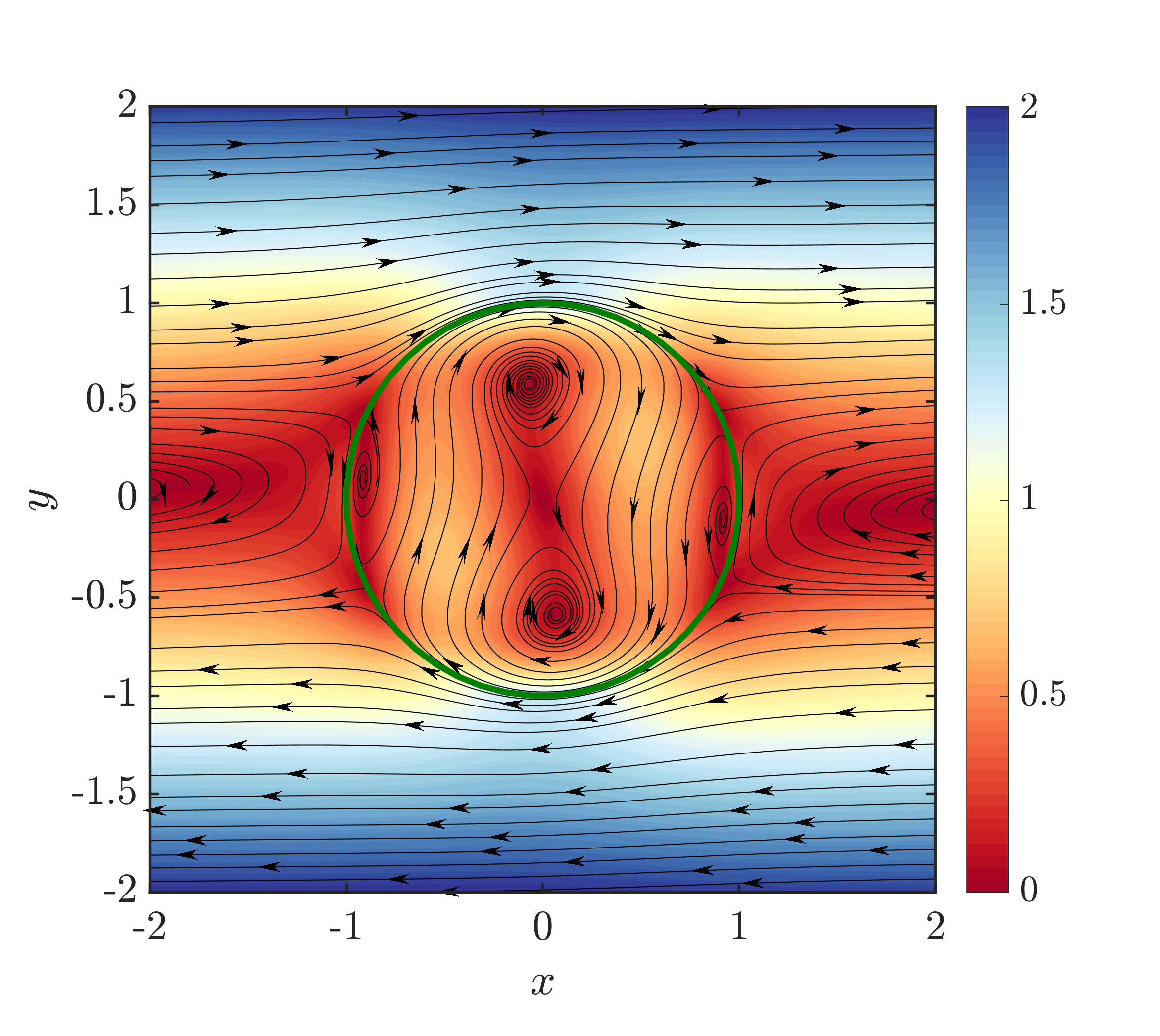}		
		\vspace{5.5ex}
		\caption{Surfactant-coated, system-B}
		\label{fig:stream_B_M_5_tilt_piby4_surf}
	\end{subfigure}
	\begin{subfigure}{0.45\textwidth}
		\centering
		\includegraphics[width=01.2\textwidth]{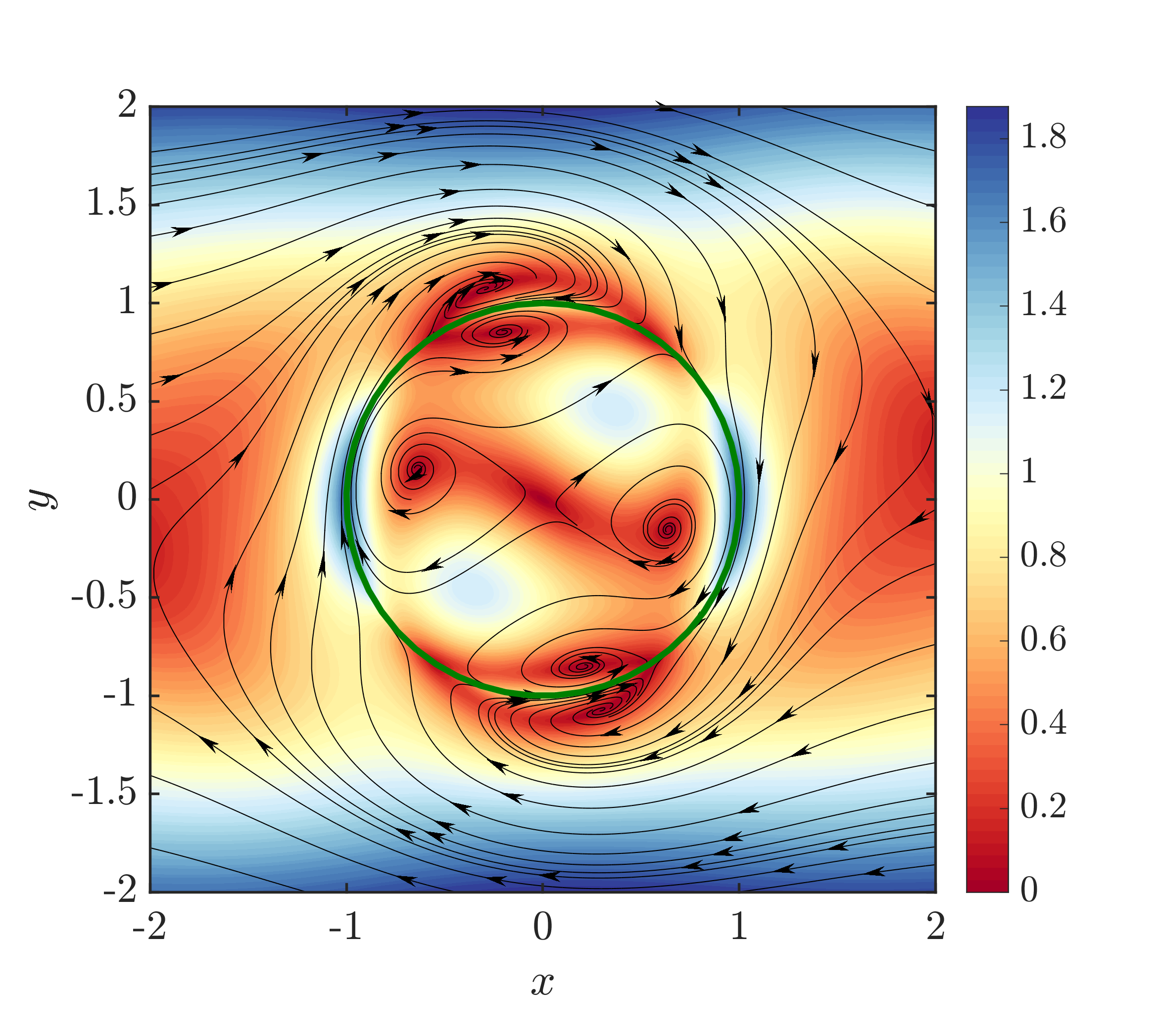}
		\vspace{5.5ex}
		\caption{Clean, system-C}
		\label{fig:stream_C_M_5_tilt_piby4_clean}
	\end{subfigure}	
	\qquad
	\begin{subfigure}{0.45\textwidth}
		\centering
		\includegraphics[width=1.2\textwidth]{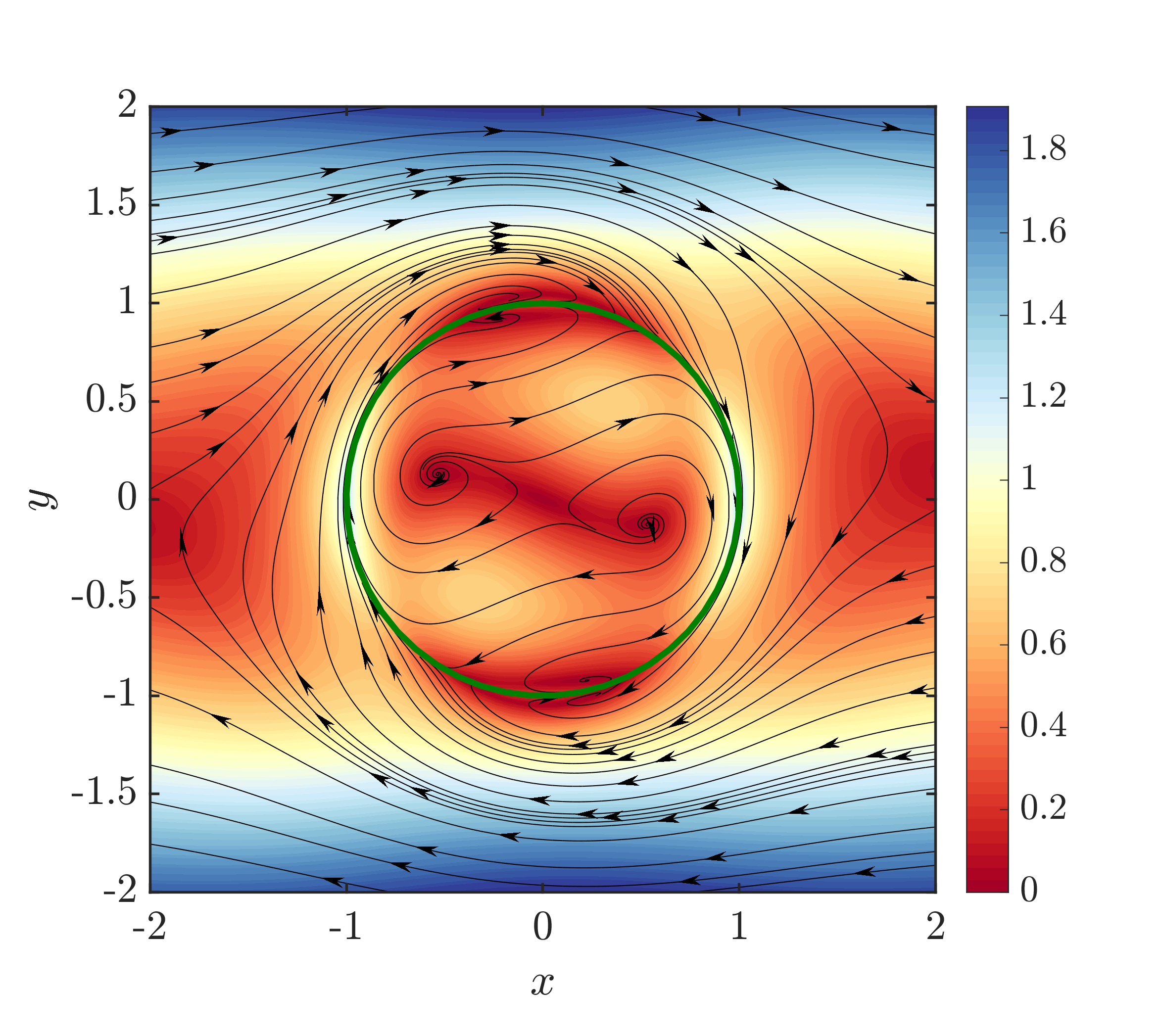}		
		\vspace{5.5ex}
		\caption{Surfactant-coated, system-C}
		\label{fig:stream_C_M_5_tilt_piby4_surf}
	\end{subfigure}
	\vspace{-2ex}
	\caption{Comparison of leading order flow fields in and around a drop for clean and surfactant-coated case. The streamlines are shown in the plane of shear and with respect to a body-fitted reference frame. The color-bars show the variation in velocity magnitude, while the green circle in each figure denotes the drop-matrix interface. Subplots (a),(b) are for $ M=5,$ system-B and (c),(d) are for $ M=5,$ system-C. The parameters are chosen as $ \Phi_t=\pi/4, \beta=0.8$ and $k=1$.} 
	\label{fig:B_C_streamline}
\end{figure}

In order to visualize the effect of surfactant-coating on the drop surface, in \ref{fig:B_C_streamline}(a) to (d) we  demonstrate the flow fields for some representative cases. These figures show that the orientation of the vorticity poles for system-B (figures~\ref{fig:B_C_streamline}(a), (b)) are vertical while it becomes horizontal for system-C (figures~\ref{fig:B_C_streamline}(c), (d)). The surfactants create further alterations in both the velocity magnitude and the pattern of the disturbance flow  around the drop. This subsequently causes an enhancement or reduction in the viscous dissipation of the bulk mixture and finally a corresponding change in the effective viscosity results.

\begin{figure}[!htbp]
	\centering
	\begin{subfigure}[!h]{0.4\textwidth}
		\centering
		\includegraphics[width=1.1\textwidth]{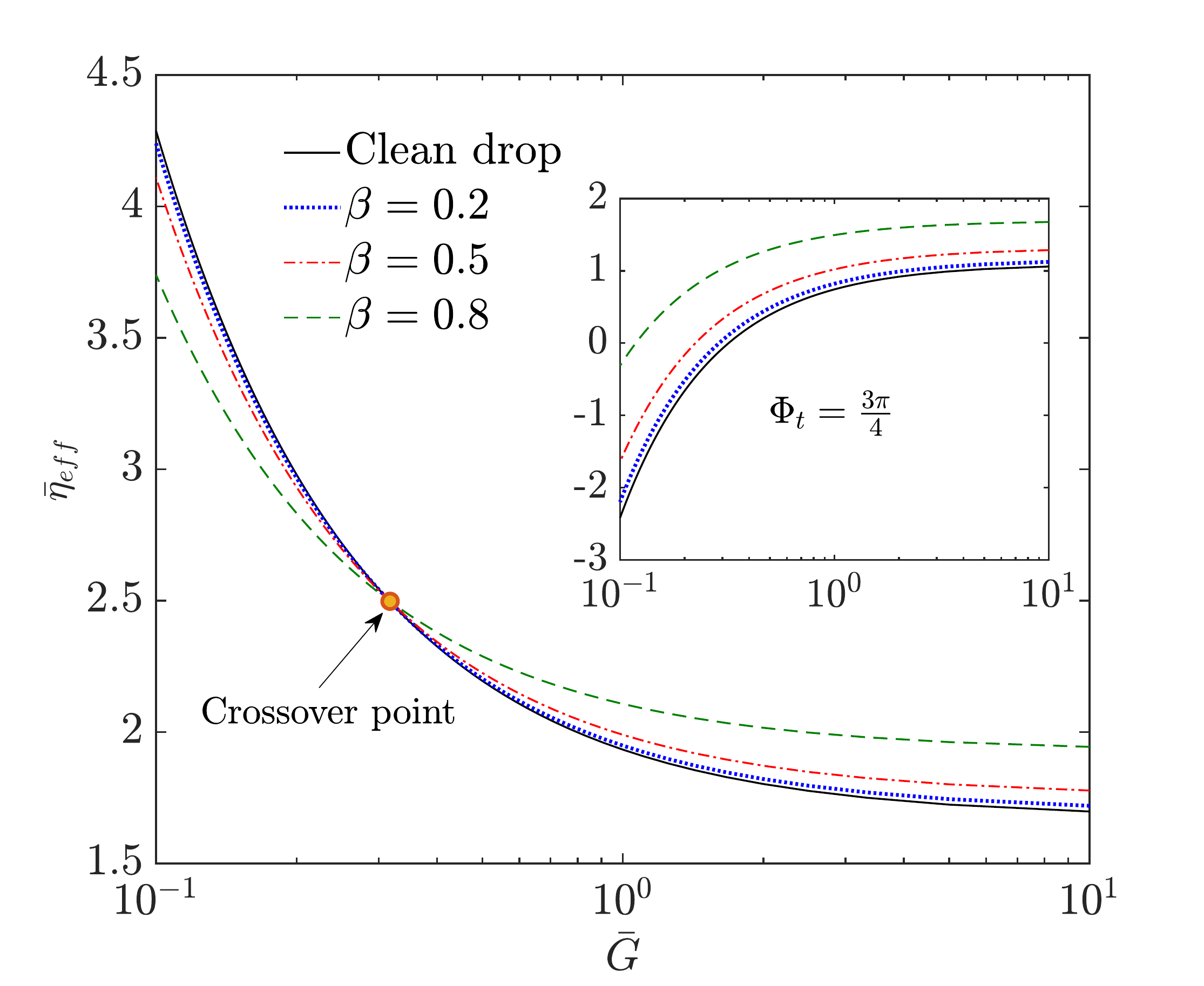}		
		\vspace{7ex}
		\caption{System-A}
		\label{fig:eff_corr_vs_M_tilt_piby4_HY_A}
	\end{subfigure}
	\qquad \quad
	\begin{subfigure}[!htbp]{0.4\textwidth}
		\centering
		\vspace{0.5ex}
		\includegraphics[width=1.17\textwidth]{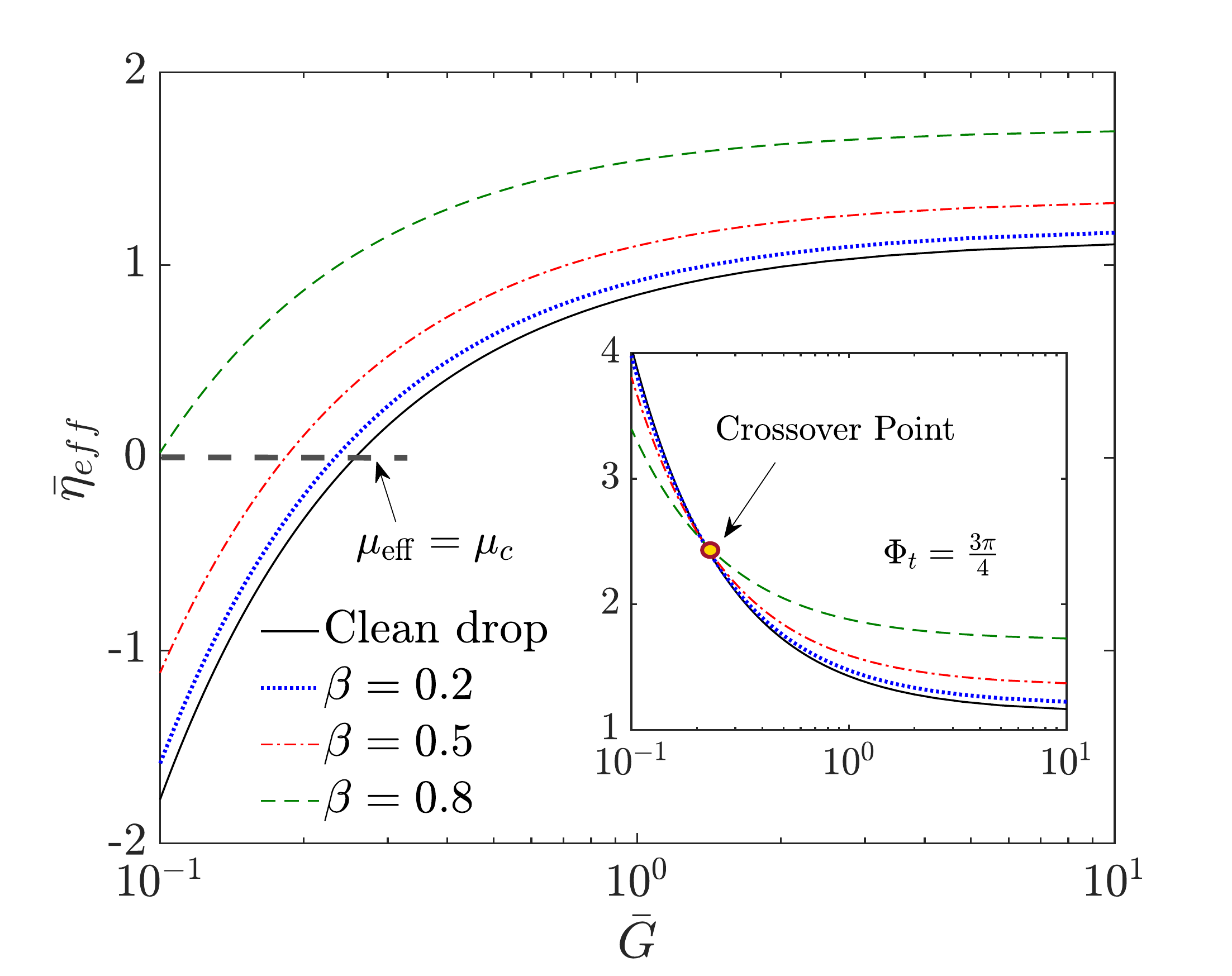}
		\vspace{7ex}
		\caption{System-B}
		\label{fig:eff_vs_M_vary_beta_B}
	\end{subfigure}
	\\
	\begin{subfigure}[!htbp]{0.4\textwidth}
		\centering
		%	\hspace{15ex}
		\includegraphics[width=1.06\textwidth]{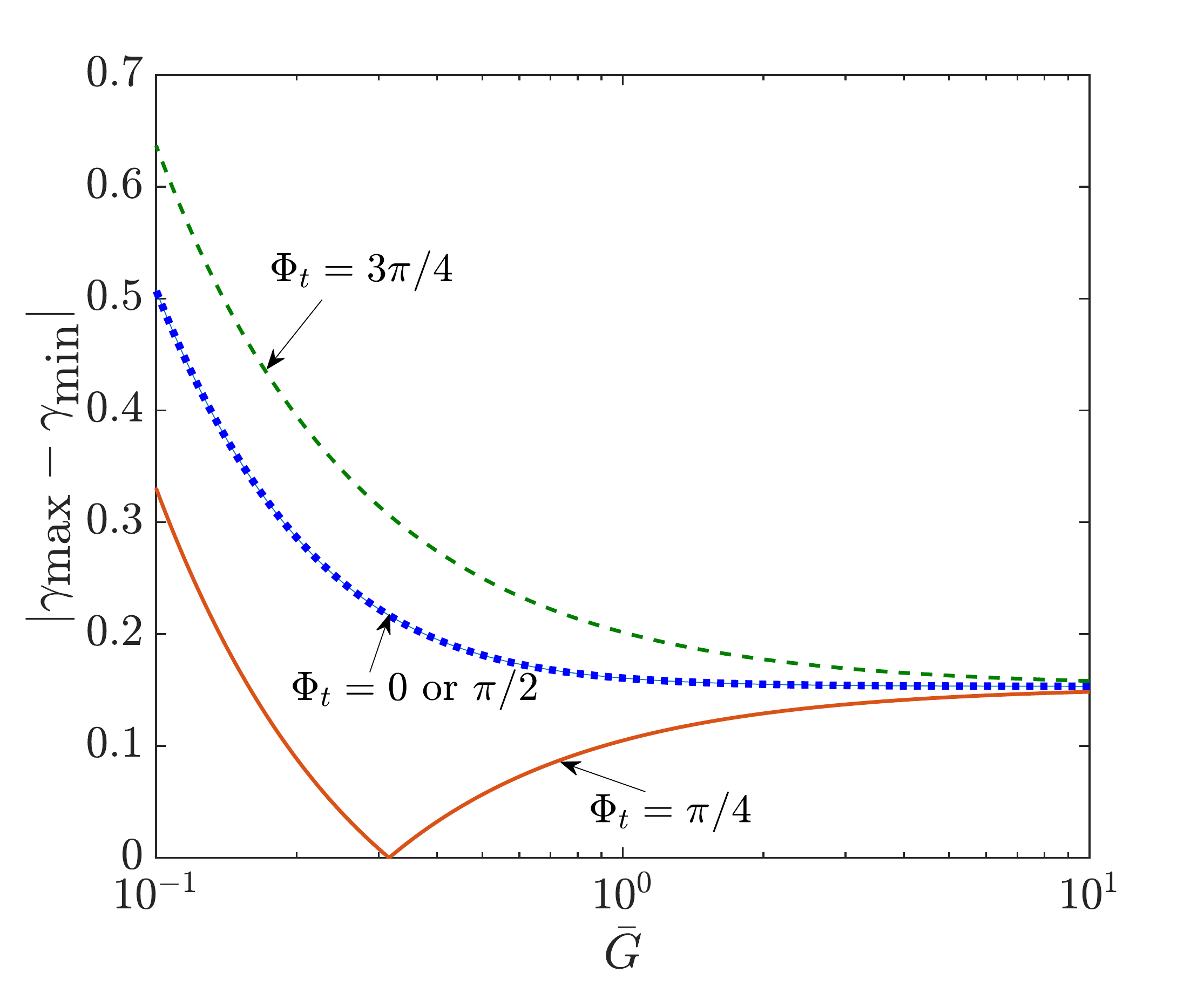}
		\vspace{7ex}
		\caption{}
		\label{fig:grad_T_vs_G_vary_tilt_HY_A}
	\end{subfigure}
\\
\begin{subfigure}[!htbp]{0.4\textwidth}
	\centering
%		\hspace{ex}
	\includegraphics[width=1.1\textwidth]{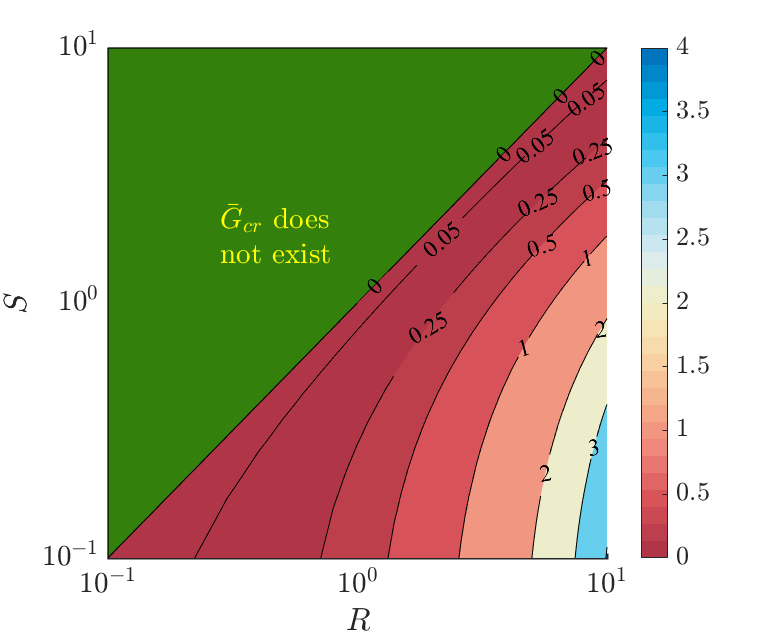}	\vspace{7ex}
	\caption{$\Phi_t={\pi}/{4}$}
	\label{fig:M_cr_R_S_contour}
\end{subfigure}
\qquad %\quad
\begin{subfigure}[!htbp]{0.4\textwidth}
	\centering
%	\hspace{15ex}
	\includegraphics[width=1.06\textwidth]{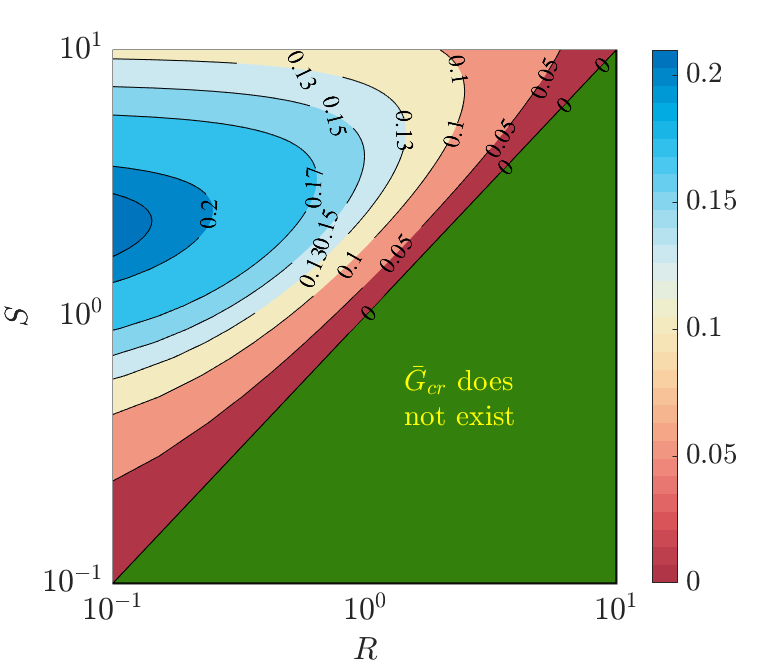}
	\vspace{7ex}
	\caption{$\Phi_t={3\pi}/{4}$}
	\label{fig:M_cr_R_S_contour_3pi_4}
\end{subfigure}
	\caption{Subplot (a) and (b): Variation of normalized effective shear viscosity $ (\bar{\eta}_\mathrm{eff}) $ with relative shear rate $ \bar{G}=1/M $ for different values of the elasticity parameter, $ \beta$. The subplots (a) and (b) are for system-A and B, respectively. In both the subplots $ \Phi_t={\pi}/{4}$ is chosen, while the inset to each subplot shows the corresponding variations for $ \Phi_t={3\pi}/{4}$. Subplot (c): Variation of surface tension gradient $ (|\gamma_\text{max}-\gamma_\text{min}|) $ with relative shear rate for various $ \Phi_t $. Here the demonstration is for system-A. Subplot (d) and (e) show the maps of the critical relative shear rate on the $ (R,S) $ plane for electric field tilt angle, $ \Phi_t={\pi}/{4} $ and $ \Phi_t={3\pi}/{4}$, respectively. } 
	\label{fig:eff_vs_shear_rate}
\end{figure}

The dependence of the normalized effective shear viscosity, $( \bar{\eta}_\text{eff}= (\eta^{(0)}_\text{eff}-1)/\nu) $ on the shear rate is shown in figures~\ref{fig:eff_corr_vs_M_tilt_piby4_HY_A} and \ref{fig:eff_vs_M_vary_beta_B} for two different combinations of electrohydrodynamic parameters. First we define a relative hydrodynamic shear rate with respect to the electrical stress as $\, \bar{G}=1/M=\dfrac{\mu_c \widetilde{G}}{\epsilon_c \widetilde{E}_\infty^2}$.  It was previously reported \citep{Mandal2017a} that in the leading  order, the effective viscosity becomes a function of  relative shear rate $ \bar{G}$ only when the electric field is tilted with respect to direction of the applied shear flow, while the emulsion shows a Newtonian behavior when the electric field is applied in either along the direction of shear flow or its gradient. Now a careful inspection these figures reveals that the non-uniformity surfactant distribution, has a potential in altering the non-Newtonian shear rate dependency of the effective shear rate. It is quite interesting to find from figure~\ref{fig:eff_corr_vs_M_tilt_piby4_HY_A} that for a tilt angle of external electric field, $\Phi_t=\pi/4$  there exists a crossover point at $ \bar{G}\sim 0.2 $ before which an increase in $ \beta $ causes the effective shear viscosity to decrease in comparison to an uncontaminated drop, while beyond this point a corresponding increase in $\bar{\eta}_\text{eff}$ results with similar changes in $\beta$. However  for $ \Phi_t=3\pi/4 $, $\bar{\eta}_\text{eff}$ shows a steady enhancement with $ \beta$.  In this case, for a fixed relative shear rate, the variation in $ \beta $ can lead to a change in  effective shear viscosity in such an extent that it becomes greater than  the   viscosity of the continuous medium $ \bar{\eta}_\text{eff}>0 \;\text{or},\; \mu_\text{eff}>\mu_c$, getting increased from a condition of $ \bar{\eta}_\text{eff}<0 \;\text{or},\; \mu_\text{eff}<\mu_c$ for a clean drop.
 The role of the electric field tilt angle gets reversed for system-B, as shown in figure~\ref{fig:eff_vs_M_vary_beta_B}.

In the EHD literature the flow behavior is often shown to be of opposite nature when the electrical property combinations give rise to a positive ($ R>S $) or negative ($ R<S $) polarity \citep{Taylor1966,Tsukada1993,Salipante2010,Mandal2017a}. Based on the polarity, the direction of the electrical traction at the interface gets altered. However,  comparing the present results in figure~\ref{fig:eff_vs_beta_vary_k} for different electrohydrodynamic systems, it is observed that the consequence of the Marangoni effect cannot be simply categorized based on the condition of $ R>S $ or $ R<S $. Although system-A and B have properties $ R<S $ and $ R>S $, respectively, they exhibit similar trend of Marangoni effect, while both systems-A, C show opposite behavior despite having same polarity. Along similar lines, in figures~\ref{fig:eff_corr_vs_M_tilt_piby4_HY_A} and \ref{fig:eff_vs_M_vary_beta_B} the Marangoni effect is observed to be dependent on the electric field tilt angle ($ \Phi_t $) as well. Inspired by such observations we  analyze the expression of the leading order effective shear viscosity, $ {\eta}^{(0)}_\text{eff} $  in	  \eqref{eq:leading--eta}.  
		We obtain a unique discriminating factor in terms of the crossover relative shear rate $\bar{G}_\mathrm{cr}$ given as
	\begin{equation}
	\label{eq:M_cr_dnv}
	\bar{G}^{(0)}_\mathrm{cr}=\frac{9}{5}\,{\frac {\sin(2\Phi_t) \left( S-R\right)}{\left( R+2 \right) ^{2}}}
	\end{equation}
	Similar a discriminating factors have been previously obtained by others in the context to drop deformation \citep{Taylor1966} and surface charge convection \citep{Xu2006a}. 
	If $\Phi_t<\dfrac{\pi}{2}$ then $\sin(2\Phi_t)$ is a positive quantity. Thus for a realistic  $\bar{G}_\mathrm{cr}^{(0)}$ to exist, the condition which must be satisfied is  $R<S$. Again for $\Phi_t>\dfrac{\pi}{2}$ the property ratios must fulfill the condition $R>S$. Hence the Marangoni effect on the electrically induced emulsion viscosity can be suitably controlled by finely tuning the combination of the electric field direction and electrical properties of the drop-surrounding fluid pair. 
To understand such a behavior of Marangoni effect we plot the surface tension gradient $ (|\gamma_\text{max}-\gamma_\text{min}|) $ with relative shear rate for different values of the tilt angle in figure~\ref{fig:grad_T_vs_G_vary_tilt_HY_A} for system-A. It shows that the surface tension gradient vanishes at a relative shear rate which is exactly equal to the crossover point observed in figure~\ref{fig:eff_corr_vs_M_tilt_piby4_HY_A}.  This indicates that due to competitive effects of the hydrodynamic and electrical stresses on the surface tension variation, for a specific relative shear rate, the Marangoni effect vanishes completely and the drop surface behaves as an uncontaminated one. The corresponding competing effects can be realized by referring back to \eqref{eq:leading--eta} where distinct functional forms of the hydrodynamic and electric surfactant contribution terms  $\left( \eta^{(0)}_\text{h,surf},  \eta^{(0)}_\text{e,surf} \right)$ appeared in the expression of $ \eta^{(0)}_\text{eff} $. The lesser extent of variation of $ \eta^{(0)}_\text{eff} $ with $ \bar{G}$, after the crossover point $ \bar{G}_{cr} $ is reached (observation from figure~\ref{fig:eff_corr_vs_M_tilt_piby4_HY_A}), can also be explained from figure~\ref{fig:grad_T_vs_G_vary_tilt_HY_A} where the range of surface tension gradient $ (|\gamma_\text{max}-\gamma_\text{min}|)$ becomes   lower after $ \bar{G}_{cr} $  than it was before $ \bar{G}_{cr} $.

 The value range of $ \bar{G}_{cr} ^{(0)}$ is observed to be  different in the two set of electrohydrodynamic parameters. In this context, the parametric mapping of the critical relative shear rate on the $ R-S $ plane is shown for two different tilt angles in figures~\ref{fig:M_cr_R_S_contour} and \ref{fig:M_cr_R_S_contour_3pi_4}, respectively. 
	It is to be noted that a critical normalized shear rate was obtained previously by \cite{Mandal2017a} in relation to a surfactant-free drop under the combined influence of electric field and shear flow. In that case, the critical parameter had been used to characterize whether the effective shear viscosity is greater (or lesser) than the suspending medium. In stark  contrast, the presently obtained critical relative shear rate is related to the surfactant induced Marangoni effect. To the best of our knowledge, for the first time we have obtained a discriminating factor in terms of a relative shear rate for which the Marangoni effect will have no contribution in altering the emulsion rheology.

\begin{figure}[!htb]	
	\centering
	\includegraphics[width=0.8\textwidth]{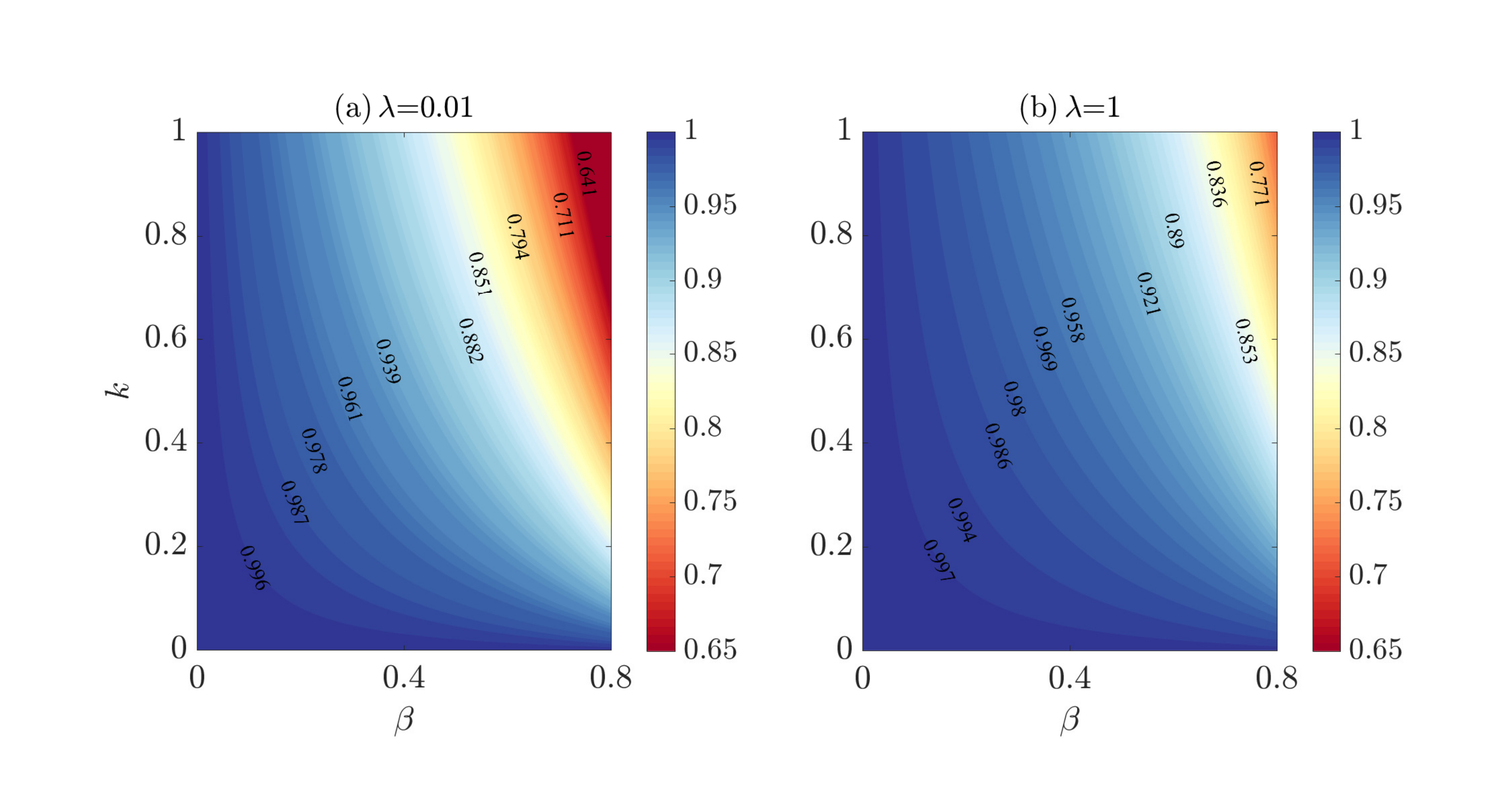}
		\vspace{-25pt} 
	\caption{Surfactant modification parameter for the normal stress differences of the emulsion $ (\mathcal{N}^{(0)}) $ on $ \beta - k $ plane for  different viscosity ratios ($ \lambda $).}
	\label{fig:correction-fact-N}
\end{figure}

%\FloatBarrier
The another important aspect of bulk rheology, i.e. the first $ (N_1^{(0)}) $ and second $ (N_2^{((0))}) $ normal stress differences are significantly affected by the presence of surfactant contamination on the drop surface. Equation~(38) shows that in the leading order the normal stress differences in the bulk rheology arise only from the electrohydrodynamic origins even when the surfactant effects are considered.  From the functional form of the $ \mathcal{N}^{(0)}_{surf} $ it is evident that for all the realistic parametric ranges of the parameters $ (0 \le \beta <1, k >0 \text{\,\;and\;} \lambda >0) $ the surfactant induced Marangoni effect always acts to reduce the normal stress differences, i.e. $\mathcal{N}^{(0)}_{surf}<1$. In figure~\ref{fig:correction-fact-N} a parametric map of the  common surfactant contribution factor $ \mathcal{N}^{(0)}_{surf} $ (defined in \eqref{eq:corr_N1_N2}) is shown on $ \beta-k $ plane for different viscosity ratios of the drop-matrix fluid pair. It shows that for lower viscosity of the drop phase as compared to the matrix fluid, the surfactant contribution is more prominently observed. In the limit of $ \lambda \to 0 $, when the drop tends  behave as a bubble, the surfactant effect is the maximum and we obtain $\displaystyle{\lim_{\lambda \to 0}} \; \mathcal{N}^{(0)}_{surf} = \dfrac {(1-\beta)}{
	\left(k-5 \right) \beta+5}.$ On the other limit of $ \lambda \to \infty $, the drop behaves like a particle and its surface becomes completely immobilized. In that case the normal stress differences vanish themselves for a spherical drop.
\subsection{Interplay between charge convection and surfactant transport}
\label{ssec:ReE_result}
\begin{figure}
	\centering
	\begin{subfigure}[!htb]{0.4\textwidth}
	\centering
	\includegraphics[width=01.0425\textwidth]{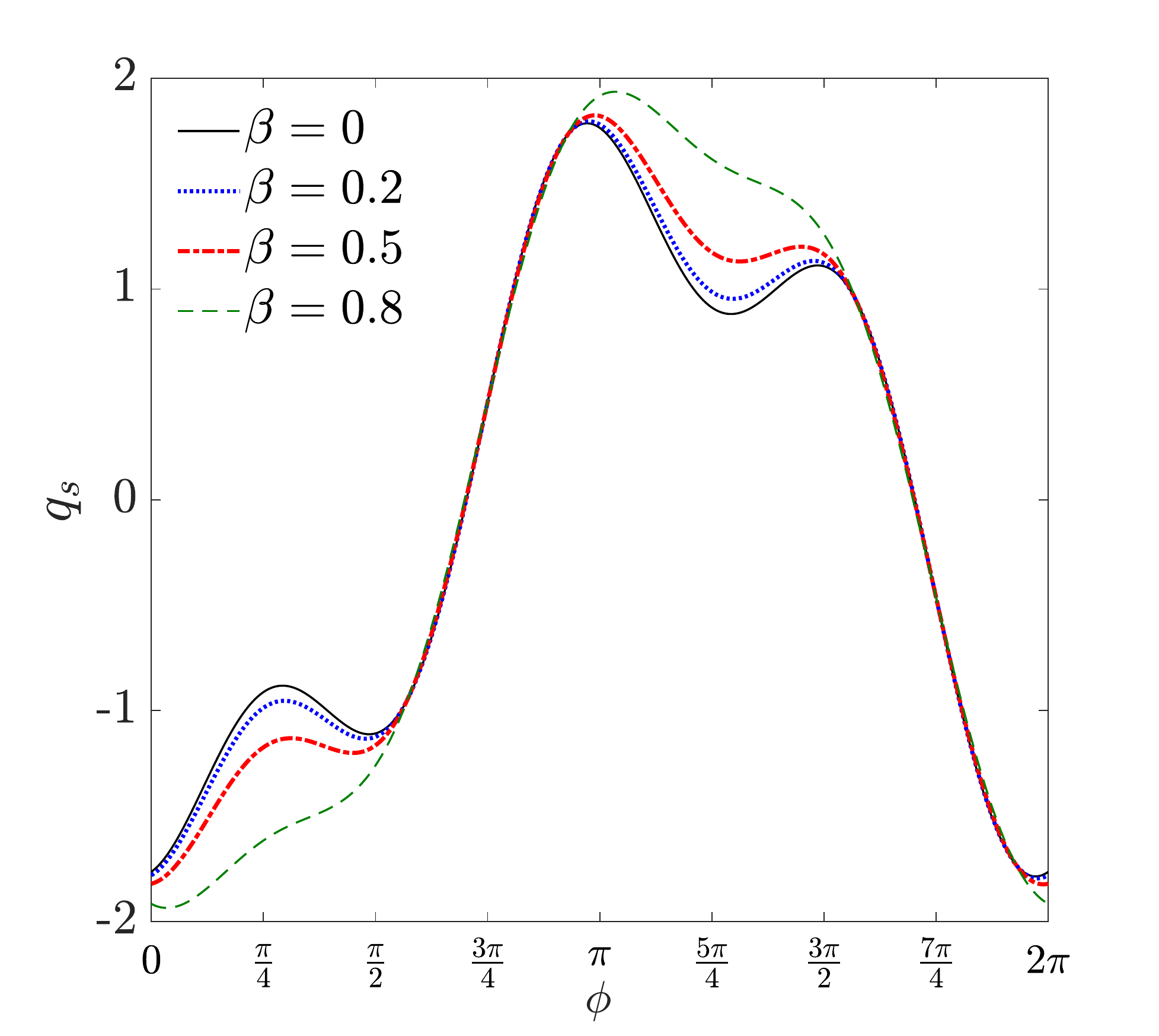}
	\vspace{5.5ex}
	\caption{}
	\label{fig:Qs_vs_PHI_vary_BETA_C}
\end{subfigure}	
	\qquad
\begin{subfigure}[!htb]{0.4\textwidth}
	\centering
	\includegraphics[width=1.1\textwidth]{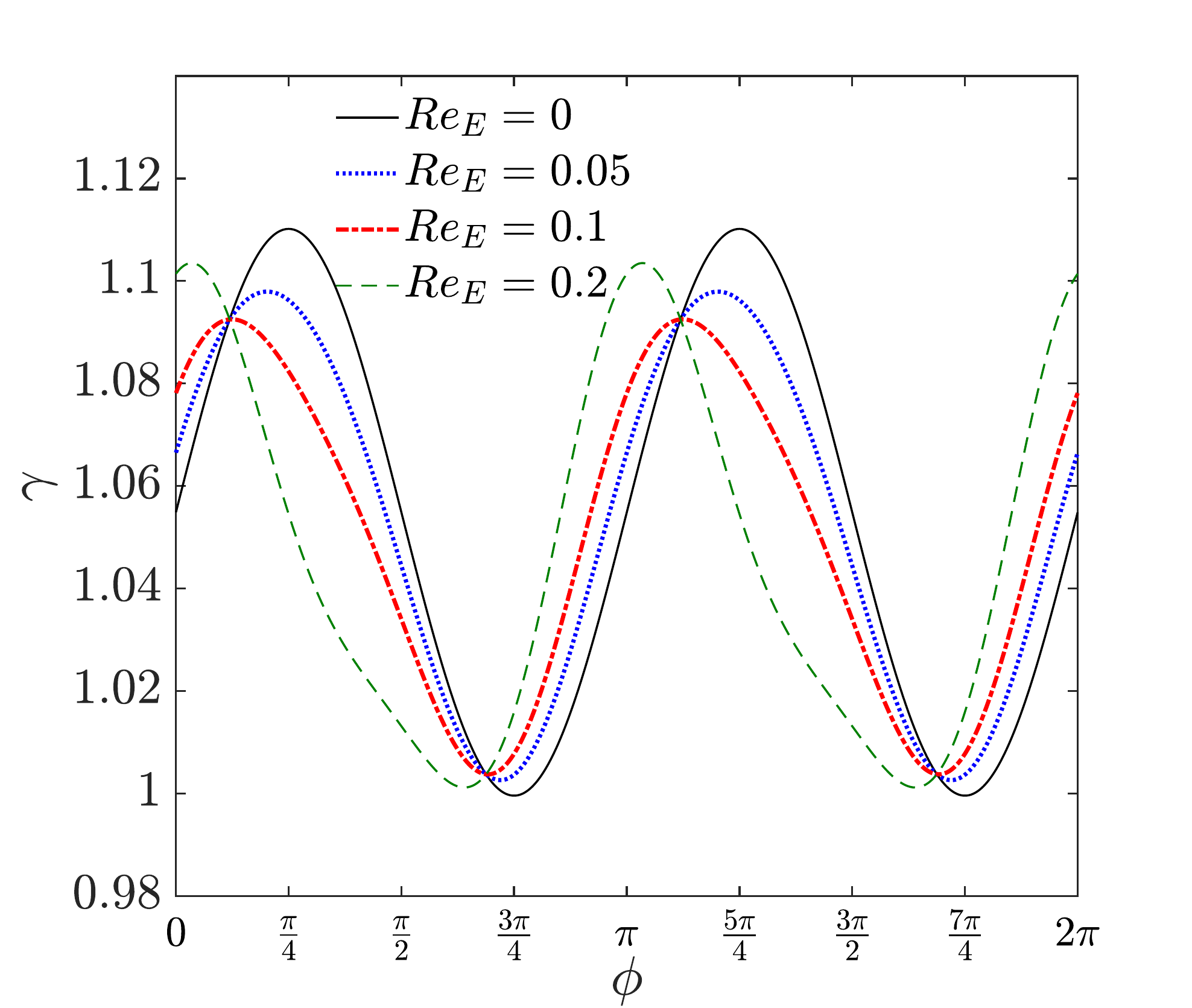}		
	\vspace{5.5ex}
	\caption{}
	\label{fig:ST_vs_PHI_vary_ReE_C}
\end{subfigure}
\vspace{-2ex}
	\caption{(a) Azimuthal variation of the surface charge density $ (q_S=q_S^{(0)}+Re_E\,q_S^{(Re_E)})$ for different values of the surface elasticity parameter, $\beta $. (b)  variation of the surface tension $ (\gamma=\gamma^{(0)}+Ca\,\gamma^{(Ca)}+CaRe_E\,\gamma^{(Re_E)}) $ for different values of electric Reynolds number $(Re_E)$.   The demonstrations are for system-C. Parameters are chosen as $ Re_E=0.2,M=2,k=1,\Phi_t=\pi/4$ and $ Ca=0.2$.} 
\end{figure}

For a clean drop, the presence of finite charge convection creates a redistribution of charges.
Following the discussion in section~\ref{ssec:sol_Re_E},  the $ O(Re_E) $ electrical potential and in turn the accumulated surface charge density become dependent on the leading order interfacial fluid velocity $(\mathbf{u}_S^{(0)})$, which is already affected by the surfactant effects (please refer to equation~\ref{eq:leading_us} for full expression). In figure~\ref{fig:Qs_vs_PHI_vary_BETA_C} we demonstrate the variations in the surface charge distribution $ (q_S=q_S^{(0)}+Re_E\,q_S^{(Re_E)}) $ for different surface elasticity parameter $ \beta $. It is observed that for a surfactant-contaminated drop, the redistribution of charges due to the presence of finite charge convection at the interface is affected by the Marangoni effects. On the other hand, the modulations in the electrohydrodynamic flow created by the finite charge convection, interacts with the  Marangoni convection and variations in the surface tension result. This behavior is shown in figure~\ref{fig:ST_vs_PHI_vary_ReE_C}.

\subsubsection{Order $\boldsymbol{Re_E}$ bulk rheology}
\label{sssec:eff_vis_Re_E}
\begin{figure}[!htb]
	\centering
	\begin{subfigure}{0.4\textwidth}
		\centering
		\includegraphics[width=1\textwidth]{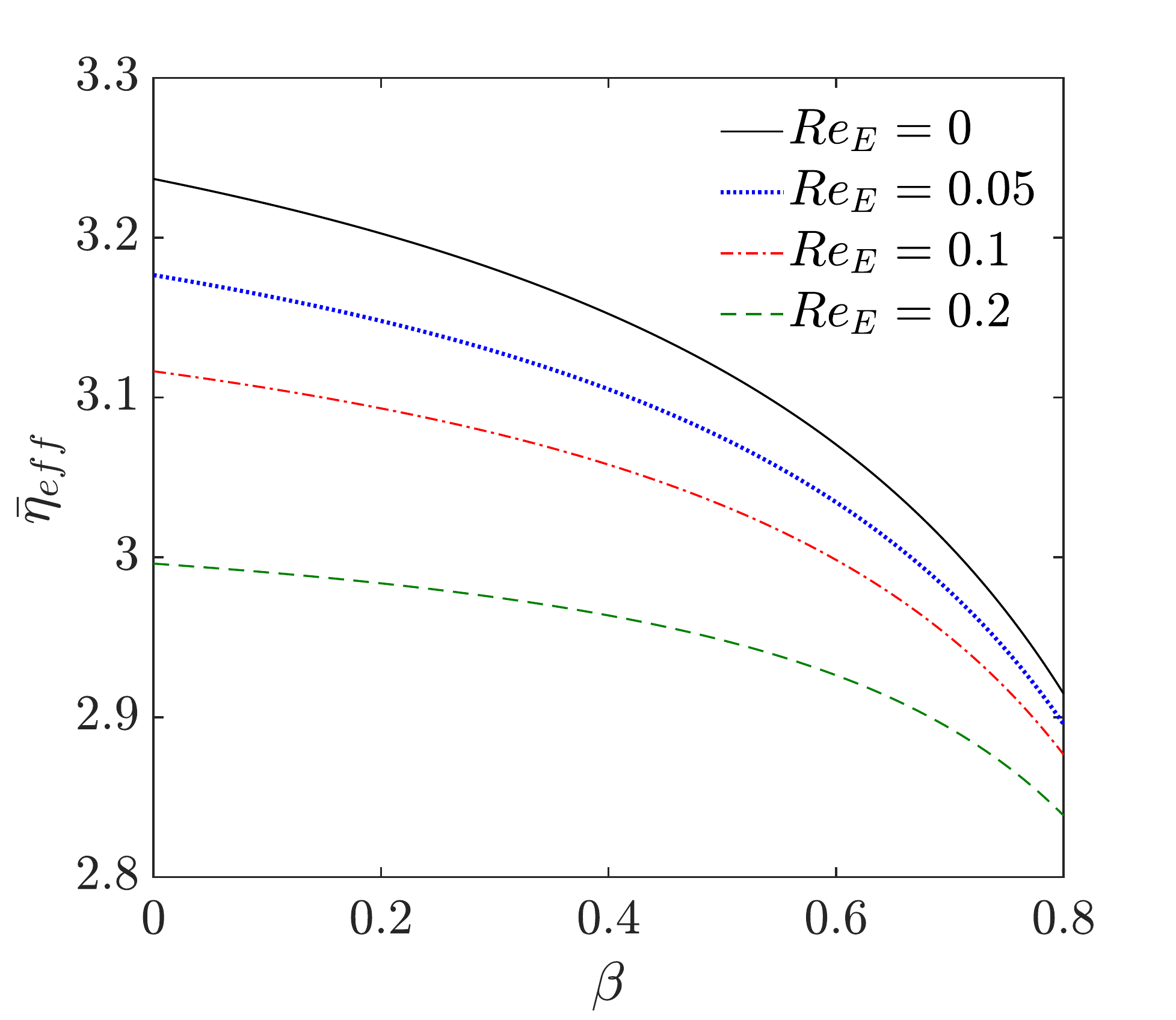}
		\vspace{10ex}
		\caption{}
		\label{fig:eff_vs_beta_vary_ReE_C}
	\end{subfigure}
\\
	\begin{subfigure}{0.4\textwidth}
		\centering
		\includegraphics[width=1\textwidth]{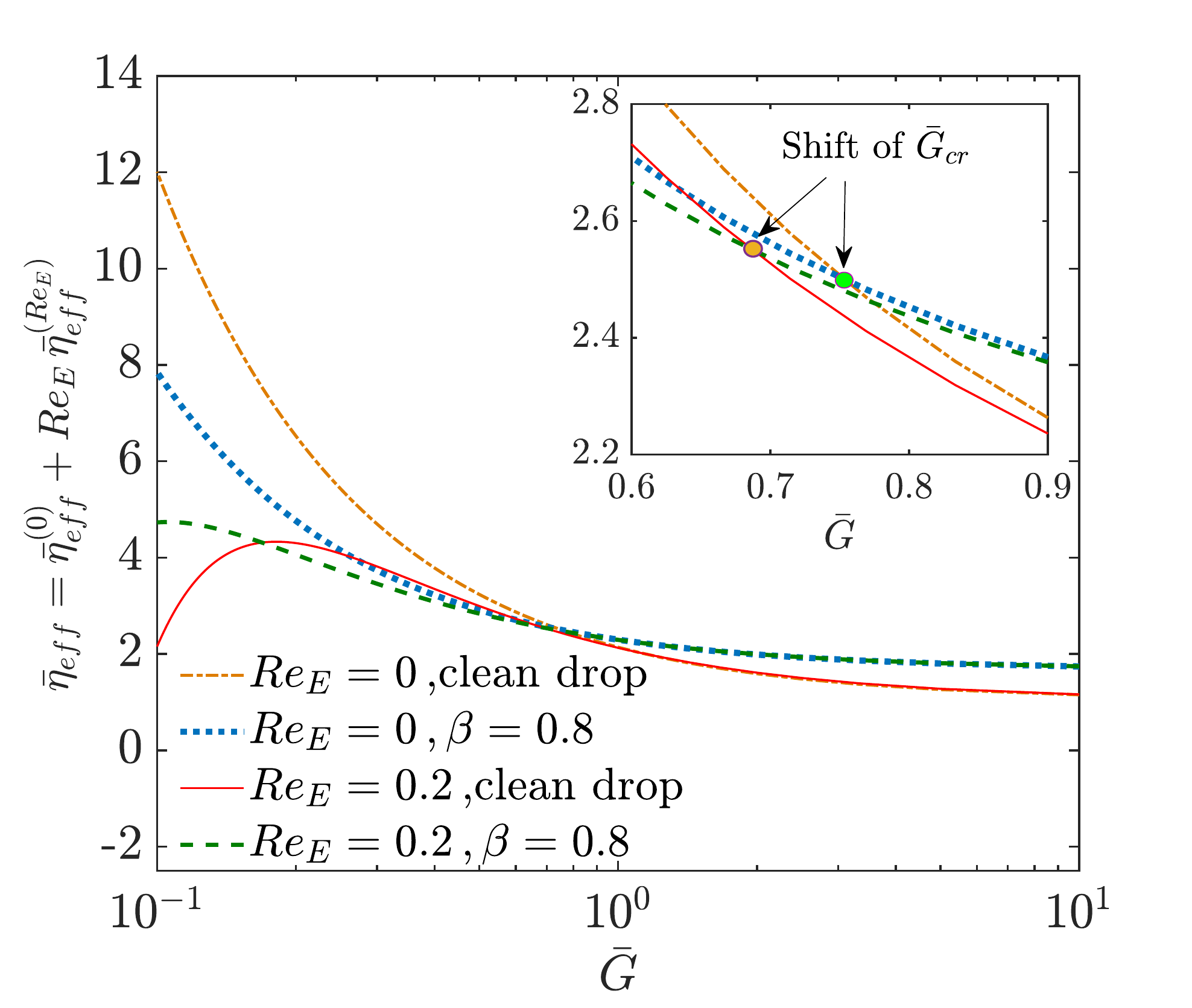}		
		\vspace{5ex}
		\caption{$ \Phi_t=\pi/4 $}
		\label{fig:eff_vs_M_vary_beta_C_ReE_added}
	\end{subfigure}
	\qquad
	\begin{subfigure}{0.4\textwidth}
		\centering
		\includegraphics[width=1\textwidth]{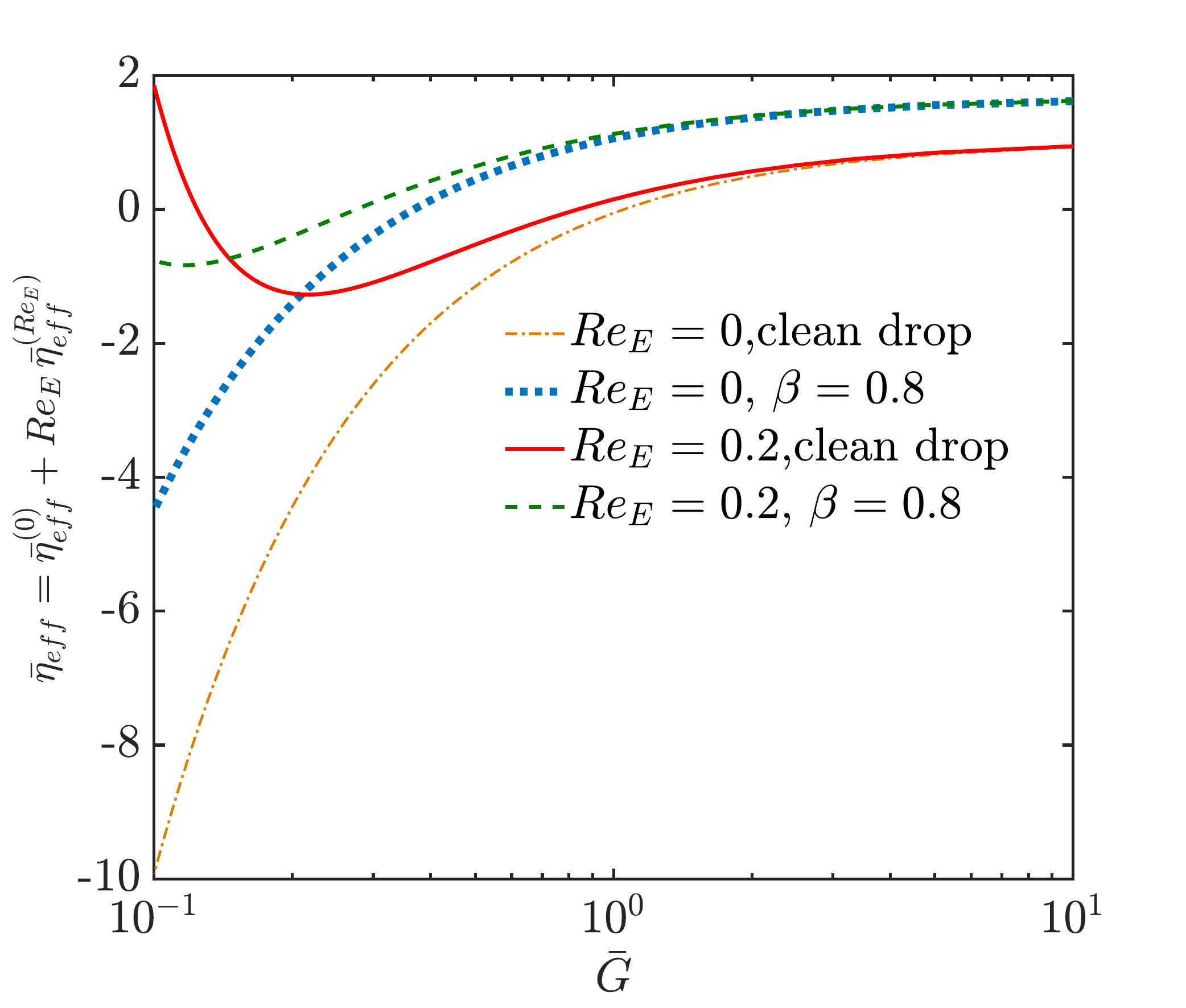}
		\vspace{5ex}
		\caption{$ \Phi_t=3\pi/4 $}
		\label{fig:eff_vs_M_vary_beta_C_ReE_added_3pi4}
	\end{subfigure}
	\vspace{-1.7ex}
	\caption{(a): Normalized effective shear viscosity ($\bar{\eta}_\text{eff}=\bar{\eta}^{(0)}_\text{eff}+Re_E\,\bar{\eta}^{(Re_E)}_\text{eff}$) variation with elasticity parameter $ \beta $ for different electric Reynolds number. Here $ M=2$ and $ \Phi_t=\pi/4 $. (b),(c): $\bar{\eta}_\text{eff}$ variation with relative shear rate, $(\bar{G})$ for different elasticity parameter, $ \beta $ and electric Reynolds number, $Re_E$. In the inset  of subplot (b) the shift of critical shear rate, $ \bar{G}_\text{cr} $ with $Re_E$ is highlighted. Here $ \Phi_t=\pi/4 $ and  $ \Phi_t=3\pi/4 $ in subplots (b) and (c), respectively. The other parameters in all the sub-figures are chosen following system-C.}
	\label{fig:eff-vs-M-ReE} 
\end{figure}

The coupled effects of charge convection and surfactant non-uniformity is described in  figure
\ref{fig:eff-vs-M-ReE}. Figure~\ref{fig:eff-vs-M-ReE}(a) depicts than increase in the parameter $ \beta $ suppresses the $ O(Re_E) $ modifications in $ \bar{\eta}_\text{eff}$.  To justify such phenomena we plot the variations in $ O(Re_E) $ surface velocity and the electrical traction at the interface in figure~\ref{fig:us_te} for different values of $\beta$. The figure shows that the charge convection triggered modifications in the flow characteristics (exemplified by modulations in $  u_{S,\phi}^{(Re_E)}$ and  $ [[T^E_\phi]]^{(Re_E)}$) are severely altered due to the surface elasticity parameter, $\beta$. 

The above interplay between the charge convection and Marangoni flow at the interface, also modifies the shear rate dependent response of the emulsion viscosity, as shown in figure~\ref{fig:eff-vs-M-ReE}. The critical relative shear rate corresponding to the crossover point of the Marangoni effect $ (\bar{G}_{cr}) $, gets shifted from its  leading  order counterpart (shown in figure~\ref{fig:eff-vs-M-ReE}(a)).
This effect can be substantiated by carefully inspecting the $ O(Re_E) $ effective shear viscosity ($\eta^{(Re_E)}_\mathrm{eff}$). Similar to the leading order, there exists a specific range of relative shear rate for which the Marangoni effect has hardly any role in modifying the $O(Re_E)$ effective viscosity of the medium ($\eta^{(Re_E)}_\mathrm{eff}$). However the critical shear rate in this case is also a function of the surfactant parameters and the viscosity ratio, as shown below: 
	\begin{equation}\label{key}
	\bar{G}^{(Re_E)}_{\mathrm{cr},\bar{\eta}}=\frac{R-S}{(R+2)^2}\frac{\sin(2\Phi_t)}{\delta_1\cos(2\Phi_t)+\delta_2},\; \mathrm{where}
	\end{equation}
	\begin{equation}\label{key}
	\delta_1=\frac{175}{36}{\frac { \left( 3R+4 \right) \left( \lambda
			+1 \right)  \left( R-2S-2 \right)  \left( \beta{k}-5\beta
			\lambda-5\beta+5\lambda+5 \right) }{  
			\left( 69{R}^{2}-54RS+130R-80S+40 \right)  \left( \beta{
				k}-10\beta\lambda-10\beta+10\lambda+10 \right) }}
	\end{equation}
	and
	\begin{equation}\label{key}
	\delta_2=-\frac{5}{108}{\frac { \left( 701\,{R}^{2}-506\,RS+1410\,R-760\,S+520 \right) 
		}{  \left( 69
			\,{R}^{2}-54\,RS+130\,R-80\,S+40 \right)   }}.
	\end{equation}

 In addition, comparing figures~\ref{fig:eff-vs-M-ReE}(a) and (b), we observe that the non-Newtonian shear thickening (shear  thinning) rheology   at  low shear rates (corresponding to the curve $ Re_E=0.2,$ clean drop), becomes a shear thinning (shear thickening) one in the presence of surfactant (refer to the curve $ Re_E=0.2, \beta=0.8$). This reversal in non-Newtonian behavior is absent when charge convection effects are insignificant.
%\begin{figure}[!htb]	
%	\centering
%	\includegraphics[width=0.45\textwidth]{eff_vs_M_vary_beta_C_ReE_added-eps-converted-to.pdf}
%	\vspace{-10pt} 
%	\caption{Normalized effective shear viscosity variation with relative shear rate for different elasticity parameter, $ \beta $ and electric Reynolds number, $Re_E$. In the inset the shift of critical shear rate, $ \bar{G}_\text{cr} $ with $Re_E$ is highlighted. Here $ \Phi_t=\pi/4 $ and the other parameters are chosen following system-C.}
%	\label{fig:eff-vs-M-ReE}
%\end{figure}

\begin{figure}
	\centering
	\begin{subfigure}[!htb]{0.4\textwidth}
		\centering
		\includegraphics[width=1\textwidth]{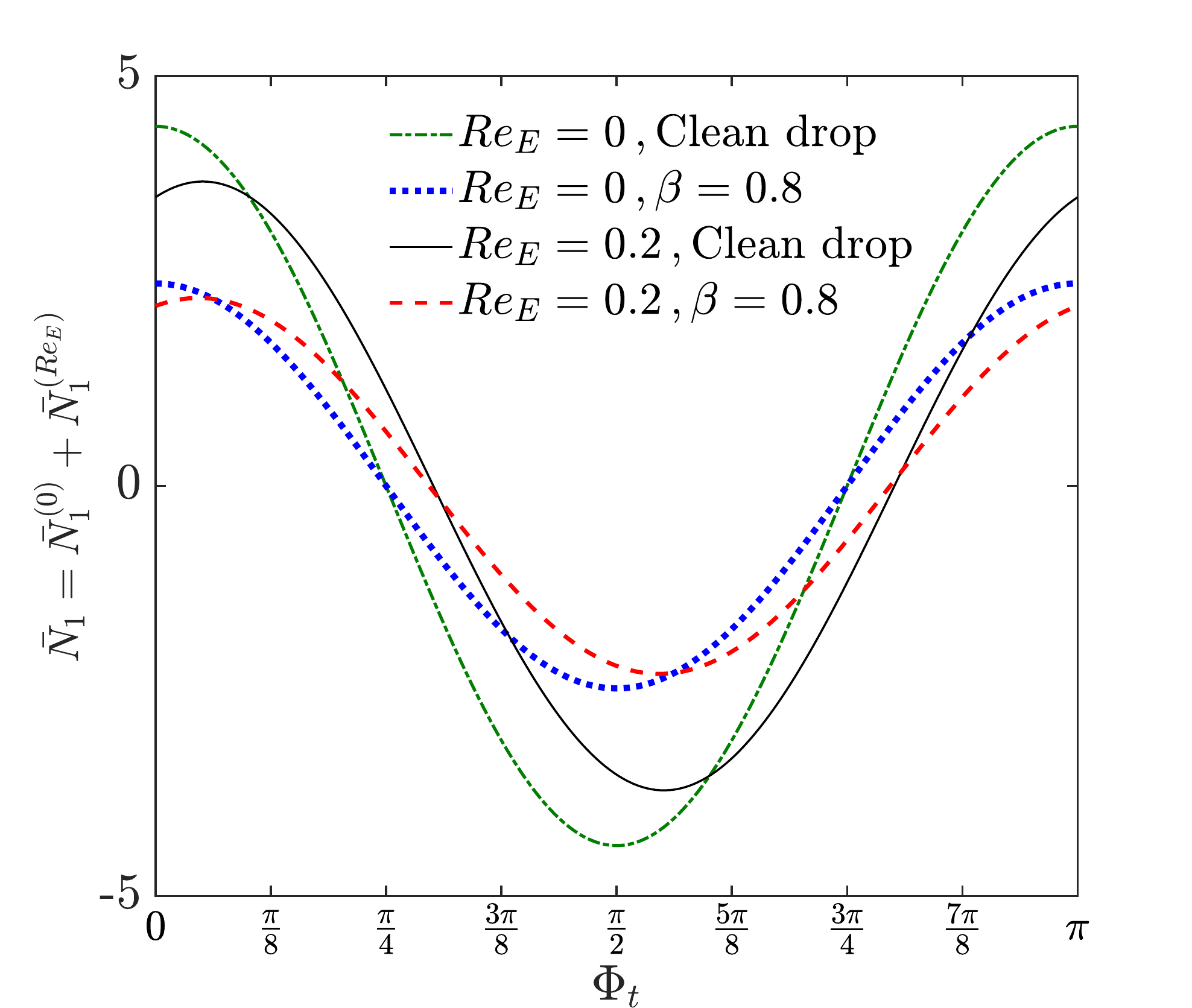}		
		\vspace{6ex}
		\caption{}
		\label{fig:N1_vs_tilt_C_ReE_0p2}
	\end{subfigure}
	\qquad
	\begin{subfigure}[!htb]{0.4\textwidth}
		\centering
		\includegraphics[width=1\textwidth]{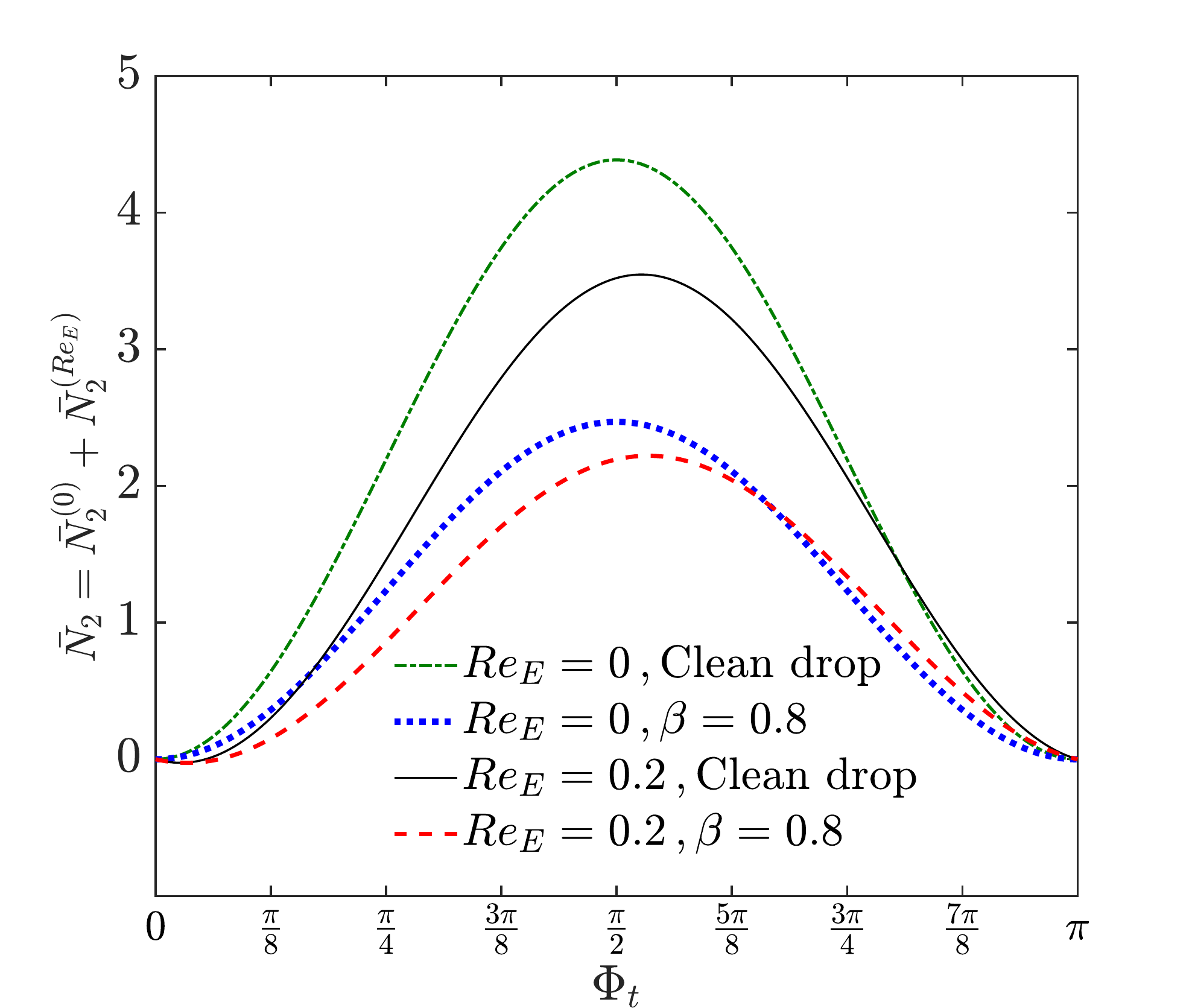}
		\vspace{6ex}
		\caption{}
		\label{fig:N2_vs_tilt_C_ReE_0p2}
	\end{subfigure}
\vspace{-2ex}
	\caption{ Variation of normalized first $ (\bar{N}_1=N_1/\nu) $ and second $ (\bar{N}_2=N_2/\nu) $ normal stress differences in subplots (a) and (b), respectively with the tilt angle of the applied electric field, $\Phi_t$ for different values of elasticity parameter $ \beta $ and electrical Reynolds number, $Re_E$. Here the electrohydrodynamic parameters are chosen according to system-C. Other parameters are $\,M=2 $ and $ k=1 $.} 
	\label{fig:N1_N2_ReE}
\end{figure}

In the absence  of charge convection, the surfactant contributions to the first  and second  normal stress differences $ (\Delta \bar{N}_1=|\bar{N}_1-\bar{N}_{1,\text{clean}}|,\Delta \bar{N}_2=|\bar{N}_2-\bar{N}_{2,\text{clean}}|) $   increase with the tilt angle and are symmetric about $ \Phi_t=\pi/2$, as depicted in figures~\ref{fig:N1_vs_tilt_C_ReE_0p2} and
\ref{fig:N2_vs_tilt_C_ReE_0p2}. $ \Delta \bar{N}_1$ takes a maximum value for $ \Phi_t=0,\pi/2 $ and vanishes  for $\Phi_t=\pi/4$; while   $ \Delta \bar{N}_1$ increases from zero to maximum from $\Phi_t=0$ to $ \pi/2$. However, owing to the presence of the charge convection, such symmetric nature is broken, although the increasing trends of both  $\Delta \bar{N}_1$ and $\Delta \bar{N}_2$ are preserved.

%\textcolor{red}{$ \bar{G}^{(Re_E)}_{\mathrm{cr},N_1}=\delta_3 (\beta, k, \lambda, R, S)\tan(2\Phi_t) $ and
%	$ \bar{G}^{(Re_E)}_{\mathrm{cr},N_2}=\delta_4 (\beta, k, \lambda, R, S)\cot(\Phi_t). $
%	The expressions of $\delta_3$ and $\delta_4$ being too cumbersome to present here, are provided in the supplementary MATLAB file.}

%\label{sssec:N1-N2_Re_E}

\subsection{Bulk rheology of deformed drops}
\label{ssec:Ca_result}
\begin{figure}[!htb]
	\centering
\begin{subfigure}{0.4\textwidth}
	\centering
	\includegraphics[width=1.0\textwidth]{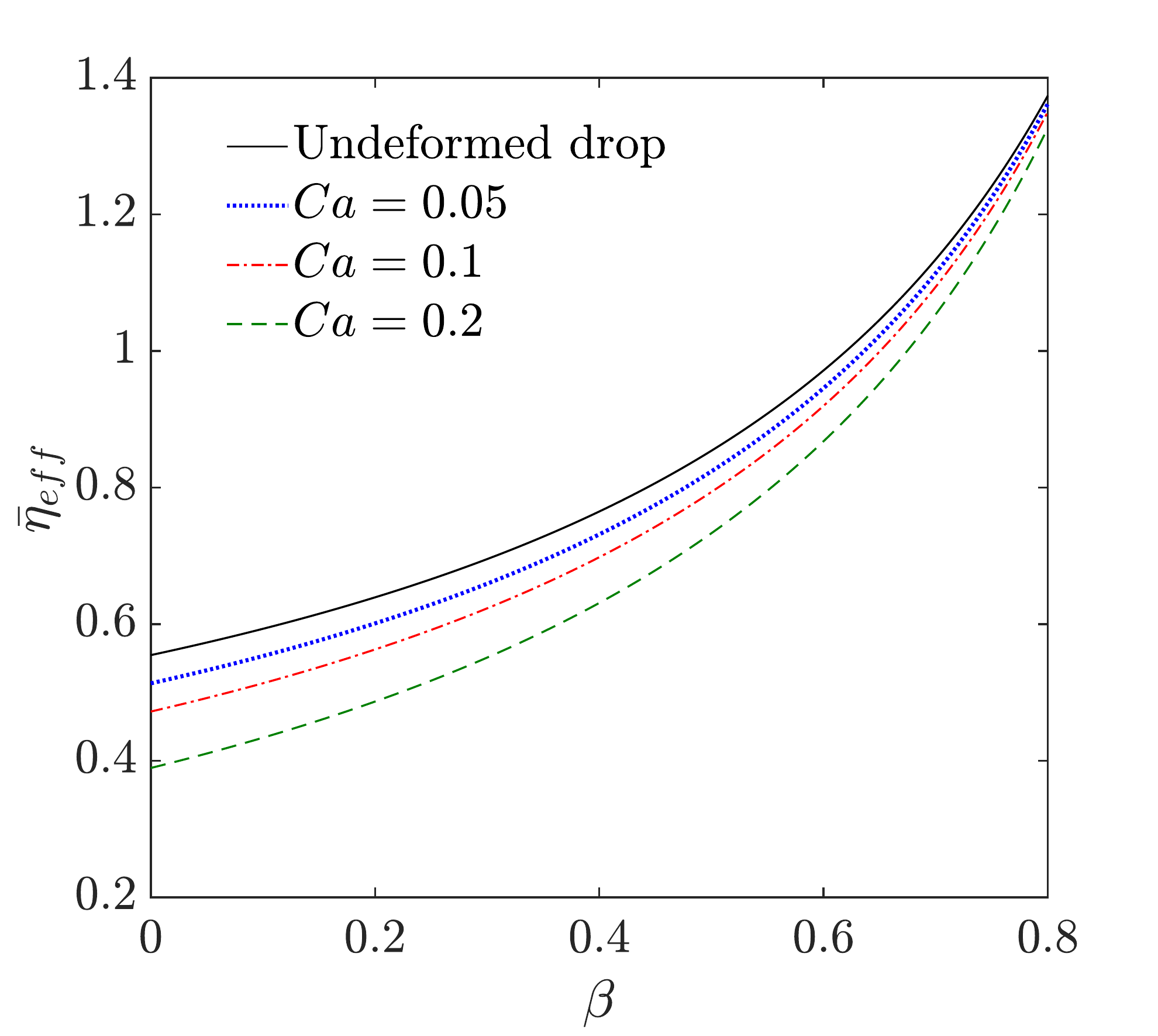}
	\vspace{6ex}
	\caption{System-B}
	\label{fig:eff_vs_beta_vary_CA_B}
\end{subfigure}
	\qquad \quad
\begin{subfigure}{0.4\textwidth}
		\centering
		\includegraphics[width=1.0\textwidth]{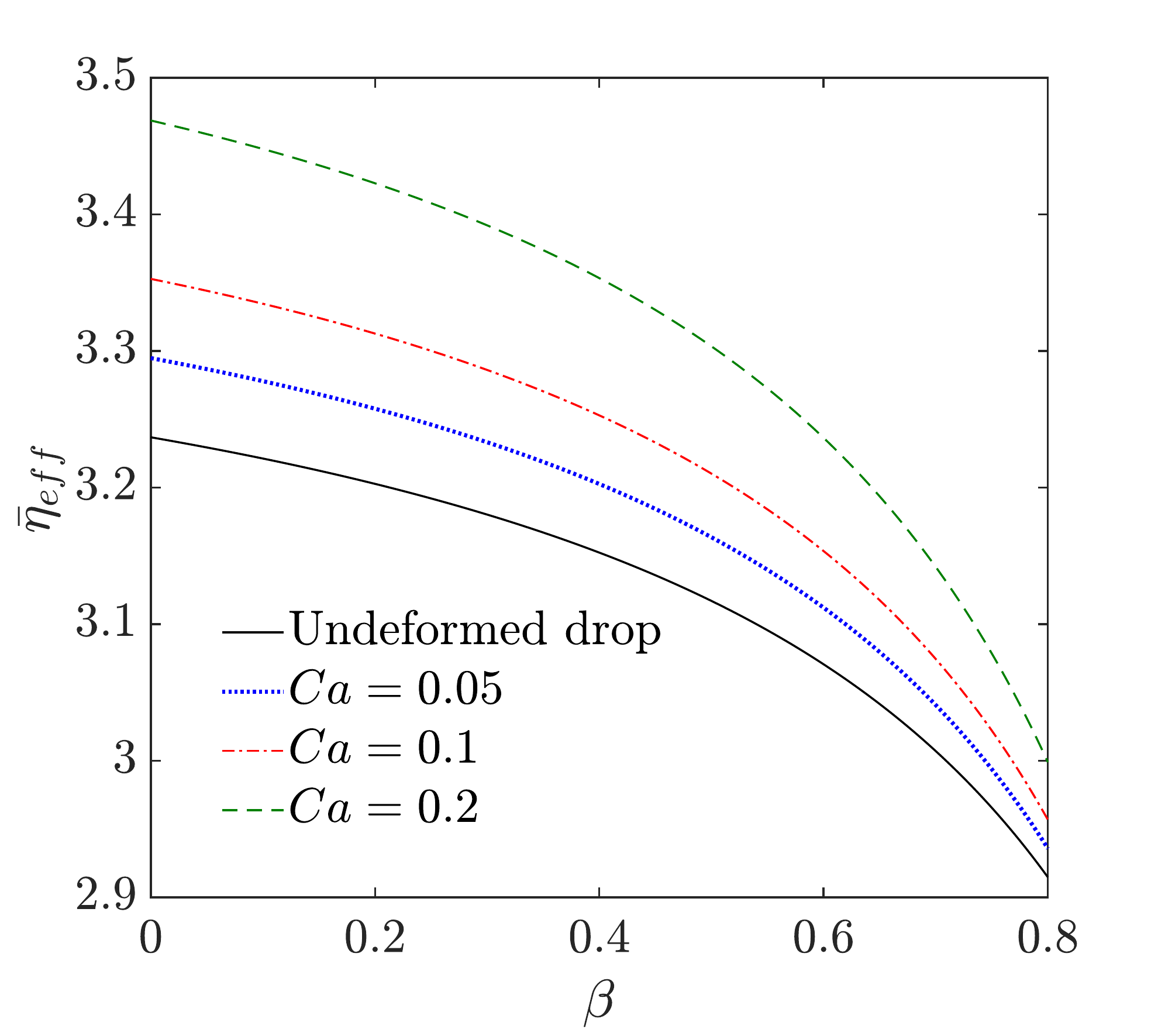}		
		\vspace{6ex}
		\caption{System-C}
		\label{fig:eff_vs_beta_vary_CA_C}
\end{subfigure}
\\
\begin{subfigure}{0.45\textwidth}
	\centering
	\includegraphics[width=1.0\textwidth]{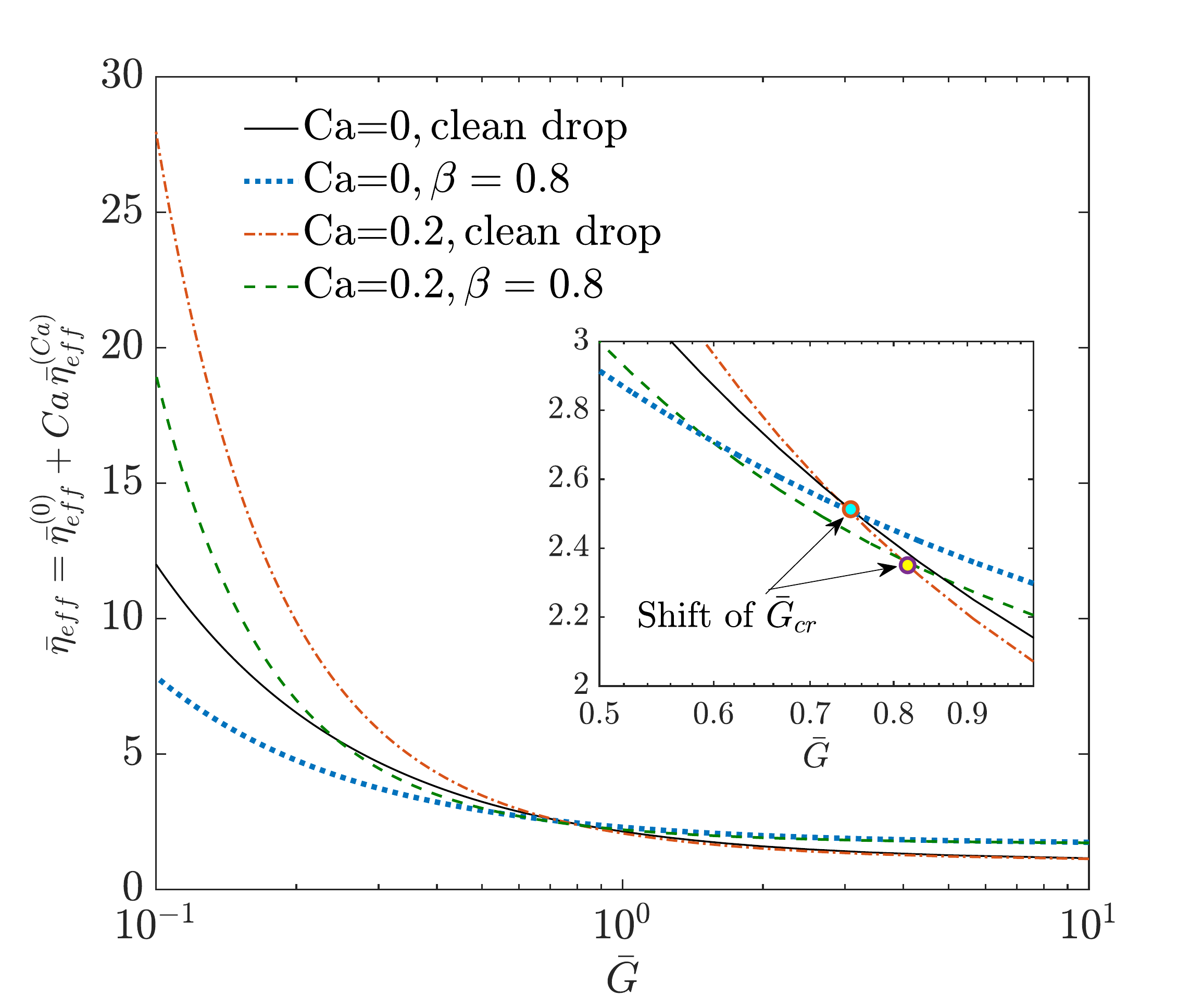}		
	\vspace{6ex}
	\caption{System-C}
	\label{fig:eff_vs_G_vary_beta_Ca_C}
\end{subfigure}
\vspace{-2ex}
\caption{(a),(b): Normalized effective shear viscosity ($\bar{\eta}_\text{eff}=\bar{\eta}^{(0)}_\text{eff}+Ca\,\bar{\eta}^{(Ca)}_\text{eff}$) vs.  elasticity parameter, $ \beta $ for different capillary numbers, $ Ca $. The subplots (a) and (b) are for system-B and C, respectively. Here $M=2 $. Subplot-(c): Normalized effective shear viscosity vs. relative shear rate for different elasticity parameter, $ \beta $ and Capillary number, $Ca$. In the inset, the shift of critical shear rate, $ \bar{G}_\text{cr} $ with $Ca$, is highlighted. Here  other parameters are chosen following system-C. In all the subplots $\Phi_t={\pi}/{4}$ and $ k=1$.} 
	\label{fig:eff-Ca-plots}
\end{figure}

Here we analyze the consequences of the coupled interaction between the deformable nature of the drop and   surfactant coating for different electrohydrodynamic systems. From figures~\ref{fig:eff-Ca-plots}(a) and (b) it can be observed that the deformation induced modification in the Marangoni effect is much pronounced for lower values of the elasticity parameter $ \beta $. Also the  effect of capillary number becomes smaller as $ \beta \to 0.8.$ Most interestingly, the increase in capillary number causes a fall in the effective viscosity for system-B, while the trend is opposite for system-C. To understand the physical origin of such behavior we investigate the nature  of drop deformation in different systems and the consequence of surfactant coating on them. For this we appeal to the   normal stress balance \eqref{eq:n_stress_bal} where the Marangoni effect is seen to play its role. From physical perspective, the localized accumulation of surfactant molecules reduces the local surface tension and makes that region more prone to deformation in order to balance the normal stresses. This  is known as the `tip stretching' phenomena in the literature \citep{Pawar1996}. Again due to increased surface area of the deformed drop, the overall surfactant concentration gets diluted and the local surface tension tends to fall. 

In order to quantify the deformation behavior of the drop we define a deformation parameter in the plane of shear as follows,
\begin{equation}\label{key}
\mathcal{D}=\frac{\text{max}(r_S (\theta=\dfrac{\pi}{2},\phi))-\text{min}(r_S (\theta=\dfrac{\pi}{2},\phi))}{\text{max}(r_S (\theta=\dfrac{\pi}{2},\phi))+\text{min}(r_S (\theta=\dfrac{\pi}{2},\phi))}.
\end{equation}
Another important quantification of the drop deformation is the drop inclination angle  which can be defined in the $ x-y $ plane as the angle $ \phi $ corresponding to the  maximum value of $ r_{\!_S} $ for $ \theta=\pi/2$ and is given as 
%\begin{equation}\label{eq:incline-angle}
$\varphi_d=\frac{1}{2}\tan^{-1}({\hat{L}_{2,2}}/{{L}_{2,2}}).$
In comparison to the case when only a simple shear flow is present \citep{Li1997}, the  electrical effects $ \left( \text{with}\, \Phi_t=\pi/4,M=2\right)$ increases the deformation parameter $ (\mathcal{D}) $ for system-B. In contrast, for system-C the initial prolate shape (with respect to the applied electric field direction along $ \Phi_t=\pi/4 $) of the drop under simple shear flow becomes a complete oblate one when electrical effects are present. This phenomenon is portrayed in the insets of the figures~\ref{fig:Drop_shape}(a) and (b)). This condition for a surfactant-free drop surface is   further modified by surface contamination and the deformation parameter shows a steady increase with $ \beta $ as shown in  figures~\ref{fig:Drop_shape}(a) and (b).  Thus the nature as well as intensity of drop deformation is modified. As a result the drop poses different kinds of resistances to the imposed shear flow, thereby creating diverse modulations in the effective shear viscosity of the emulsion.  
%
%\begin{figure}[!htb]
%	\centering
%	\begin{subfigure}[!htb]{0.45\textwidth}
%		\centering
%		\includegraphics[width=1.0\textwidth]{eff_vs_beta_vary_CA_HY_A-eps-converted-to.pdf}		
%		\vspace{8ex}
%		\caption{System-A}
%		\label{fig:eff_vs_beta_vary_CA_T_A}
%	\end{subfigure}
%	\qquad \quad
%	\begin{subfigure}[!htb]{0.45\textwidth}
%		\centering
%		\includegraphics[width=1.0\textwidth]{eff_vs_beta_vary_CA_B-eps-converted-to.pdf}
%		\vspace{8ex}
%		\caption{System-B}
%		\label{fig:eff_vs_beta_vary_CA_B}
%	\end{subfigure}
%	\\
%	\begin{subfigure}[!htb]{0.45\textwidth}
%		\centering
%		\includegraphics[width=1.1\textwidth]{eff_vs_beta_vary_CA_C-eps-converted-to.pdf}		
%		\vspace{10ex}
%		\caption{System-C}
%		\label{fig:eff_vs_beta_vary_CA_C}
%	\end{subfigure}
%	\caption{The $ O(Ca) $ effective viscosity is not only decided by the nature of the deformed shape. It is evident from these figures that the different sign of the Taylor discriminating factor does  not ensure an opposite behavior of the emulsion rheology.} 
%	\label{fig:eff-Ca-plots}
%\end{figure}

The deformation of the drop surface also influences the non-Newtonian rheology of the emulsion as shown in figure~\ref{fig:eff-Ca-plots}(c). The critical shear rate corresponding to the vanishing Marangoni effect  $ (\bar{G}_{cr}) $ gets shifted  to a higher value when the capillary number becomes 0.2 from an undeformed state, for the specific parametric choices of system-C. Till $\bar{G}_{cr} $ is reached, the shape deformation causes significant enhancement in the effective shear viscosity. However, with an increase in $\bar{G}$ beyond this critical point, the impact of deformation effects on the effective viscosity, becomes vanishingly small. 
%\begin{figure}[!htb]	
%	\centering
%	\includegraphics[width=0.45\textwidth]{eff_vs_G_vary_beta_Ca_C-eps-converted-to.pdf}
%\vspace{-2ex}
%	\caption{Normalized effective shear viscosity vs. relative shear rate for different elasticity parameter, $ \beta $ and Capillary number, $Ca$. In the inset the shift of critical shear rate, $ \bar{G}_\text{cr} $ with $Ca$ is highlighted. Here $ \Phi_t=\pi/4 $ and the other parameters are chosen following system-C.}
%	\label{fig:eff-vs-M-Ca}
%\end{figure}

\begin{figure}[!htb]
	\centering
	\begin{subfigure}{0.4\textwidth}
		\centering
		\includegraphics[width=1\textwidth]{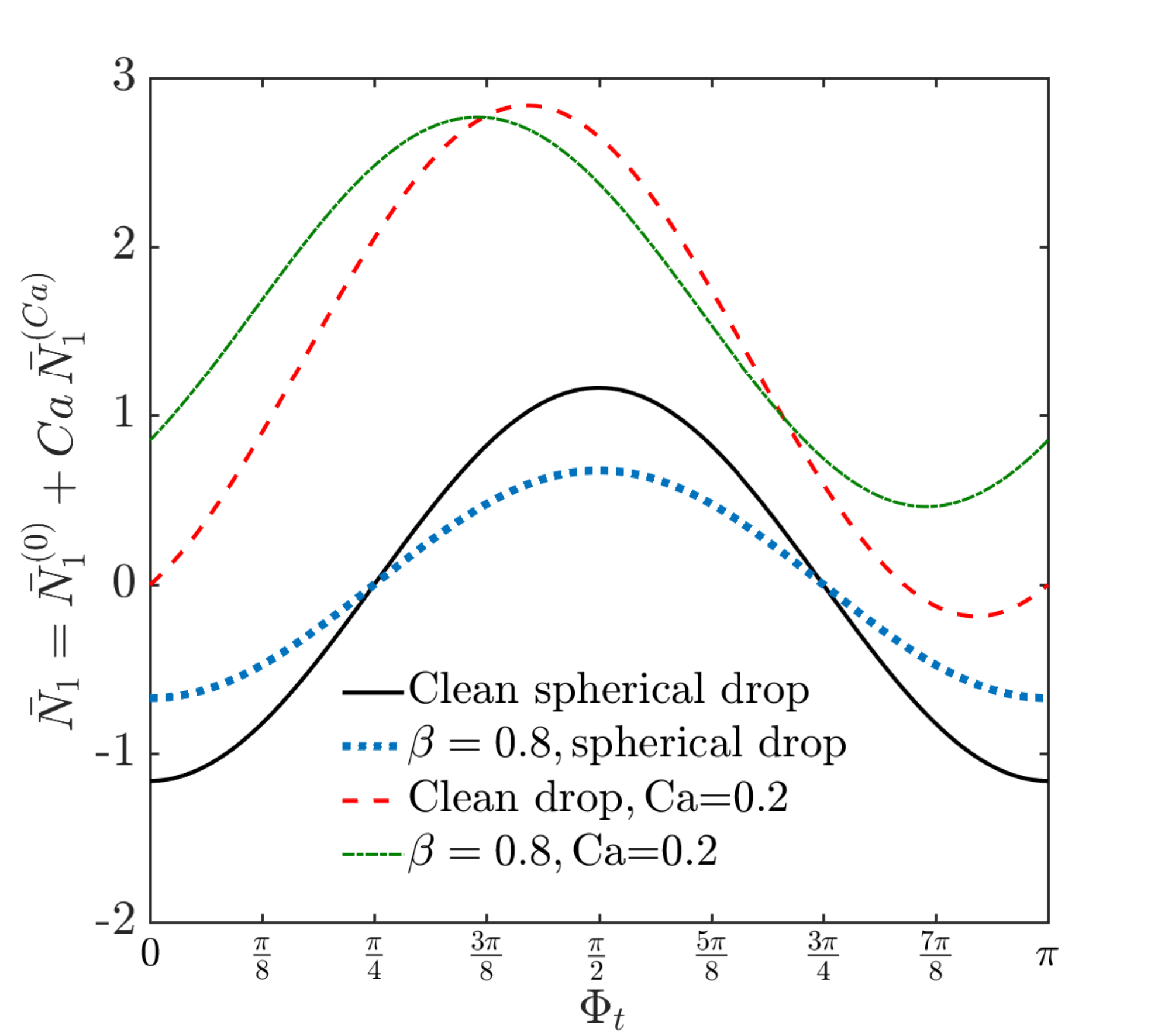}		
		\vspace{6ex}
		\caption{}
		\label{fig:N1_vs_tilt_B_CA_0p2}
	\end{subfigure}
	\qquad
	\begin{subfigure}{0.4\textwidth}
		\centering
		\includegraphics[width=1\textwidth]{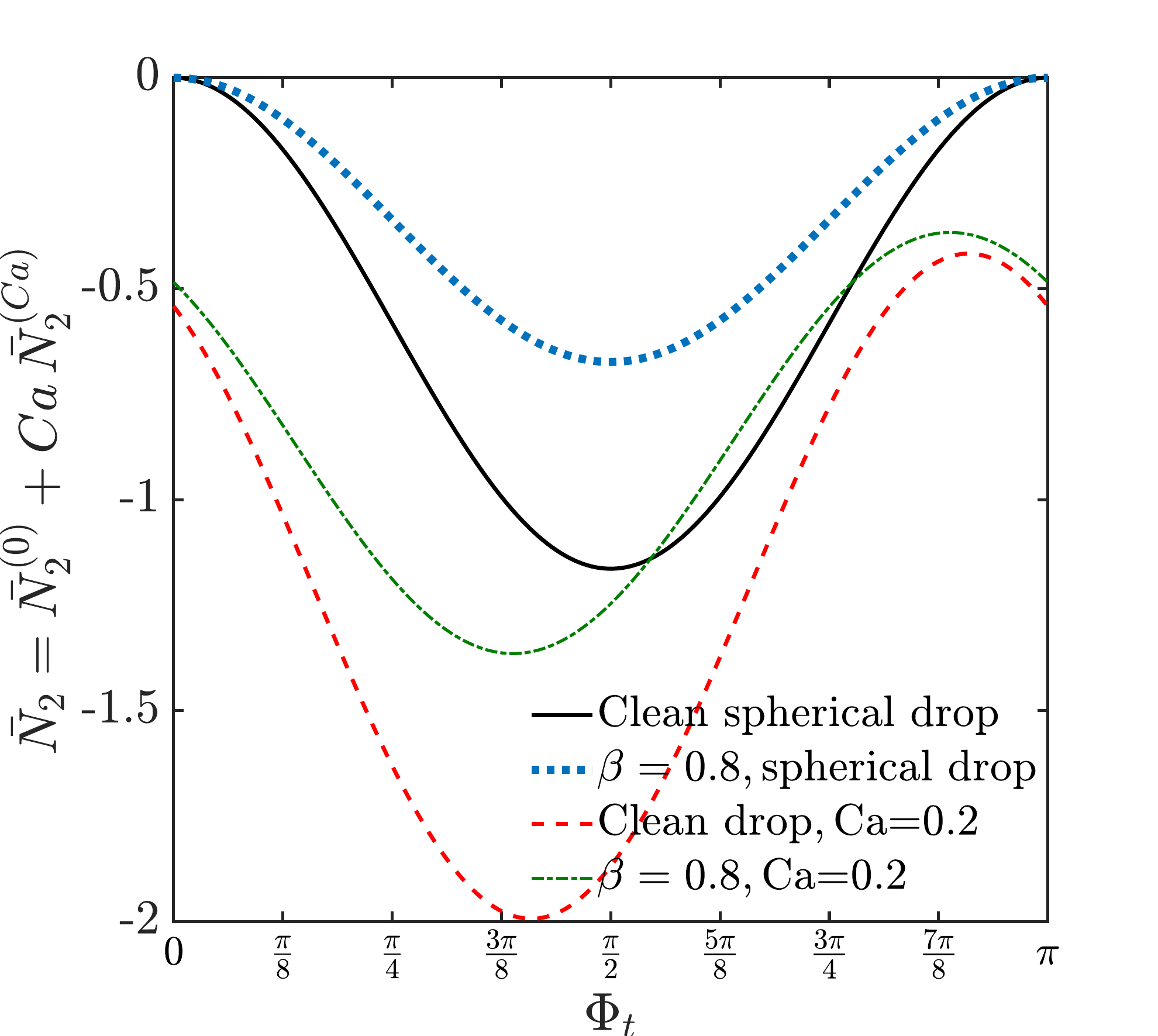}
		\vspace{6ex}
		\caption{}
		\label{fig:N2_vs_tilt_B_CA_0p2}
	\end{subfigure}
\vspace{-2ex}
	\caption{ Normalized first $ (\bar{N}_1=N_1/\nu) $ and second $ (\bar{N}_2=N_2/\nu) $ normal stress differences vs. tilt angle of the applied electric field, $\Phi_t$ for various elasticity parameter $ \beta $ and Capillary number, $Ca$. Here the electrohydrodynamic parameters are chosen according to system-B. Other parameters are $\,M=2 $ and $ k=1 $.} 
	\label{fig:N1_N2_Ca}
\end{figure}

In figure~\ref{fig:N1_N2_Ca} we demonstrate the variations in the bulk normal stress differences with the tilt angle of the applied electric field for different values of the surface elasticity number and capillary number. It is found that the leading order symmetry in both $ \bar{N}_1 $ and $ \bar{N}_2$ about the tilt angle $ \Phi_t=\pi/2$ is  destroyed  owing to a deformed drop surface.  In the deformed condition, the contribution of the surfactants in modifying the first normal stress difference $ (\Delta \bar{N}_1=|\bar{N}_1-\bar{N}_{1,\text{clean}}|) $, is observed to be positive for certain values of the tilt angle, while the reverse happens for other tilt angles (please refer to figure~\ref{fig:N1_N2_Ca}(a)). However, figure~\ref{fig:N1_N2_Ca}(b) depicts that under similar circumstances, the surfactant-induced modifications in the second normal stress difference $ (\Delta \bar{N}_2=|\bar{N}_2-\bar{N}_{2,\text{clean}}|) $ show a steady decrease for all the tilt angles between $ \Phi_t=0 $ and $ \pi$. Variations in the orientation of the applied electric field not only alters the behavior of the drop deformation but also cause severe changes the electrohydrodynamic flow pattern around the drop. In turn, the surface tension gradient is redistributed. Finally, an intricate interplay between the physical mechanisms of electrohydrodynamic  and Marangoni flow dictates  the  elastic characteristics of the emulsion, resulting in corresponding changes in the normal stress differences. 

%\FloatBarrier
\section{Conclusions}
In the above sections we presented a detailed analysis of the surfactant coating effects on the electrorheology of a dilute emulsion. We have considered a linear shear flow and a tilted electric field acting in tandem on a deformable viscous drop which is covered with non-ionic, bulk insoluble surfactants. The electrical effects have been modeled through the Taylor-Melcher leaky dielectric framework. Also the surfactant transport at the interface is considered to be diffusion dominated and the equations are obtained in the limit of small surface Peclet number, $Pe_S$. Next we have employed a double asymptotic perturbation in terms of small dimensional parameters of electrical Reynolds number $(Re_E)$ and capillary number $(Ca)$. In the absence of drop deformation, the tangential stress balance at the interface (\eqref{eq:t_stress_bal}) is  interacted by the Marangoni stresses at the interface due to non-uniform surfactant distribution. This creates alterations in the tangential flow around the drop and subsequently the nature of viscous dissipation due to the presence of the drops is changed. On the other hand, the drop deformation is also severely influenced by the Marangoni stresses (\eqref{eq:n_stress_bal}), and finally the drop shape is distorted in a varying extent. The deformation of the drop in or against the flow direction, can make the emulsion shear thinning or shear thickening, respectively. Not only the viscosity, but also the effective tension along the flow streamlines, is affected by the rich physical interaction of the Marangoni, hydrodynamic and electrical Maxwell stresses.  We can conclude the below mentioned points  from the presented work:

\begin{enumerate}[label=(\roman*),leftmargin=0pt,itemindent=3em]
	\item 
The surfactant distribution on the drop surface not only depends on the imposed shear flow, but also the tilt angle of the applied field plays a major role in dictating the same. In some cases, this modification is so severe that the peak surfactant accumulation along the poles of the straining axis in the plane of shear, is changed to out of the plane locations. The relative intensity of the hydrodynamic and electrical stresses, quantified by the Mason number, determines the  points of peak surfactant accumulation and in turn, the nature of  surface tension gradient $ (|\gamma_\text{max}-\gamma_\text{min}|) $ on the interface is also influenced.
\item Both the hydrodynamic and electrical components of the effective shear viscosity are modulated greatly due to surface contamination, beyond just a linear superposition to the  existing expressions for a clean drop. Depending on the drop-matrix combination of the electrohydrodynamic parameters,  the overall surfactant contribution can  either increase or decrease the effective shear viscosity.
\item We report the unique existence of a critical hydrodynamic shear rate relative to the electric stress for which the surface tension gradient vanishes on the drop surface. At this critical relative shear rate, the corresponding surfactant effect on the effective shear viscosity, gets vanished.  This critical shear rate strongly depends on the tilt angle of the applied electric field,the electrical conductivity ratio and the permittivity ratio of the drop-matrix fluid pair.
\item The bulk normal stress differences, describing the anisotropy in the normal stress of the effective fluid mixture,  always reduce due to surfactant effect in the leading order. This decrement becomes much pronounced for low drop viscosities. 
\item The interaction between surfactant-induced Marangoni stress and the surface charge convection, can make the shear response of the emulsion, from shear thinning to shear thickening or vice-versa, depending on the tilt angle of electric field. The charge convection also creates a shift in the critical shear rate for vanishing surfactant effect, as observed in the leading order.   
\item In the deformed condition, for a specific choice of the electrohydrodynamic parameters, the drop shape may become ellipsoidal along  the direction of the applied electric field or perpendicular to it. Now the surfactant effects cause the deformation intensity to rise in both cases. Finally the resistance to flow is affected in such a way that  the deformation-induced rise or fall in the effective shear viscosity, gets suppressed due to Marangoni effects. Similar to the charge convection, the shape deformation can also create a shift in the critical shear rate. 
    
\qquad To summarize, the  detailed analysis presented above shows that the disturbance in drop-scale flow dynamics,  created due to the presence of dispersed drops in a combined action of shear flow and uniform electric field, can be severely modulated by surfactant coating on the interface. As a consequence, the bulk rheological properties of a dilute emulsion may exhibit distinct non-Newtonian behaviors depending upon the highly coupled interaction between the surfactant distribution on the interface and  electrohydrodynamic flow pattern.  The understanding of the micro-scale flow physics, depicted by the above results and its impact on the rheology, thus become a stepping stone towards analyzing  more complex rheological behavior of  emulsions where multi-drop interactions  become important.  

\end{enumerate}

\begin{appendices}
	\renewcommand{\thesection}{Appendix \Alph{section}}
	\renewcommand{\thesubsection}{\thesection.\arabic{subsection}}
	\renewcommand\thefigure{A-\arabic{figure}}    
	\setcounter{figure}{0} 
	\renewcommand{\theequation}{A-\arabic{equation}}
	% redefine the command that creates the equation no.
	\setcounter{equation}{0}  % reset counter 
	
	\section{Detailed expression of the leading order surface velocity $ \mathbf{u}_S^{(0)} $}
	\label{sec:leading_us}
	The leading order surface velocity can be expressed in the spherical coordinate system as $ \mathbf{u}_S^{(0)}= u^{(0)}_{S,\theta}\, \mathbf{i}_\theta+u^{(0)}_{S,\phi}\, \mathbf{i}_\phi$, where the scalar components are of the form:		
	%\newpage
	\begin{equation}
	\label{eq:leading_us}
	\begin{aligned} 
	u^{(0)}_{S,\theta}=
	\frac { \Biggl\{
		\splitdfrac{
			( 1-\beta )  (  ( 9M ( R-S
			) \sin ( 2\Phi_{{t}} ) +5 ( 2+R ) ^{2
			} ) \sin ( 2\phi )}
		{+9M ( 1+\cos ( 2
			\Phi_{{t}} ) \cos ( 2\phi )  )} 
		\Biggr \}
		\sin ( 2\theta )
	}
	{ 4(R+2)^2  (k\beta-5\beta\lambda-5\beta
		+5\lambda+5 )}, \text{\;\, and }\\
	u_{S,\phi}^{(0)}=
	\frac {\sin ( \theta ) }{ (  ( k-5
		\lambda-5 ) \beta+5\lambda+5 )  ( 2+R ) ^{2}}
	\biggl\{ \biggl( \frac{9M}{2} ( R-S ) (\beta-1 ) \sin ( 2(\Phi_t-\phi)) \\
	+\frac {5 ( 2+R ) ^2  (\beta-1 ) \cos ( 2\phi ) }{2} \biggr )    
	- \left( \frac{1}{2} ( k-5\lambda-
	5 ) \beta+{\frac {5\lambda}{2}}+{\frac {5}{2}}\right) 
	( 2+R ) ^{2} \biggr\}.
	\end{aligned}
	\end{equation}
	
	\section{Comparison with previous works}
	\label{sec:validation}
In different steps of  calculation, the  expressions of  important physical variables have been validated against various limiting cases of the present physical situation, as elaborated below:
	\begin{enumerate}[label=(\roman*),leftmargin=0pt,itemindent=3em]
		\item 
\textit{Surfactant-free drop in shear flow:}\vspace{0.55em}\\
 This simplistic case can be obtained in the present study by substituting $ M=0 $ and $ \beta=0 $ in the resulting expressions. Under such a condition the leading order effective shear viscosity was first obtained by \citet{Taylor1932}. Our leading order results in \eqref{eq:leading--eta}, i.e. $ 	\eta^{(0)}_\text{h,clean} = 1+\nu{\dfrac {\left( 5\,\lambda+2 \right) }{\left(2\,\lambda+2 \right)}
} $, corroborates with that classic result.   

In the same limit, we obtained the $ O(Ca) $   normal stress differences as given below:
\begin{subequations} 
	\begin{equation}
%	\label{eq:leading-N1}
	N_{1}^{(Ca)}=\nu{\frac { \left( 16+19\,\lambda \right) ^{2}}{ 40\left( \lambda+
			1 \right) ^{2}}}
	\end{equation}
	and
	\begin{equation}
%	\label{eq:leading-N2}
	N_{2}^{(Ca)}=-\nu{\frac {551\,{\lambda}^{3}+1623\,{\lambda}^{2}+1926\,\lambda+800}{280
			\, \left( \lambda+1 \right) ^{3}}}.
	\end{equation}
	\label{eq:leading_N1N2_M0_beta0}
\end{subequations}  
Also in $ O(Ca) $, the hydrodynamic contribution to the effective shear viscosity is obtained as zero   $ (\eta_\text{eff}^{(Ca)}=0)$. These observations are in agreement with the seminal work of 
\citet{Schowalter1968}.  For similar conditions, the $ O(Ca) $ shape function $ f^{(Ca)} $ matches with the expressions obtained by \citet{Barthes-Biesel1973}
and \citet{Vlahovska2011}, respectively. 
			\item 	
\textit{Surfactant-coated drop in shear flow:}\vspace{0.55em}\\
The expressions in the vanishing diffusion regime or the high surface P\'eclet number regime ($Pe_S\gg1$) \citep{Vlahovska2009} can be obtained by taking a limit $ k \to \infty$ and substituting $M=0$ in the present expressions, as shown below:
\begin{subequations} 
	\begin{equation}
	%	\label{eq:leading-N1}
	N_{1}^{(Ca)}= {\frac { 5\,\nu\left( 3\beta+1 \right) }{2\beta}}
	\end{equation}
	and
	\begin{equation}
	%	\label{eq:leading-N2}
	N_{2}^{(Ca)}=-{\frac { 5 \, \nu \left( 6\,\beta+7 \right) }{28\,\beta}}.
	\end{equation}
	\label{eq:leading_N1N2_M0_k_inf}
\end{subequations}  
\qquad In the diffusion dominated limit (i.e. $ Pe_S \ll 1$) if we substitute $ M=0 $  in the presently obtained   surfactant concentration $\Gamma=\Gamma^{(0)}+Ca\,\Gamma^{(Ca)}$ and $ O(Ca) $ drop shape  correction $ f^{(Ca)}$, they become similar to those obtained by \citet{Mandal2017c}. In all these our calculations consistently capture  the  result of $ N_2<0 $ in the absence of electrical effects. This is commonly observed in polymeric solutions and has been reported by earlier studies \cite{Li1997}.		\item 
\textit{Uncontaminated drop in combined shear flow and electric field:}\vspace{0.55em}\\
 In the limit $ \beta \to 0 $ the surface tension becomes uniform along the drop surface. For that situation, the present results e.g. the drop deformation, effective shear viscosity and normal stress differences in various orders of perturbation, match exactly with those of \cite{Mandal2017a}. When the electric field is applied along the direction of velocity gradient $ \Phi_t=\pi/2$, we find that in 
\eqref{eq:leading--eta} the clean drop effective viscosity due to electric field $(\eta^{(0)}_\text{e,clean})$ vanishes. A similar observation was also made in another reported work \citep{Vlahovska2011}. The electric field contribution, to the normal stress differences in the leading order (\eqref{eq:leading_N1N2}), are also in agreement with them. 
  
	\end{enumerate}

 Hence we obtained a reasonable confidence on the correctness of the theoretical analysis presented in this work.	 
\section{Effect of Mason number on surface tension}
\label{sec:gamma_vs_Phi_vary_G}
\begin{figure}[!htbp]	
	\centering
	\includegraphics[width=0.5\textwidth]{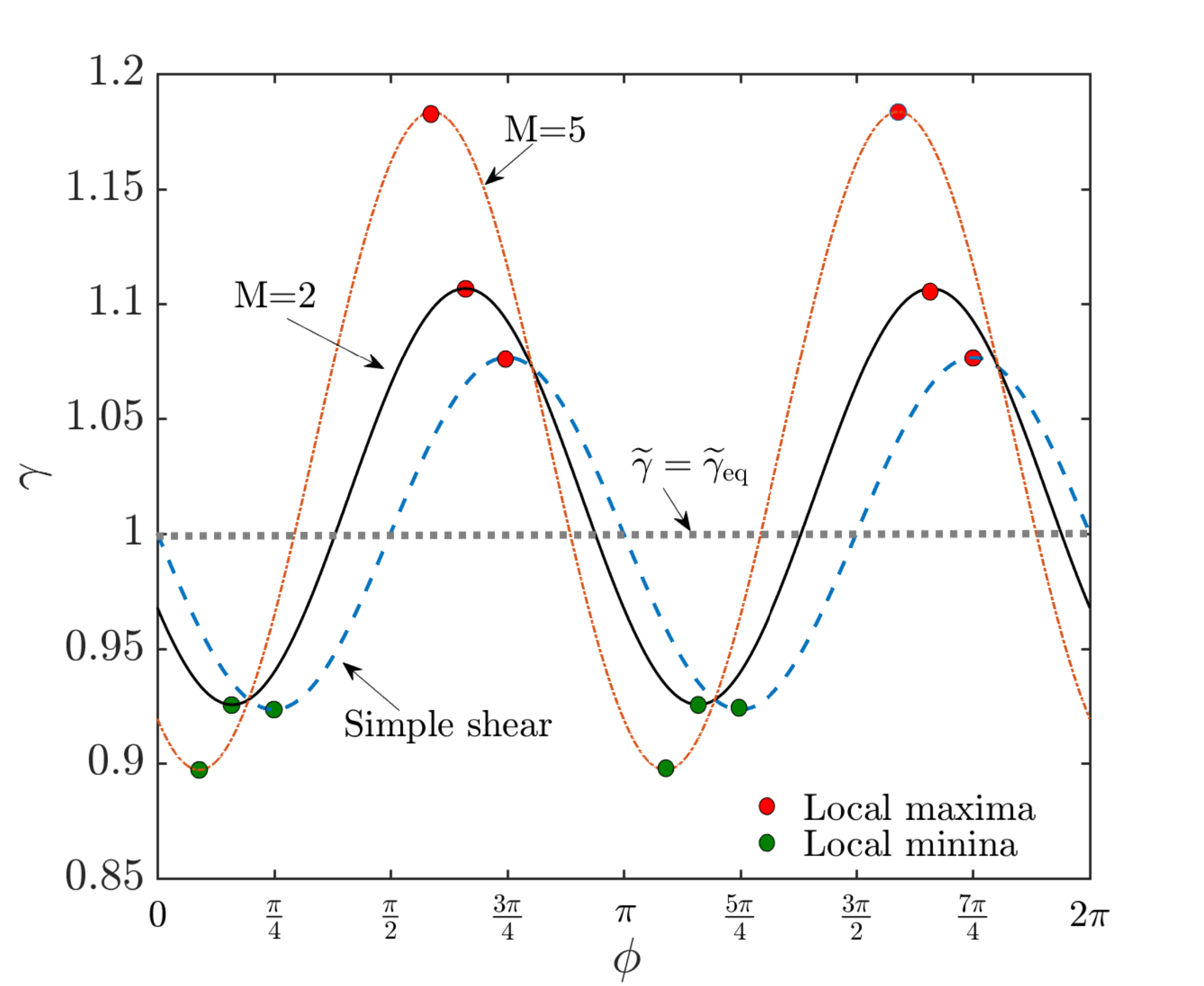}
\vspace{-2ex}
	\caption{Interfacial tension variation $ (\gamma) $ in the $x-y$ plane for different Mason number $ (M) $.  Here  $\beta=0.8,k=1$ and  $\Phi_t=\pi/2 $. Other parameters are chosen following system-A.}
	\label{fig:gamma_vs_Phi_vary_G}
\end{figure}

In this section in figure~\ref{fig:gamma_vs_Phi_vary_G} we delineate the effects of  strength of electrical stresses relative to the hydrodynamic stresses, by varying the Mason number, define as $ M=\dfrac{\epsilon_c \widetilde{E}_\infty^2}{\mu_c \widetilde{G}}$. 

%%\FloatBarrier
	\section{Effect of $ \beta $ and $ k $ on surface concentration and surface tension}
\label{sec:beta_k_Gamma}
\begin{figure}[!htb]
	\centering
	\begin{subfigure}{0.4\textwidth}
		\centering
		\includegraphics[width=1.0\textwidth]{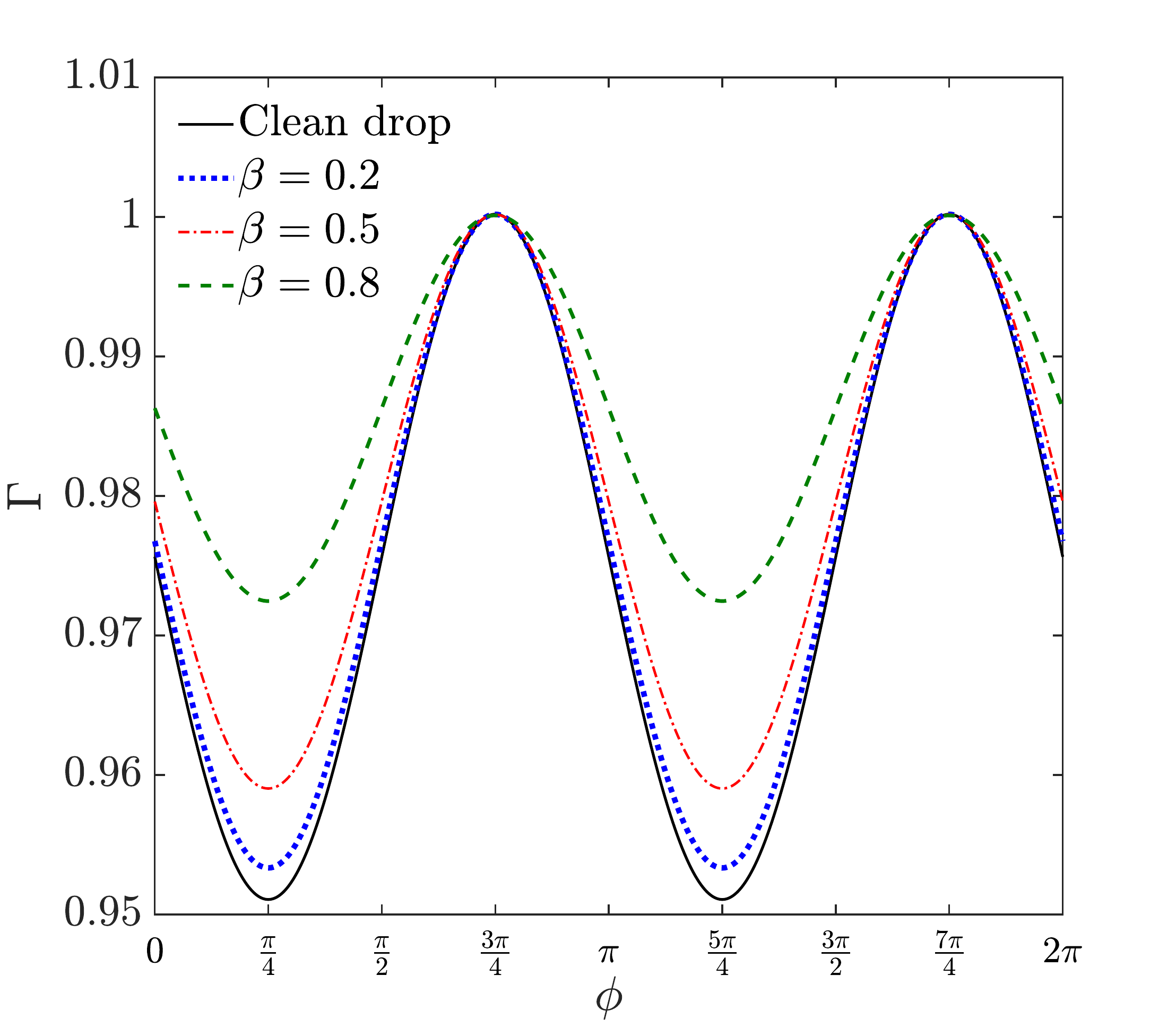}		
		\vspace{5ex}
		\caption{}
		\label{fig:SC_vs_theta_vary_beta}
	\end{subfigure}
	\qquad \qquad
	\begin{subfigure}{0.4\textwidth}
		\centering
		\includegraphics[width=1.0\textwidth]{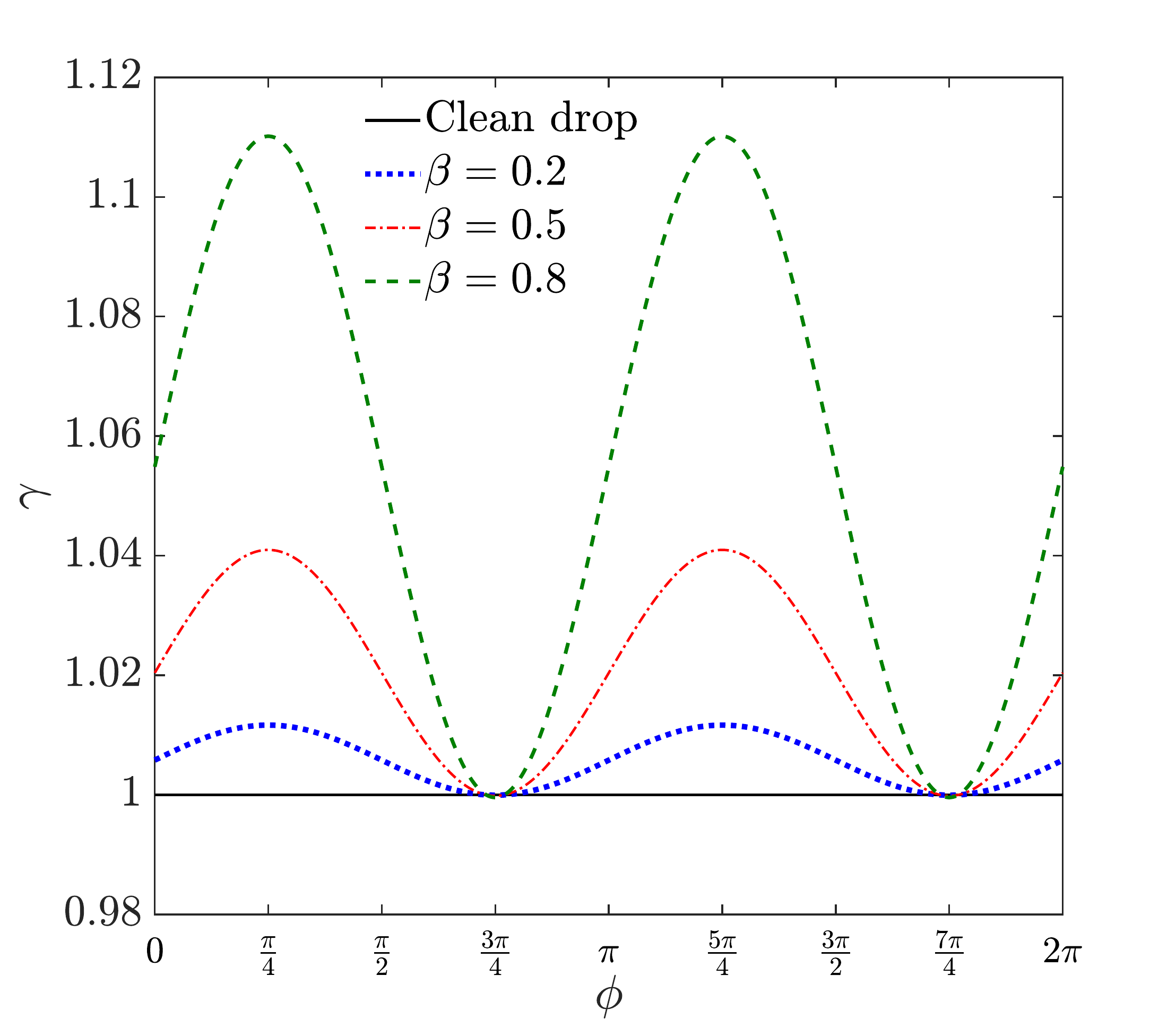}	
		\vspace{5ex}
		\caption{}
		\label{fig:ST_vs_theta_vary_beta}
	\end{subfigure}
	%	\\[-12ex] 
	\begin{subfigure}{0.4\textwidth}
		\centering
			\includegraphics[width=1.0\textwidth]{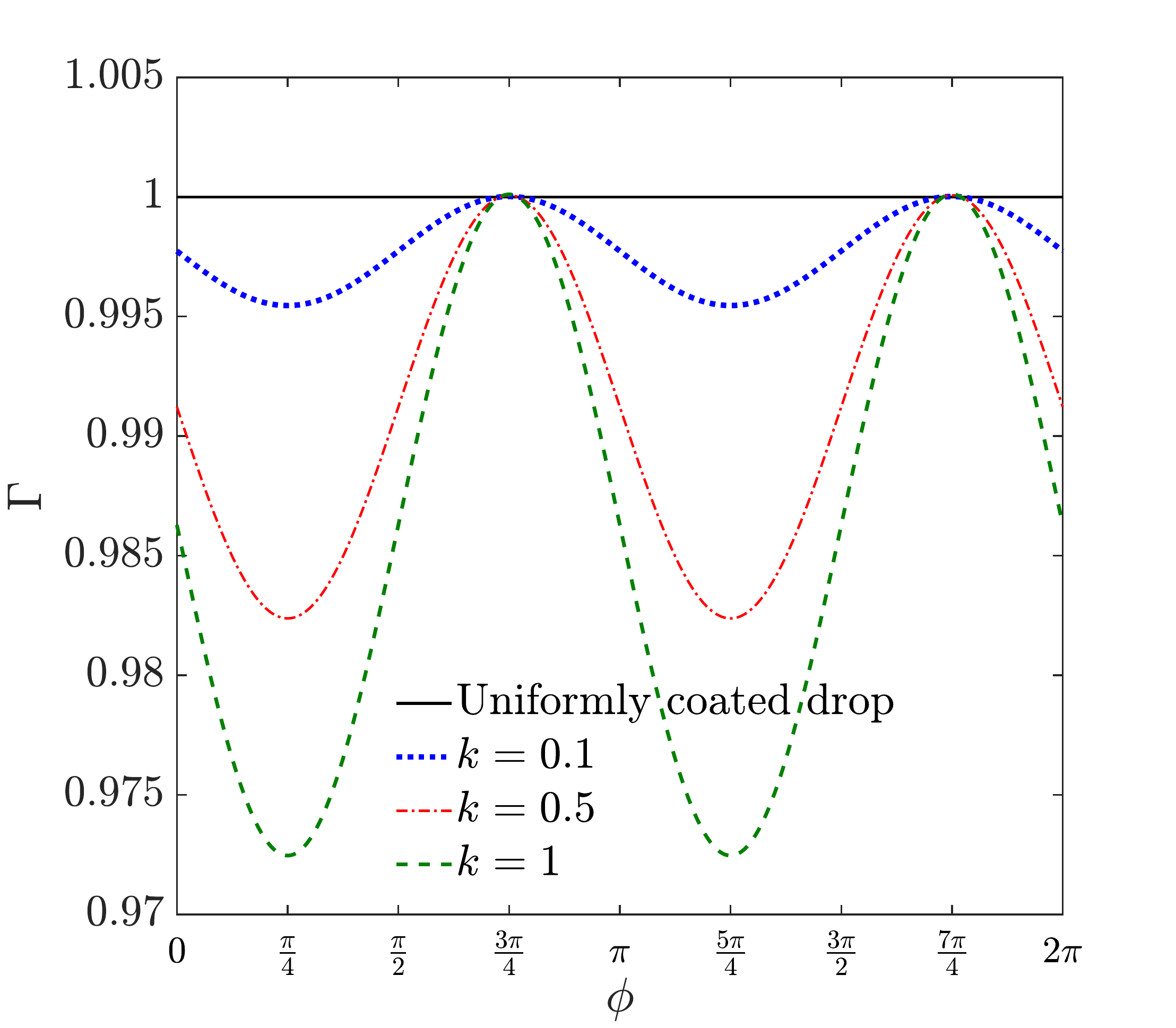}		
		\vspace{5ex}
		\caption{}
		\label{fig:SC_vs_theta_vary_k}
	\end{subfigure}
	\qquad \qquad 
	\begin{subfigure}{0.4\textwidth}
		\centering
%		\vspace{1.8ex}
		\includegraphics[width=1.0\textwidth]{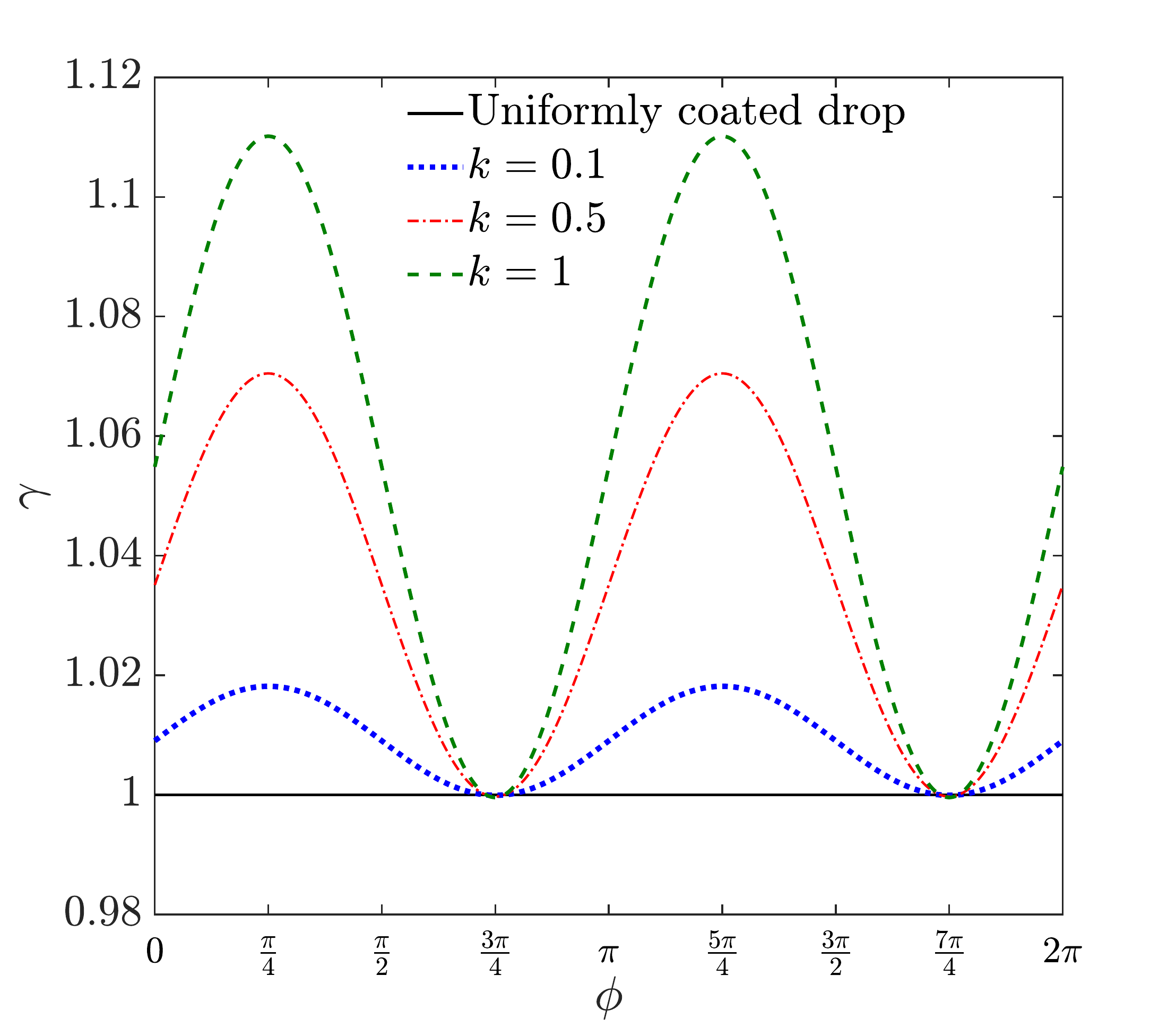}	
		\vspace{5ex}
		\caption{}
		\label{fig:ST_vs_theta_vary_k}
	\end{subfigure}
\vspace{-2ex}
	\caption{Variation of  surfactant concentration $\left(\Gamma=\Gamma^{(0)}+Ca\,\Gamma^{(Ca)} \right)$ and interfacial tension $\left(\gamma=\gamma^{(0)}+Ca\,\gamma^{(Ca)} \right)$ in the $ x-y $ plane for different values of the parameters $ \beta $ and $ k$. In sub-figures (a) and (b), $ k=1 $ is chosen while $ \beta=0.8 $ is taken for (c) and (d). Here the tilt angle of the electric field is $\Phi_t=\pi/4 $ , Mason number $ M=2$ and other parameters are as per system-C.} 
	\label{fig:SC_ST_var_beta_k}
\end{figure}

Here in figure~\ref{fig:SC_ST_var_beta_k}, we explore the role of the surfactant characterization parameters, namely the elasticity number $ \beta $ and physicochemical constant $ k$, in altering the surfactant concentration and surface tension variation on the drop surface.

\section{Variation of interfacial velocity and electrical traction}
	\label{sec:Us_and_TE}
\begin{figure}[!htb]
	\centering
	\begin{subfigure}{0.4\textwidth}
		\centering
		\includegraphics[width=1\textwidth]{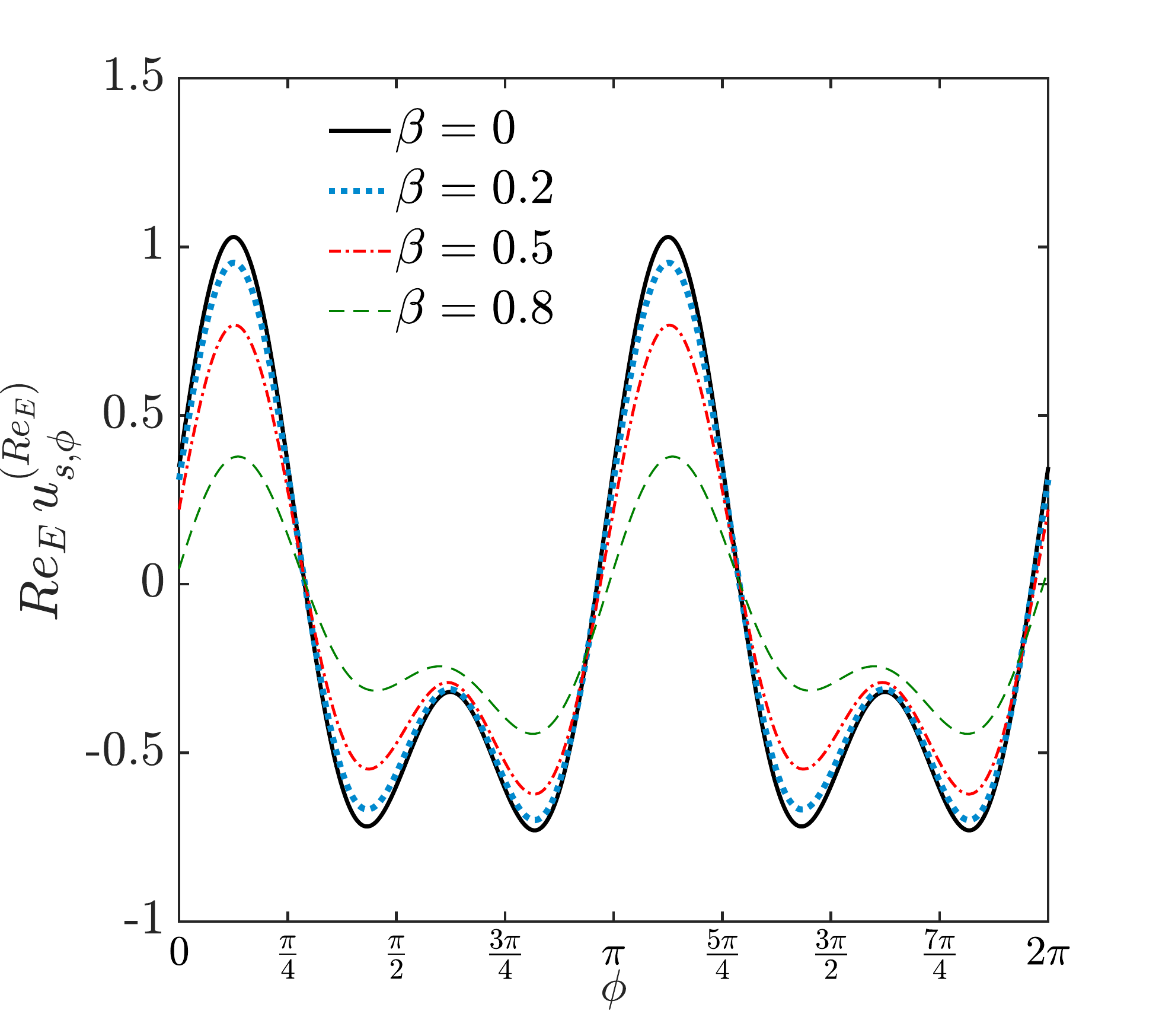}
		\vspace{5ex}
		\caption{}
		\label{fig:us_RE_vs_PHI_vary_beta_C}
	\end{subfigure}
	\quad \quad
	\begin{subfigure}{0.4\textwidth}
		\centering
		\includegraphics[width=1\textwidth]{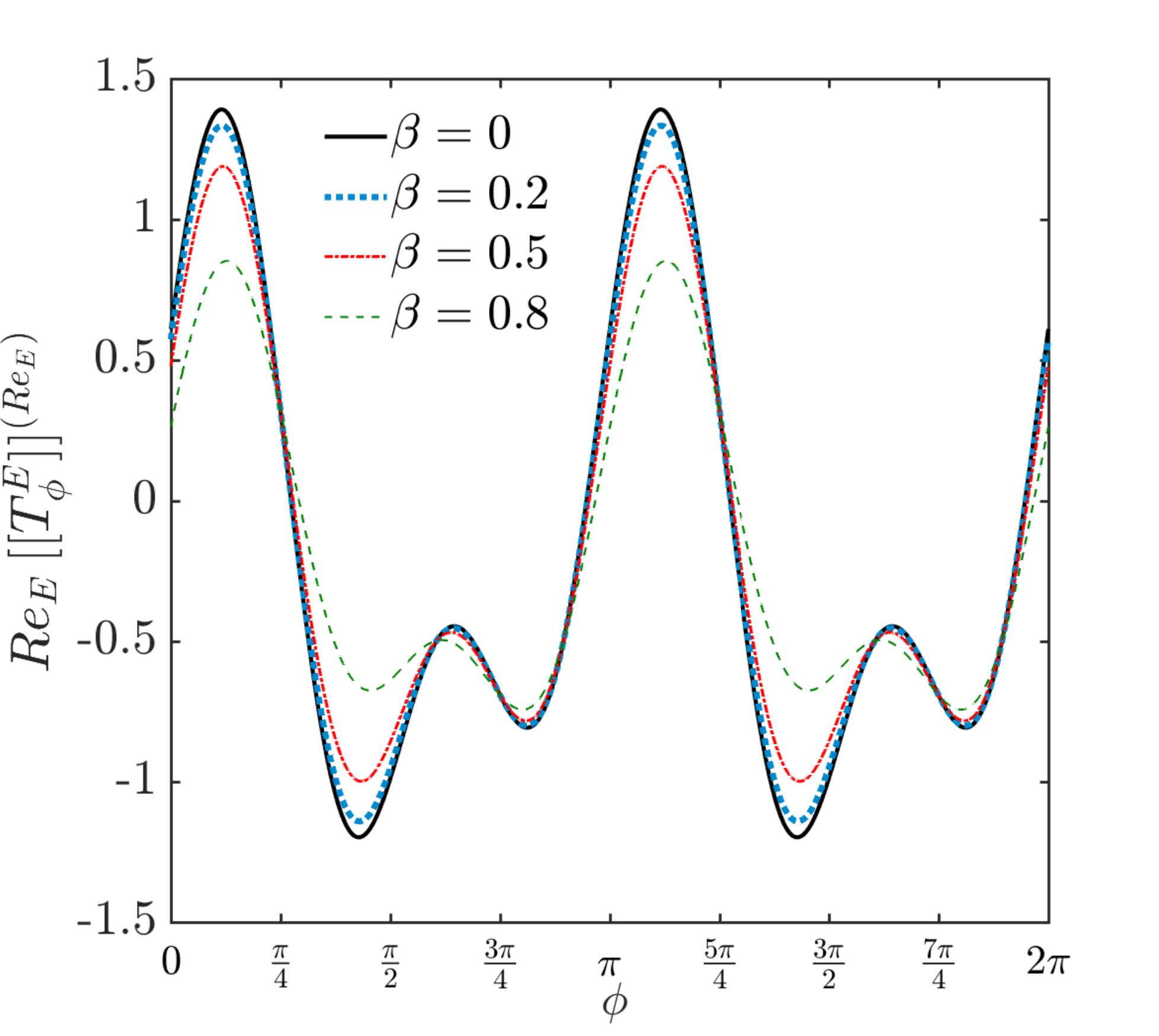}
		\vspace{5ex}
		\caption{}
		\label{fig:TE_RE_vs_PHI_vary_beta_C}
	\end{subfigure}
\vspace{-2ex}
	\caption{Variation of $ O(Re_E) $ correction to the surface velocity component, $Re_E \, u_{S,\phi}^{(Re_E)}$ and electrical traction component at the drop surface, $ [[T_\phi]]^{(Re_E)}={T_\phi^{E(Re_E)}}\big{|}_{c} -{T_\phi^{E(Re_E)}}\big{|}_{d} $ for different surface elasticity parameter, $\beta$. Here the demonstration os for system-C, while rest of the parameters are $ M=2, Re_E=0.2,\Phi_t=\pi/4$ and $ k=1$.} 
	\label{fig:us_te}
\end{figure}

	Here we exemplify the coupled consequence of charge convection and Marangoni stress on the flow field by showing the variation in the azimuthal components the drop velocity and electrical traction at the interface in figures~\ref{fig:us_te}. 
\section{Surfactant effect on drop deformation}
\label{sec:deform_surf}
\begin{figure}[!htbp]
	\centering
\begin{subfigure}{0.4\textwidth}
		\centering
\includegraphics[width=1.16\textwidth]{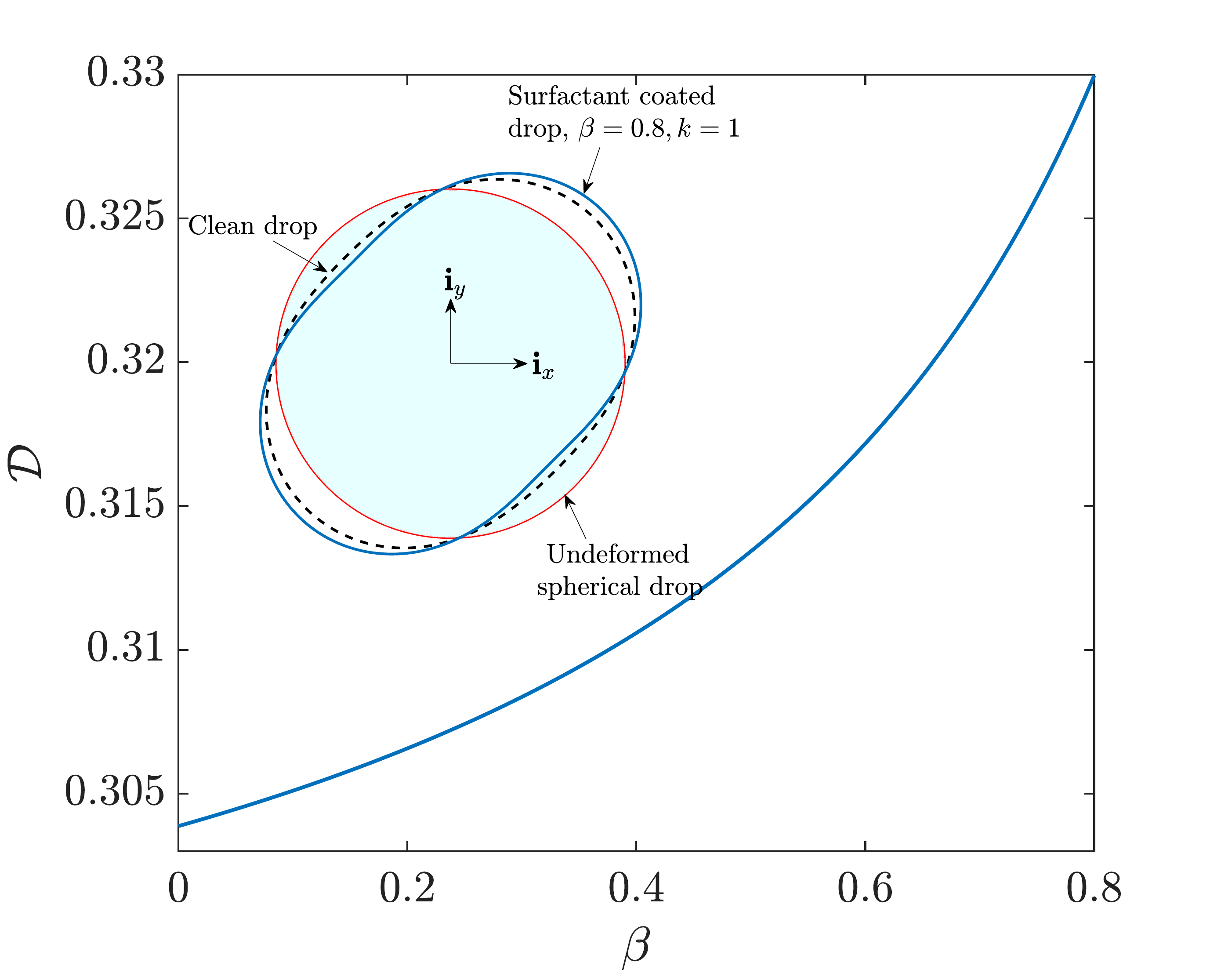}		
		\vspace{6ex}
		\caption{System-B}
		\label{fig:Drop_Shape_B}
\end{subfigure}
\quad
\begin{subfigure}{0.4\textwidth}
	\centering
\includegraphics[width=1.16\textwidth]{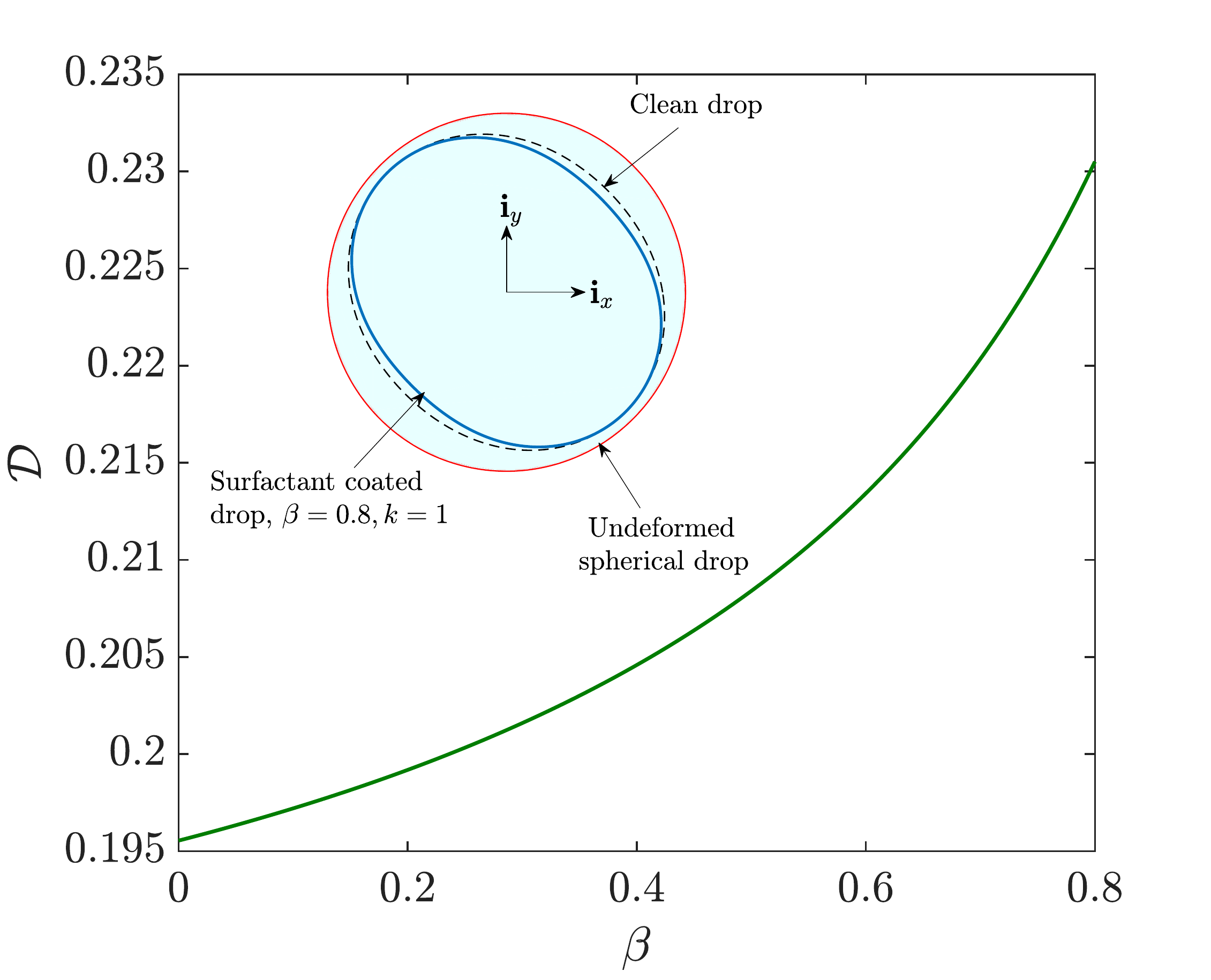}		
	\vspace{6ex}
	\caption{System-C}
	\label{fig:Drop_Shape_C}
\end{subfigure}
\vspace{-2ex}
	\caption{Effect of non-uniform surfactant distribution on the drop shape deformation in the plane of shear for systems B and C, respectively. The other parameters are $ M=2,\,Ca=0.2,\,\Phi_t={\pi}/{4}$.} 
	\label{fig:Drop_shape}
\end{figure}

In this section (figure~\ref{fig:Drop_shape}) we plotted the drop shapes for different electrohydrodynamic systems. Also the variation of deformation parameter $ \mathcal{D} $ with the surface elasticity number $ \beta $ is shown.

\FloatBarrier
\end{appendices}
%\bigskip
%\newpage
%\bibliographystyle{jfm}


\begin{thebibliography}{69}
	\expandafter\ifx\csname natexlab\endcsname\relax\def\natexlab#1{#1}\fi
	\def\au#1{#1} \def\ed#1{#1} \def\yr#1{#1}\def\at#1{#1}\def\jt#1{\textit{#1}}
	\def\bt#1{#1}\def\bvol#1{\textbf{#1}} \def\vol#1{#1} \def\pg#1{#1}
	\def\publ#1{#1}\def\arxiv#1{#1}\def\org#1{#1}\def\st#1{\textit{#1}}
	
	\bibitem[Allan \& Mason(1962)]{Allan1962}
	{\sc \au{Allan, R.~S.} \& \au{Mason, S.~G.}} \yr{1962}  \at{Particle behaviour
		in shear and electric fields. i. deformation and burst of fluid drops}.
	\jt{Proc. R. Soc. A}  \bvol{267},  \pg{45}.
	
	\bibitem[Anna {\em et~al.\/}(2003)Anna, Bontoux \& Stone]{Anna2003}
	{\sc \au{Anna, Shelley~L.}, \au{Bontoux, Nathalie} \& \au{Stone, Howard~A.}}
	\yr{2003}  \at{Formation of dispersions using “flow focusing” in
		microchannels}.  \jt{Applied Physics Letters}  \bvol{82}~(3),  \pg{364--366}.
	
	\bibitem[Bandopadhyay {\em et~al.\/}(2016)Bandopadhyay, Mandal, Kishore \&
	Chakraborty]{Bandopadhyay2016}
	{\sc \au{Bandopadhyay, A.}, \au{Mandal, S.}, \au{Kishore, N.~K.} \&
		\au{Chakraborty, S.}} \yr{2016}  \at{{Uniform electric-field-induced lateral
			migration of a sedimenting drop}}.  \jt{J. Fluid Mech.}  \bvol{792}~(2016),
	\pg{553--589}.
	
	\bibitem[Barnes(1994)]{Barnes1994}
	{\sc \au{Barnes, Howard~A.}} \yr{1994}  \at{Rheology of emulsions — a
		review}.  \jt{Colloids and Surfaces A: Physicochemical and Engineering
		Aspects}  \bvol{91},  \pg{89 -- 95}, a selection of papers presented at the
	First World Congress on Emulsions.
	
	\bibitem[Barthès-Biesel \& Acrivos(1973)]{Barthes-Biesel1973}
	{\sc \au{Barthès-Biesel, D.} \& \au{Acrivos, A.}} \yr{1973}  \at{Deformation
		and burst of a liquid droplet freely suspended in a linear shear field}.
	\jt{J. Fluid Mech.}  \bvol{61},  \pg{1}.
	
	\bibitem[Batchelor(1970)]{Batchelor1970}
	{\sc \au{Batchelor, G.~K.}} \yr{1970}  \at{The stress system in a suspension of
		force-free particles}.  \jt{J. Fluid Mech.}  \bvol{41}~(3),  \pg{545--570}.
	
	\bibitem[Bentley \& Leal(1986)]{Bentley1986}
	{\sc \au{Bentley, B.~J.} \& \au{Leal, L.~G.}} \yr{1986}  \at{An experimental
		investigation of drop deformation and breakup in steady, two-dimensional
		linear flows}.  \jt{J. Fluid Mech.}  \bvol{167},  \pg{241--283}.
	
	\bibitem[Bird {\em et~al.\/}(1987)Bird, Armstrong \& Hassager]{Bird1987}
	{\sc \au{Bird, R.~B.}, \au{Armstrong, R.~C.} \& \au{Hassager, O.}} \yr{1987}
	\at{Dynamics of polymeric liquids. vol. 1: Fluid mechanics} .
	
	\bibitem[Chan \& Leal(1979)]{Chan1979}
	{\sc \au{Chan, P. C.-H.} \& \au{Leal, L.~G.}} \yr{1979}  \at{{The motion of a
			deformable drop in a second-order fluid}}.  \jt{J. Fluid Mech.}
	\bvol{92}~(01),  \pg{131----170}.
	
	\bibitem[Das \& Saintillan(2016)]{Das2016}
	{\sc \au{Das, D.} \& \au{Saintillan, D.}} \yr{2016}  \at{A nonlinear
		small-deformation theory for transient droplet electrohydrodynamics}.  \jt{J.
		Fluid Mech.}  \bvol{810},  \pg{225}.
	
	\bibitem[Eow \& Ghadiri(2002)]{Eow2002}
	{\sc \au{Eow, John~S.} \& \au{Ghadiri, Mojtaba}} \yr{2002}  \at{Electrostatic
		enhancement of coalescence of water droplets in oil: a review of the
		technology}.  \jt{Chemical Engineering Journal}  \bvol{85}~(2),  \pg{357 --
		368}.
	
	\bibitem[Ervik {\em et~al.\/}(2018)Ervik, Penne, Helles{\o}, Munkejord \&
	M{\"u}ller]{Ervik2018}
	{\sc \au{Ervik, {\AA}.}, \au{Penne, T.~E.}, \au{Helles{\o}, S.~M.},
		\au{Munkejord, S.~T.} \& \au{M{\"u}ller, B.}} \yr{2018}  \at{Influence of
		surfactants on the electrohydrodynamic stretching of water drops in oil}.
	\jt{Int. J. Multiphase Flow}  \bvol{98},  \pg{96--109}.
	
	\bibitem[Esmaeeli \& Sharifi(2011)]{Esmaeeli2011}
	{\sc \au{Esmaeeli, Asghar} \& \au{Sharifi, Payam}} \yr{2011}  \at{Transient
		electrohydrodynamics of a liquid drop}.  \jt{Physical Review E}
	\bvol{84}~(3),  \pg{036308}.
	
	\bibitem[Feng(1999)]{Feng1999}
	{\sc \au{Feng, J.~Q.}} \yr{1999}  \at{{Electrohydrodynamic behaviour of a drop
			subjected to a steady uniform electric field at finite electric Reynolds
			number}}.  \jt{Proc. R. Soc. Lond. A}  \bvol{455}~(1986),  \pg{2245--2269}.
	
	\bibitem[Fernández(2008)]{Fernandez2008}
	{\sc \au{Fernández, A.}} \yr{2008}  \at{Response of an emulsion of leaky
		dielectric drops immersed in a simple shear flow: Drops less conductive than
		the suspending fluid}.  \jt{Phys. Fluids}  \bvol{20},  \pg{043304}.
	
	\bibitem[Fischer \& Erni(2007)]{Fischer2007}
	{\sc \au{Fischer, P.} \& \au{Erni, P.}} \yr{2007}  \at{Emulsion drops in
		external flow fields--the role of liquid interfaces}.  \jt{Curr. Opin.
		Colloid Interface Sci.}  \bvol{12}~(4-5),  \pg{196--205}.
	
	\bibitem[Guazzelli \& Morris(2011)]{Guazzelli2011}
	{\sc \au{Guazzelli, E.} \& \au{Morris, J.~F.}} \yr{2011} {\em A physical
		introduction to suspension dynamics\/}, ,  \vol{vol.~45}.  \publ{Cambridge
		University Press}.
	
	\bibitem[Ha {\em et~al.\/}(1999)Ha, Moon \& Yang]{Ha1999}
	{\sc \au{Ha, Jong-Wook}, \au{Moon, Jun~Hyuk} \& \au{Yang, Seung-Man}} \yr{1999}
	\at{Effect of nonionic surfactants on the electrorheology of emulsions}.
	\jt{Korea-Australia Rheology Journal}  \bvol{11}~(3),  \pg{241--246}.
	
	\bibitem[Ha \& Yang(1995)]{Ha1995}
	{\sc \au{Ha, J.-W.} \& \au{Yang, S.-M.}} \yr{1995}  \at{Effects of surfactant
		on the deformation and stability of a drop in a viscous fluid in an electric
		field}.  \jt{J. Colloid Interface Sci.}  \bvol{175}~(2),  \pg{369--385}.
	
	\bibitem[Ha \& Yang(2000)]{Ha2000}
	{\sc \au{Ha, Jong-Wook} \& \au{Yang, Seung-Man}} \yr{2000}  \at{Rheological
		responses of oil-in-oil emulsions in an electric field}.  \jt{Journal of
		Rheology}  \bvol{44}~(2),  \pg{235--256}.
	
	\bibitem[Haber \& Hetsroni(1971)]{Haber1971}
	{\sc \au{Haber, S.} \& \au{Hetsroni, G.}} \yr{1971}  \at{{The dynamics of a
			deformable drop suspended in an unbounded Stokes flow}}.  \jt{J. Fluid Mech.}
	\bvol{49}~(02),  \pg{257--277}.
	
	\bibitem[Haber \& Hetsroni(1972)]{Haber1972}
	{\sc \au{Haber, S.} \& \au{Hetsroni, G.}} \yr{1972}  \at{{Hydrodynamics of a
			drop submerged in an unbounded arbitrary velocity field in the presence of
			surfactants}}.  \jt{Appl. Sci. Res.}  \bvol{25}~(1),  \pg{215--233}.
	
	\bibitem[Haliburton {\em et~al.\/}(2017)Haliburton, Kim, Clark, Sperling, Weitz
	\& Abate]{Haliburton2017}
	{\sc \au{Haliburton, J.~R.}, \au{Kim, S.~C.}, \au{Clark, I.~C.}, \au{Sperling,
			R.~A.}, \au{Weitz, D.~A.} \& \au{Abate, A.~R.}} \yr{2017}  \at{Efficient
		extraction of oil from droplet microfluidic emulsions}.
	\jt{Biomicrofluidics}  \bvol{11}~(3),  \pg{034111}.
	
	\bibitem[Happel \& Brenner(1981)]{Happel1981}
	{\sc \au{Happel, J} \& \au{Brenner, H}} \yr{1981} {\em {Low Reynolds number
			hydrodynamics}\/}.  \publ{Springer}.
	
	\bibitem[Hu \& Lips(2003)]{Hu2003}
	{\sc \au{Hu, Y.} \& \au{Lips, A.}} \yr{2003}  \at{Estimating surfactant surface
		coverage and decomposing its effect on drop deformation}.  \jt{Phys. Rev.
		Lett.}  \bvol{91},  \pg{044501}.
	
	\bibitem[Jeon \& Macosko(2003)]{Jeon2003}
	{\sc \au{Jeon, H.~K.} \& \au{Macosko, C.~W.}} \yr{2003}  \at{Visualization of
		block copolymer distribution on a sheared drop}.  \jt{Polymer}  \bvol{44},
	\pg{5381}.
	
	\bibitem[Kim \& Karila(1991)]{Kim1991}
	{\sc \au{Kim, S} \& \au{Karila, S.~J.}} \yr{1991} {\em Microhydrodynamics:
		Principles and Selected Applications\/}.  \publ{, Butterworth-Heinemann,
		Boston}.
	
	\bibitem[Lanauze {\em et~al.\/}(2015)Lanauze, Walker \& Khair]{Lanauze2015}
	{\sc \au{Lanauze, J.~A.}, \au{Walker, L.~M.} \& \au{Khair, A.~S}} \yr{2015}
	\at{{Nonlinear electrohydrodynamics of slightly deformed oblate drops}}.
	\jt{J. Fluid Mech.}  \bvol{774},  \pg{245--266}.
	
	\bibitem[Leal(2007)]{Leal2007}
	{\sc \au{Leal, L.~G.}} \yr{2007} {\em Advanced transport phenomena: fluid
		mechanics and convective transport processes\/}.  \publ{Cambridge University
		Press}.
	
	\bibitem[Lequeux(1998)]{Lequeux1998}
	{\sc \au{Lequeux, F.}} \yr{1998}  \at{Emulsion rheology}.  \jt{Curr. Opin.
		Colloid Interface Sci.}  \bvol{3}~(4),  \pg{408--411}.
	
	\bibitem[Li \& Pozrikidis(1997)]{Li1997}
	{\sc \au{Li, X.} \& \au{Pozrikidis, C.}} \yr{1997}  \at{{The effect of
			surfactants on drop deformation and on the rheology of dilute emulsions in
			Stokes flow}}.  \jt{J. Fluid Mech.}  \bvol{341}~(1997),  \pg{165--194}.
	
	\bibitem[Maehlmann \& Papageorgiou(2009)]{Maehlmann2009}
	{\sc \au{Maehlmann, Stefan} \& \au{Papageorgiou, Demetrios~T}} \yr{2009}
	\at{Numerical study of electric field effects on the deformation of
		two-dimensional liquid drops in simple shear flow at arbitrary reynolds
		number}.  \jt{Journal of Fluid Mechanics}  \bvol{626},  \pg{367--393}.
	
	\bibitem[Mandal {\em et~al.\/}(2016{\natexlab{{\em a\/}}})Mandal, Bandopadhyay
	\& Chakraborty]{Mandal2016a}
	{\sc \au{Mandal, S.}, \au{Bandopadhyay, A.} \& \au{Chakraborty, S.}}
	\yr{2016{\natexlab{{\em a\/}}}}  \at{{Dielectrophoresis of a surfactant-laden
			viscous drop}}.  \jt{Phys. Fluids}  \bvol{28}~(6).
	
	\bibitem[Mandal {\em et~al.\/}(2016{\natexlab{{\em b\/}}})Mandal, Bandopadhyay
	\& Chakraborty]{Mandal2016}
	{\sc \au{Mandal, S.}, \au{Bandopadhyay, A.} \& \au{Chakraborty, S.}}
	\yr{2016{\natexlab{{\em b\/}}}}  \at{{The effect of uniform electric field on
			the cross-stream migration of a drop in plane Poiseuille flow}}.  \jt{J.
		Fluid Mech.}  \bvol{809}~(2016),  \pg{726--774}.
	
	\bibitem[Mandal {\em et~al.\/}(2017{\natexlab{{\em a\/}}})Mandal, Chakrabarti
	\& Chakraborty]{Mandal2017}
	{\sc \au{Mandal, S.}, \au{Chakrabarti, S.} \& \au{Chakraborty, S.}}
	\yr{2017{\natexlab{{\em a\/}}}}  \at{Effect of nonuniform electric field on
		the electrohydrodynamic motion of a drop in poiseuille flow}.  \jt{Physics of
		Fluids}  \bvol{29}~(5),  \pg{052006}.
	
	\bibitem[Mandal \& Chakraborty(2017{\natexlab{{\em a\/}}})]{Mandal2017b}
	{\sc \au{Mandal, S.} \& \au{Chakraborty, S.}} \yr{2017{\natexlab{{\em a\/}}}}
	\at{Effect of uniform electric field on the drop deformation in simple shear
		flow and emulsion shear rheology}.  \jt{Physics of Fluids}  \bvol{29}~(7),
	\pg{072109}.
	
	\bibitem[Mandal \& Chakraborty(2017{\natexlab{{\em b\/}}})]{Mandal2017a}
	{\sc \au{Mandal, S.} \& \au{Chakraborty, S.}} \yr{2017{\natexlab{{\em b\/}}}}
	\at{Uniform electric-field-induced non-newtonian rheology of a dilute
		suspension of deformable newtonian drops}.  \jt{Phys. Rev. Fluids}  \bvol{2},
	\pg{093602}.
	
	\bibitem[Mandal {\em et~al.\/}(2017{\natexlab{{\em b\/}}})Mandal, Das \&
	Chakraborty]{Mandal2017c}
	{\sc \au{Mandal, S.}, \au{Das, S.} \& \au{Chakraborty, S.}}
	\yr{2017{\natexlab{{\em b\/}}}}  \at{Effect of marangoni stress on the bulk
		rheology of a dilute emulsion of surfactant-laden deformable droplets in
		linear flows}.  \jt{Phys. Rev. Fluids}  \bvol{2},  \pg{113604}.
	
	\bibitem[Melcher \& Taylor(1969)]{Melcher1969}
	{\sc \au{Melcher, J.~R.} \& \au{Taylor, G.~I.}} \yr{1969}
	\at{Electrohydrodynamics: a review of the role of interfacial shear
		stresses}.  \jt{Annu. Rev. Fluid Mech.}  \bvol{1}~(1),  \pg{111--146}.
	
	\bibitem[Milliken {\em et~al.\/}(1993)Milliken, Stone \& Leal]{Milliken1993}
	{\sc \au{Milliken, W.~J.}, \au{Stone, H.~A.} \& \au{Leal, L.~G.}} \yr{1993}
	\at{The effect of surfactant on transient motion of newtonian drops}.
	\jt{Phys. Fluids A}  \bvol{5},  \pg{69}.
	
	\bibitem[Na {\em et~al.\/}(2009)Na, Aida, Sakai, Kakuchi \& Orihara]{Na2009}
	{\sc \au{Na, Yang~Ho}, \au{Aida, Kohei}, \au{Sakai, Ryosuke}, \au{Kakuchi,
			Toyoji} \& \au{Orihara, Hiroshi}} \yr{2009}  \at{Response of shear stress to
		ac electric fields under steady shear flow in a droplet-dispersed phase}.
	\jt{Physical Review E}  \bvol{80}~(6),  \pg{061803}.
	
	\bibitem[Nganguia {\em et~al.\/}(2013)Nganguia, Young, Vlahovska,
	B{\l}awzdziewcz, Zhang \& Lin]{Nganguia2013}
	{\sc \au{Nganguia, H.}, \au{Young, Y.~N.}, \au{Vlahovska, P.~M.},
		\au{B{\l}awzdziewcz, J.}, \au{Zhang, J.} \& \au{Lin, H.}} \yr{2013}
	\at{{Equilibrium electro-deformation of a surfactant-laden viscous drop}}.
	\jt{Phys. Fluids}  \bvol{25}~(9).
	
	\bibitem[Pak {\em et~al.\/}(2014)Pak, Feng \& Stone]{Pak2014}
	{\sc \au{Pak, O.~S.}, \au{Feng, J.} \& \au{Stone, H.~A}} \yr{2014}  \at{Viscous
		marangoni migration of a drop in a poiseuille flow at low surface p{\'e}clet
		numbers}.  \jt{J. Fluid Mech.}  \bvol{753},  \pg{535--552}.
	
	\bibitem[Pan \& McKinley(1997)]{Pan1997}
	{\sc \au{Pan, X.~D.} \& \au{McKinley, G.~H.}} \yr{1997}  \at{Characteristics of
		electrorheological responses in an emulsion system}.  \jt{J. Colloid
		Interface Sci.}  \bvol{195},  \pg{101}.
	
	\bibitem[Pawar \& Stebe(1996)]{Pawar1996}
	{\sc \au{Pawar, Y.} \& \au{Stebe, K.~J.}} \yr{1996}  \at{Marangoni effects on
		drop deformation in an extensional flow: The role of surfactant physical
		chemistry. i. insoluble surfactants}.  \jt{Phys. Fluids}  \bvol{8}~(7),
	\pg{1738--1751}.
	
	\bibitem[Poddar {\em et~al.\/}(2018)Poddar, Mandal, Bandopadhyay \&
	Chakraborty]{Poddar2018}
	{\sc \au{Poddar, A.}, \au{Mandal, S.}, \au{Bandopadhyay, A.} \&
		\au{Chakraborty, S.}} \yr{2018}  \at{Sedimentation of a surfactant-laden drop
		under the influence of an electric field}.  \jt{J. Fluid Mech.}  \bvol{849},
	\pg{277--311}.
	
	\bibitem[Puyvelde {\em et~al.\/}(2001)Puyvelde, Velankar \&
	Moldenaers]{Puyvelde2001}
	{\sc \au{Puyvelde, Peter~Van}, \au{Velankar, Sachin} \& \au{Moldenaers, Paula}}
	\yr{2001}  \at{Rheology and morphology of compatibilized polymer blends}.
	\jt{Current Opinion in Colloid \& Interface Science}  \bvol{6}~(5),  \pg{457
		-- 463}.
	
	\bibitem[Rallison(1984)]{Rallison1984}
	{\sc \au{Rallison, J.~M.}} \yr{1984}  \at{The deformation of small viscous
		drops and bubbles in shear flows}.  \jt{Annu. Rev. Fluid Mech.}  \bvol{16},
	\pg{45}.
	
	\bibitem[Salipante \& Vlahovska(2010)]{Salipante2010}
	{\sc \au{Salipante, P.~F.} \& \au{Vlahovska, P.~M.}} \yr{2010}
	\at{Electrohydrodynamics of drops in strong uniform dc electric fields}.
	\jt{Phys. Fluids}  \bvol{22}~(11),  \pg{112110}.
	
	\bibitem[Saville(1997)]{Saville1997}
	{\sc \au{Saville, D.~A.}} \yr{1997}  \at{Electrohydrodynamics: the
		taylor-melcher leaky dielectric model}.  \jt{Annu. Rev. Fluid Mech.}
	\bvol{29}~(1),  \pg{27--64}.
	
	\bibitem[Schowalter {\em et~al.\/}(1968)Schowalter, Chaffey \&
	Brenner]{Schowalter1968}
	{\sc \au{Schowalter, W.~R.}, \au{Chaffey, C.~E.} \& \au{Brenner, H.}} \yr{1968}
	\at{Rheological behavior of a dilute emulsion}.  \jt{Journal of colloid and
		interface science}  \bvol{26}~(2),  \pg{152--160}.
	
	\bibitem[Stone(1990)]{Stone1990}
	{\sc \au{Stone, H.~A.}} \yr{1990}  \at{A simple derivation of the
		time-dependent convective-diffusion equation for surfactant transport along a
		deforming interface}.  \jt{Phys. Fluids A}  \bvol{2},  \pg{111}.
	
	\bibitem[Stone \& Leal(1990)]{Stone1990a}
	{\sc \au{Stone, H.~A.} \& \au{Leal, L.~G.}} \yr{1990}  \at{{The effects of
			surfactants on drop deformation and breakup}}.  \jt{J. Fluid Mech.}
	\bvol{220},  \pg{161}.
	
	\bibitem[Taylor(1966)]{Taylor1966}
	{\sc \au{Taylor, G.}} \yr{1966}  \at{Studies in electrohydrodynamics. i. the
		circulation produced in a drop by an electric field}.  \jt{Proc. R. Soc.
		Lond. A}  \bvol{291}~(1425),  \pg{159--166}.
	
	\bibitem[Taylor(1932)]{Taylor1932}
	{\sc \au{Taylor, G.~I.}} \yr{1932}  \at{The viscosity of a fluid containing
		small drops of another fluid}.  \jt{Proc. R. Soc. Lond. A}  \bvol{138}~(834),
	\pg{41--48}.
	
	\bibitem[Teigen \& Munkejord(2010)]{Teigen2010}
	{\sc \au{Teigen, K.} \& \au{Munkejord, S.~T.}} \yr{2010}  \at{{Influence of
			surfactant on drop deformation in an electric field}}.  \jt{Phys. Fluids}
	\bvol{22}~(11).
	
	\bibitem[Thaokar(2012)]{Thaokar2012}
	{\sc \au{Thaokar, RM}} \yr{2012}  \at{Dielectrophoresis and deformation of a
		liquid drop in a non-uniform, axisymmetric ac electric field}.  \jt{The
		European Physical Journal E}  \bvol{35}~(8),  \pg{76}.
	
	\bibitem[Tsukada {\em et~al.\/}(1993)Tsukada, Katayama, Ito \&
	Hozawa]{Tsukada1993}
	{\sc \au{Tsukada, T.}, \au{Katayama, T.}, \au{Ito, Y.} \& \au{Hozawa, M.}}
	\yr{1993}  \at{{Theoretical and Experimental Studies of Circulations Inside
			and Outside a Deformed Drop under a Uniform Electric Field.}}  \jt{J. Chem.
		Eng. Jpn.}  \bvol{26}~(6),  \pg{698--703}.
	
	\bibitem[Tucker \& Moldenaers(2002)]{Tucker2002}
	{\sc \au{Tucker, C.} \& \au{Moldenaers, P.}} \yr{2002}  \at{Microstructural
		evolution in polymer blends}.  \jt{Annu. Rev. Fluid Mech.}  \bvol{34},
	\pg{177}.
	
	\bibitem[Valkovska \& Danov(2000)]{Valkovska2000}
	{\sc \au{Valkovska, D.~S.} \& \au{Danov, K.~D.}} \yr{2000}  \at{Determination
		of bulk and surface diffusion coefficients from experimental data for thin
		liquid film drainage}.  \jt{Journal of colloid and interface science}
	\bvol{223}~(2),  \pg{314--316}.
	
	\bibitem[Velankar {\em et~al.\/}(2004)Velankar, Van~Puyvelde, Mewis \&
	Moldenaers]{Velankar2004}
	{\sc \au{Velankar, S.}, \au{Van~Puyvelde, P.}, \au{Mewis, J.} \&
		\au{Moldenaers, P.}} \yr{2004}  \at{Steady shear rheological properties of
		model compatibilized blends}.  \jt{J. Rheol.}  \bvol{48},  \pg{725}.
	
	\bibitem[Velankar {\em et~al.\/}(2001)Velankar, Van~Pyuvede, Mewis \&
	Moldenaers]{Velankar2001}
	{\sc \au{Velankar, S.}, \au{Van~Pyuvede, P.}, \au{Mewis, J.} \& \au{Moldenaers,
			P.}} \yr{2001}  \at{Effect of compatibilization on the breakup of polymeric
		drops in shear flow}.  \jt{J. Rheol.}  \bvol{45},  \pg{1007}.
	
	\bibitem[Vlahovska(2011)]{Vlahovska2011}
	{\sc \au{Vlahovska, P.~M.}} \yr{2011}  \at{On the rheology of a dilute emulsion
		in a uniform electric field}.  \jt{J. Fluid Mech.}  \bvol{670},
	\pg{481--503}.
	
	\bibitem[Vlahovska(2019)]{Vlahovska2019}
	{\sc \au{Vlahovska, P.~M.}} \yr{2019}  \at{Electrohydrodynamics of drops and
		vesicles}.  \jt{Annu. Rev. Fluid Mech.}  \bvol{51}~(1),  \pg{null}.
	
	\bibitem[Vlahovska {\em et~al.\/}(2009)Vlahovska, B{\l}awzdziewicz \&
	Loewenberg]{Vlahovska2009}
	{\sc \au{Vlahovska, P.~M.}, \au{B{\l}awzdziewicz, J.} \& \au{Loewenberg, M.}}
	\yr{2009}  \at{Small-deformation theory for a surfactant-covered drop in
		linear flows}.  \jt{J. Fluid Mech.}  \bvol{624},  \pg{293--337}.
	
	\bibitem[Vlahovska {\em et~al.\/}(2005)Vlahovska, Loewenberg \&
	Blawzdziewicz]{Vlahovska2005}
	{\sc \au{Vlahovska, P. M.}, \au{Loewenberg, M.} \& \au{Blawzdziewicz, J.}} \yr{2005}  \at{Deformation of a surfactant-covered drop in a linear
		flow}.  \jt{Physics of Fluids}  \bvol{17}~(10),  \pg{103103}.
	
	\bibitem[Xu {\em et~al.\/}(2006)Xu, Li, Lowengrub \& Zhao]{Xu2006}
	{\sc \au{Xu, J.~J.}, \au{Li, Z.}, \au{Lowengrub, J.} \& \au{Zhao, H.}}
	\yr{2006}  \at{{A level-set method for interfacial flows with surfactant}}.
	\jt{J. Comput. Phys.}  \bvol{212}~(2),  \pg{590--616}.
	
	\bibitem[Xu \& Homsy(2006)]{Xu2006a}
	{\sc \au{Xu, X.} \& \au{Homsy, G.~M.}} \yr{2006}  \at{{The settling velocity
			and shape distortion of drops in a uniform electric field}}.  \jt{J. Fluid
		Mech.}  \bvol{564},  \pg{395}.
	
	\bibitem[Yariv \& Almog(2016)]{Yariv2016}
	{\sc \au{Yariv, E.} \& \au{Almog, Y.}} \yr{2016}  \at{The effect of
		surface-charge convection on the settling velocity of spherical drops in a
		uniform electric field}.  \jt{J. Fluid Mech.}  \bvol{797},  \pg{536--548}.
\end{thebibliography}
\end{document}